\documentclass[aps,prd,12pt,preprint,superscriptaddress,preprintnumbers,nofootinbib,floatfix,longbibliography]{revtex4-1}

\usepackage[T1]{fontenc}
\usepackage[utf8]{inputenc}
\usepackage[english]{babel}
\usepackage{amsmath,amsthm,amssymb,amsfonts,mathrsfs,amsbsy,bm}
\usepackage{comment}
\usepackage{tensor}
\usepackage{slashed}
\usepackage{float}
 \usepackage{mathrsfs,tikz} 
\usepackage{esint}
\usepackage[a4paper, margin=1.5cm]{geometry}
\usepackage{cancel}
\usepackage{graphicx}
\usepackage{multirow}
\usepackage{array}
\usepackage{booktabs}
\usepackage{makecell}
\usepackage{xcolor}
\colorlet{BLUE}{blue}
\usepackage{tikz}
\usetikzlibrary{quotes,angles,arrows,decorations.markings,decorations.pathmorphing}
\usepackage{hyperref}
\hypersetup{colorlinks=true,breaklinks=true,citecolor=blue,linkcolor=[rgb]{0,0.5,0.9},urlcolor=blue}
\graphicspath{{figures/}}

\newcommand{\be}{\begin{equation}}
\newcommand{\ee}{\end{equation}}
\newcommand{\Be}{\begin{eqnarray}}
\newcommand{\Ee}{\end{eqnarray}}

\newcommand{\mincir}{\raise-3.truept\hbox{\rlap{\hbox{$\sim$}}\raise4.truept\hbox{$<$}\ }}
\newcommand{\magcir}{\raise-3.truept\hbox{\rlap{\hbox{$\sim$}}\raise4.truept\hbox{$>$}\ }}

\providecommand{\U}[1]{}

\newcommand{\ie}{\begin{equation}}
\newcommand{\fe}{\end{equation}}
\newcommand{\se}{\begin{eqnarray}}
\newcommand{\ff}{\end{eqnarray}}

\begin{document}
\emergencystretch=4em
\usetikzlibrary{arrows.meta,patterns}


\title{Born–Infeld black holes in bumblebee gravity}

\author{A. A. Ara\'{u}jo Filho}
\email{dilto@fisica.ufc.br}
\affiliation{Departamento de F\'isica, Universidade Federal da Para\'iba, Caixa Postal 5008, 58051--970, Jo\~ao Pessoa, Para\'iba, Brazil.}
\affiliation{Departamento de F\'isica, Universidade Federal de Campina Grande, Caixa Postal 10071, 58429--900 Campina Grande, Para\'iba, Brazil.}
\affiliation{Center for Theoretical Physics, Khazar University, 41 Mehseti Street, Baku, AZ-1096, Azerbaijan.}

\author{E. M. Siquieri}
\email{eduardosiquieri@fisica.ufmt.br}
\affiliation{Programa de Pós-graduação em Física, Instituto de Física, Universidade Federal de Mato Grosso, Cuiabá, Brasil}

\author{N.~Heidari}
\email{heidari.n@gmail.com }
\affiliation{Departamento de Física, Universidade Federal de Campina Grande Caixa Postal 10071, 58429-900 Campina Grande, Paraíba, Brazil.}
\affiliation{Center for Theoretical Physics, Khazar University, 41 Mehseti Street, Baku, AZ-1096, Azerbaijan.}

\author{V. B. Bezerra}
\email{valdir@fisica.ufpb.br}
\affiliation{Departamento de Física, Universidade Federal da Paraíba, Caixa Postal 5008, 58051-970, João Pessoa, Paraíba,  Brazil.}

\author{A. F. Santos}
\email{alesandroferreira@fisica.ufmt.br}
\affiliation{Programa de Pós-graduação em Física, Instituto de Física, Universidade Federal de Mato Grosso, Cuiabá, Brasil}


\date{\today}


\begin{abstract}

We construct exact static and spherically symmetric charged solutions in bumblebee gravity minimally coupled to Born--Infeld electrodynamics, taking into account a spacelike vector norm and a vanishing cosmological constant. An independently varied Lagrange multiplier enforces the norm constraint and acquires a radial profile determined by the transverse electromagnetic pressure. The solutions include purely radial configurations and configurations with constant or varying temporal amplitudes. At $\xi=\kappa/2$, the varying temporal amplitude introduces an independent bumblebee field strength while preserving the metric of the constant temporal configuration. Moreover, the electric field remains bounded, and the temporal configurations admit smooth extensions across a simple future horizon under explicit regularity conditions. We derive the photon sphere and shadow radii for both metric and nonlinear electromagnetic propagation. The leading nonlinear corrections enlarge the shadow at fixed geometric charge, whereas the temporal amplitude enters through the charged lapse. Comparisons with observations of Sgr~A$^*$ and M87$^*$ provide conditional bounds on the charge and nonlinear scale. In addition, the adopted planetary perihelion allowances yield an illustrative sensitivity of $|\xi|(b^2+u_0^2)\lesssim2.6\times10^{-12}$ for a neutral solar exterior.

\end{abstract}

\maketitle

\tableofcontents


\section{Introduction}
\label{sec:introduction}

Lorentz symmetry can be tested gravitationally by allowing the vacuum to carry a preferred direction whose effects follow from a covariant action. The possibility that such a direction originates from spontaneous symmetry breaking was developed in string theory and subsequently incorporated into the gravitational sector of the Standard Model Extension \cite{KosteleckySamuel1989,Kostelecky2004}. Bumblebee models, on the other hand, realize this possibility with a vector field whose expectation value is selected by a potential and whose coupling to curvature affects the geometry \cite{BluhmKostelecky2005,BluhmFungKostelecky2008}. In this setting, black hole solutions make it possible to relate the orientation and dynamics of the vector to horizons, particle motion, and the propagation of light \cite{Filho:2023etf,AraujoFilho:2025zaj,Filho:2024isd}.

Early vacuum investigations considered radial configurations together with fields possessing temporal components \cite{BertolamiParamos2005}. An exact spherical solution was later obtained by Casana et al., with a Schwarzschild temporal metric coefficient and a radial deformation controlled by the vector expectation value \cite{Casana2018}. Maluf and Neves extended this construction to a nonvanishing cosmological constant \cite{MalufNeves2021}, while solutions with a global monopole introduced an additional solid angle deficit \cite{GulluOvgun2022}. Furthermore, exact geometries have been obtained in higher dimensions and in the rotating BTZ setting \cite{DingHigher2023,DingBTZ2023,Valtancoli2023}.

Allowing more general vector configurations enlarges the spherical solution space. Xu, Liang, and Shao obtained analytical special cases and numerical black holes with different radial conditions \cite{XuLiangShao2023}. Purely timelike expectation values were examined by Li and Zhu \cite{LiZhuTimelike2026}, whereas Liu et al. constructed neutral and electrically charged geometries with temporal and radial components and either lightlike or spacelike norms \cite{LiuWu2026}. The classification by Zhu and Li further identified exceptional parameter choices and configurations in which a nontrivial vector accompanies a Schwarzschild metric \cite{ZhuLi2026}. Solutions away from the potential minimum and extensions containing both $B^2R$ and $B^\mu B^\nu R_{\mu\nu}$ interactions make explicit the dependence of these results on the potential and the curvature couplings \cite{BaileyMurrayWalter2025,ZhuLiExtended2026}.

Rotation requires particular care in this theory. Early proposals included geometries modeled on Kerr and Kerr Sen \cite{DingKerr2020,JhaRahaman2021}; however, the vector configuration accompanying the original Kerr proposal does not satisfy its field equation at general rotation \cite{MalufMuniz2022}. Ding and Chen obtained a slowly rotating geometry by solving the equations to first order in angular momentum \cite{DingChen2021}. Moreover, Poulis and Soares developed a method for generating exact vacuum solutions with a gradient bumblebee field \cite{PoulisSoares2022}. Exact Taub NUT geometries have also been derived \cite{ChenLiu2025}, and Ovcharenko extended the generating construction to include electromagnetic fields and a cosmological constant, obtaining counterparts of the Kerr Newman Taub NUT family \cite{Ovcharenko2026}.

Independent variations of the metric and connection have yielded spherical and stationary axisymmetric vacuum solutions in metric affine bumblebee gravity \cite{AraujoMetricAffine2023,AraujoAxisymmetric2024} (with consequences in particle creation \cite{AraujoFilho:2025hkm}, accreation of matter \cite{Lambiase:2023zeo} and scattering effects \cite{Heidari:2024bvd}). A more general spherical configuration with temporal and radial components has recently been obtained in the traceless formulation \cite{AraujoGeneral2026} with consequences in lensing phenomena \cite{al2026shadow}. Other extensions incorporate a Gauss Bonnet curvature term or a noncommutative deformation \cite{DingGaussBonnet2022,AraujoNoncommutative2026}.

For charged configurations, an independent electromagnetic field introduces a distinction between electric charge and an amplitude associated with the bumblebee field. Liu et al. constructed spherical and slowly rotating solutions by coupling Maxwell electrodynamics directly to the bumblebee vector, both with and without a cosmological constant \cite{LiuGuo2025}. The subsequent solutions with two vector components retained this electromagnetic interaction \cite{LiuWu2026}. Charged geometries also arise in models containing a dilaton, including asymptotically Lifshitz solutions \cite{LessaSilva2023,LessaSilva2024}, and in the presence of a global monopole \cite{AhmedKalaSilva2026}.

Static and slowly rotating bumblebee solutions have also been studied with ordinary and phantom ModMax electrodynamics \cite{SekhmaniModMax2026,SekhmaniPhantom2025}, as well as with a cloud of strings \cite{BelchiorModMax2026}. Magnetic nonlinear electrodynamics, together with a phantom scalar, has been used to support an exact Simpson Visser family containing black bounce geometries and wormholes \cite{AraujoSimpsonVisser2026}.

Born--Infeld electrodynamics bounds the electrostatic field of a point charge and renders its electromagnetic self energy finite; the Maxwell theory is recovered as the nonlinear scale tends to infinity \cite{BornInfeld1934}. Its appearance in the effective dynamics of open strings gives an additional motivation \cite{FradkinTseytlin1985,Tseytlin1999}. The gravitational problem was investigated by Hoffmann \cite{Hoffmann1935}, and exact solutions of the Einstein and Born--Infeld equations were developed by García, Salazar, and Plebański \cite{GarciaSalazarPlebanski1984}. Further studies established charged black holes with a cosmological constant, their horizon structures, and extensions to arbitrary dimension \cite{FernandoKrug2003,Dey2004,CaiPangWang2004}. Particle trajectories in these geometries were examined by Bretón \cite{Breton2002}.

Within bumblebee gravity, Li and Lü recently obtained exact charged solutions for a class of nonlinear electromagnetic theories through an auxiliary Maxwell scalar description \cite{LiLu2026}. It also exhibits the separate roles of the mass singularity and the central defect associated with the conical vacuum. The existence of these solutions motivates examining, among other aspects, how a different variational implementation of the vector norm changes the admissible electromagnetic coupling and the resulting charged geometries.

A Lagrange multiplier is especially relevant to this question. Enforcing the vector norm makes the multiplier potential vanish on the solution, while its derivative with respect to that norm can still contribute to the field equations \cite{BluhmFungKostelecky2008}. Setting this derivative to zero in advance imposes an additional restriction. Keeping the multiplier as an independent field allows its value to be determined together with the geometry and matter distribution. Furthermore, a prescribed spacelike norm does not require (madatorally) the temporal component to vanish. Taking both components into account permits us to examine whether the vector itself extends smoothly across a metric horizon, a condition that cannot be inferred from its norm alone.

The optical analysis must also contain the distinction between the spacetime metric and the characteristics of the electromagnetic theory. In nonlinear electrodynamics, electromagnetic disturbances generally propagate along null directions of effective metrics determined by the background field \cite{Novello2000,Abalos2015}. The full Born--Infeld theory has a common characteristic cone for its two polarizations \cite{GibbonsHerdeiro2001}. Restricting its action to dependence on $F_{\mu\nu}F^{\mu\nu}$ alone can preserve the electric background while changing the polarization characteristics: the second electromagnetic invariant vanishes on that background, but its squared contribution still enters the second variation of the action. In this sense, a shadow calculation must specify the electromagnetic theory beyond its background solution. Comparing nonlinear electromagnetic propagation with a separate Maxwell probe identifies which optical effects arise from the geometry and which arise from the electromagnetic response.

The Event Horizon Telescope results for M87* and Sgr A* permit comparisons between predicted capture boundaries and the observed emission structure \cite{EHTM872019,EHTSgr2022}, with applications to charged spacetimes and bumblebee black holes \cite{Kocherlakota2021,XuEHT2023}. Such comparisons require an emission model and independent information on mass and distance; the angular shadow also depends on the observer and the asymptotic geometry \cite{PerlickTsupko2022}. Moreover, stellar orbits near Sgr A* and planetary ephemerides supply complementary information through the motion of massive bodies \cite{GRAVITY2024,FiengaMinazzoli2024}.

In this work, we solve the metric bumblebee equations in the presence of an independent, minimally coupled Born--Infeld field, obtaining exact charged geometries with static spherical symmetry. The vector norm is spacelike and the cosmological constant vanishes. We determine the multiplier from the field equations and obtain radial configurations together with configurations possessing constant or varying temporal amplitudes. At $\xi=\kappa/2$, the varying amplitude supports a nonzero bumblebee field strength without altering the metric obtained for a constant temporal component. We establish the conditions under which the temporal configurations admit a smooth continuation through a simple future horizon and identify their neutral and Maxwell limits. We then derive geodesics, photon spheres and shadow radii for metric and nonlinear electromagnetic propagation. Comparisons with M87*, Sgr A*, stellar precession, and Solar System observables reveal which parameter combinations each measurement can probe.



\section{Bumblebee gravity coupled to Born--Infeld electrodynamics}
\label{sec:theory}

Spontaneous Lorentz symmetry breaking gives a geometric setting in which a dynamical field acquires a nonvanishing vacuum expectation value and selects a preferred direction in spacetime. Bumblebee gravity realizes this mechanism through a vector field whose norm is controlled by a potential, while its interaction with curvature modifies the gravitational equations~\cite{KosteleckySamuel1989,Kostelecky2004,BluhmKostelecky2005}. We consider this vector coupled to an independent electromagnetic field governed by Born--Infeld electrodynamics~\cite{BornInfeld1934}. The distinction between the two fields allows the electromagnetic charge and the bumblebee configuration to enter the solution as separate physical quantities.

We work in four dimensions with signature $(-,+,+,+)$, set $c=\hbar=1$, and write $\kappa=8\pi G$. The connection is the Levi--Civita connection of the metric. With a vanishing cosmological constant, the action is
\begin{equation}
S=\int \mathrm{d}^{4}x\sqrt{-g}\left[\frac{R+\xi B^{\mu}B^{\nu}R_{\mu\nu}}{2\kappa}-\frac14 H_{\mu\nu}H^{\mu\nu}-\frac{\lambda}{2}\left(B_{\mu}B^{\mu}-b^{2}\right)+\mathcal L_{\mathrm{BI}}\right],
\label{eq:action}
\end{equation}
where $b^{2}>0$, $\xi$ is the nonminimal gravitational coupling, and $\lambda(x)$ is an independent auxiliary scalar. The antisymmetric tensors and the electromagnetic Lagrangian are defined by
\begin{equation}
\begin{aligned}
H_{\mu\nu}&=2\nabla_{[\mu}B_{\nu]},\qquad F_{\mu\nu}=2\nabla_{[\mu}A_{\nu]},\qquad \mathcal I=F_{\mu\nu}F^{\mu\nu}, \\ \mathcal L_{\mathrm{BI}}&=\beta_{\mathrm{BI}}^{2}\left(1-s\right),\qquad s=\sqrt{1+\frac{\mathcal I}{2\beta_{\mathrm{BI}}^{2}}}.
\end{aligned}
\label{eq:definitions}
\end{equation}
Here, $\beta_{\mathrm{BI}}>0$ sets the nonlinear electromagnetic scale and is unrelated to any parametrization of the temporal bumblebee amplitude. Equation~\eqref{eq:definitions} is the electric restriction of Born--Infeld theory. The additional invariant $F_{\mu\nu}\widetilde F^{\mu\nu}$ of the determinant formulation vanishes for the configurations studied below, and its squared contribution has vanishing first variation on this background. In particular, the electromagnetic normalization is fixed by
\begin{equation}
\mathcal L_{\mathrm{BI}}=-\frac14\mathcal I+\frac{\mathcal I^{2}}{32\beta_{\mathrm{BI}}^{2}}+\mathcal O\!\left(\beta_{\mathrm{BI}}^{-4}\right).
\label{eq:maxwell-lagrangian}
\end{equation}

This convention makes the Maxwell limit explicit and fixes the charge coefficient when comparing the solution with Einstein gravity coupled to nonlinear electrodynamics~\cite{CaiPangWang2004}. The electromagnetic sector couples minimally to $g_{\mu\nu}$, while the direct interaction between the bumblebee field and the Ricci tensor is retained in Eq.~\eqref{eq:action}.

The independent variations are performed with respect to $g^{\mu\nu}$, $B_{\mu}$, $A_{\mu}$, and $\lambda$ before imposing the norm constraint. Defining $X=B_{\alpha}B^{\alpha}-b^{2}$, the metric equation takes the form
\begin{equation}
G_{\mu\nu}=\kappa\left(T^{H}_{\mu\nu}+T^{\mathrm{BI}}_{\mu\nu}+\lambda B_{\mu}B_{\nu}-\frac{\lambda X}{2}g_{\mu\nu}\right)+\xi C_{\mu\nu},
\label{eq:metric-eom}
\end{equation}
with
\begin{equation}
T^{H}_{\mu\nu}=H_{\mu\alpha}H_{\nu}{}^{\alpha}-\frac14g_{\mu\nu}H_{\alpha\rho}H^{\alpha\rho},\qquad T^{\mathrm{BI}}_{\mu\nu}=\frac{F_{\mu\alpha}F_{\nu}{}^{\alpha}}{s}+g_{\mu\nu}\mathcal L_{\mathrm{BI}}.
\label{eq:stress-tensors}
\end{equation}

The curvature interaction contributes both algebraic and differential terms,
\begin{equation}
\begin{aligned}
C_{\mu\nu}={}&\frac12g_{\mu\nu}B^{\alpha}B^{\rho}R_{\alpha\rho}-B_{\mu}B^{\alpha}R_{\alpha\nu}-B_{\nu}B^{\alpha}R_{\alpha\mu} \\ &+\frac12\nabla_{\alpha}\nabla_{\mu}\left(B^{\alpha}B_{\nu}\right)+\frac12\nabla_{\alpha}\nabla_{\nu}\left(B^{\alpha}B_{\mu}\right)-\frac12\Box\left(B_{\mu}B_{\nu}\right) \\ &-\frac12g_{\mu\nu}\nabla_{\alpha}\nabla_{\rho}\left(B^{\alpha}B^{\rho}\right),\qquad \Box=\nabla_{\alpha}\nabla^{\alpha}.
\end{aligned}
\label{eq:curvature-tensor}
\end{equation}

These terms arise from the metric dependence of the raised vector indices and from the variation of the Ricci tensor. Their derivative ordering is retained explicitly in Eq.~\eqref{eq:curvature-tensor}, which also fixes the conventions used in the radial calculation~\cite{BertolamiParamos2005,Casana2018,MalufNeves2021}. The bumblebee and multiplier equations are
\begin{equation}
\nabla_{\mu}H^{\mu\nu}+\frac{\xi}{\kappa}B^{\mu}R_{\mu}{}^{\nu}-\lambda B^{\nu}=0,\qquad B_{\mu}B^{\mu}=b^{2},
\label{eq:bumblebee-eom}
\end{equation}
whereas the electromagnetic equations read
\begin{equation}
\nabla_{\mu}\left(\frac{F^{\mu\nu}}{s}\right)=0,\qquad \nabla_{[\alpha}F_{\mu\nu]}=0.
\label{eq:electromagnetic-eom}
\end{equation}

The linear multiplier potential imposes the norm while allowing its derivative to participate in the field equations~\cite{BluhmFungKostelecky2008}. Indeed, on a solution, we obetain
\begin{equation}
V(X,\lambda)=\frac{\lambda X}{2}=0,\qquad \left.\frac{\partial V}{\partial X}\right|_{X=0}=\frac{\lambda}{2}.
\label{eq:potential-derivative}
\end{equation}

In this regard, the value of the potential and its response to changes of the norm carry different information. In the charged solution, $\lambda(r)$ is determined by the curvature and the transverse electromagnetic pressure. This mechanism supplies the required bumblebee equation with minimal electromagnetic coupling. Related constructions with nonlinear electrodynamics and direct bumblebee couplings to matter have been developed by Li and L\"u~\cite{LiLu2026}; the present action instead contains the independently varied multiplier of Eq.~\eqref{eq:action}.


\section{Static charged solutions with temporal and radial bumblebee components}
\label{sec:solution}

\subsection{Common spherical equations and electromagnetic field}
\label{subsec:common-equations}

We seek static and spherically symmetric configurations within the metric class
\begin{equation}
\begin{aligned}
\mathrm{d}s^{2}&=-f(r)\mathrm{d}t^{2}+\frac{q}{f(r)}\mathrm{d}r^{2}+r^{2}\mathrm{d}\Omega^{2},\qquad \mathrm{d}\Omega^{2}=\mathrm{d}\theta^{2}+\sin^{2}\theta\,\mathrm{d}\varphi^{2}, \\ A&=a(r)\mathrm{d}t,\qquad B=u(r)\mathrm{d}t+v(r)\mathrm{d}r,
\end{aligned}
\label{eq:ansatz}
\end{equation}
where $q>0$ is constant and $r$ is the areal radius. The restriction $-g_{tt}g_{rr}=q$ defines the class being solved. Initially, the calculation is performed in a static exterior with $r>0$ and $f(r)>0$. Both components of the bumblebee expectation value are retained, as motivated by the broader spherical configurations studied in Refs.~\cite{XuLiangShao2023,ZhuLi2026,LiuWuWeiLiu2026}. The spacelike norm determines the radial component directly:
\begin{equation}
-\frac{u(r)^{2}}{f(r)}+\frac{f(r)v(r)^{2}}{q}=b^{2},\qquad v(r)=\frac{\sigma\sqrt q}{f(r)}\sqrt{b^{2}f(r)+u(r)^{2}},\qquad \sigma=\pm1.
\label{eq:norm-reconstruction}
\end{equation}

A temporal component changes the orientation of the expectation value while preserving its positive norm. The two signs in Eq.~\eqref{eq:norm-reconstruction} are kept throughout the field equations.

Primes denote derivatives with respect to $r$. For the metric~\eqref{eq:ansatz}, the temporal and radial mixed Ricci components coincide,
\begin{equation}
R^{t}{}_{t}=R^{r}{}_{r}=-\frac{1}{2q}\left(f''+\frac{2f'}r\right).
\label{eq:ricci-projection}
\end{equation}
The Einstein tensor has the corresponding components
\begin{equation}
G^{t}{}_{t}=G^{r}{}_{r}=\frac{rf'+f-q}{qr^{2}},\qquad G^{\theta}{}_{\theta}=G^{\varphi}{}_{\varphi}=\frac1q\left(\frac{f''}{2}+\frac{f'}r\right).
\label{eq:einstein-components}
\end{equation}

Since $H_{tr}=-u'$, the temporal divergence is $\nabla_{\mu}H^{\mu t}=-(r^{2}u')'/(qr^{2})$, while $\nabla_{\mu}H^{\mu r}=0$. The radial and temporal bumblebee equations reduce to
\begin{equation}
\begin{aligned}
\frac{fv}{q}\left(\frac{\xi}{\kappa}R^{r}{}_{r}-\lambda\right)&=0, \\ \frac{1}{qr^{2}}\left(r^{2}u'\right)'+\frac{u}{f}\left(\frac{\xi}{\kappa}R^{t}{}_{t}-\lambda\right)&=0.
\end{aligned}
\label{eq:bumblebee-components}
\end{equation}
In the static exterior, the positive norm implies $v\ne0$. The first of these equations determines the multiplier:
\begin{equation}
\lambda(r)=\frac{\xi}{\kappa}R^{r}{}_{r}=-\frac{\xi}{2\kappa q}\left(f''+\frac{2f'}r\right).
\label{eq:lambda-curvature}
\end{equation}

Substituting this result into the temporal equation cancels its algebraic curvature term. The remaining equation and its two integrations are
\begin{equation}
\left(r^{2}u'\right)'=0,\qquad r^{2}u'=-u_{1},\qquad u(r)=u_{0}+\frac{u_{1}}r,\qquad H_{tr}=\frac{u_{1}}{r^{2}}.
\label{eq:temporal-integration}
\end{equation}

We organize the construction according to the bumblebee expectation value. The purely radial configuration has $u_{0}=u_{1}=0$, while a constant temporal component has $u_{1}=0$ and $u_{0}\ne0$. A varying temporal component with $u_{1}\ne0$ will be included at $\xi=\kappa/2$. The electromagnetic field and its energy density can first be obtained for all three configurations.

The electromagnetic equation can be integrated before the gravitational equations. Using $\sqrt{-g}=\sqrt q\,r^{2}\sin\theta$, we obtain
\begin{equation}
\mathcal I=-\frac{2a'^{2}}q,\qquad s=\sqrt{1-\frac{a'^{2}}{q\beta_{\mathrm{BI}}^{2}}},\qquad \frac{\mathrm{d}}{\mathrm{d}r}\left(\frac{r^{2}a'}s\right)=0,\qquad \frac{r^{2}a'}s=-\sqrt q\,Q.
\label{eq:gauss-law}
\end{equation}

The signed integration constant $Q$ is the electric displacement flux per unit solid angle in this normalization. Squaring the first integral gives the algebraic relation
\begin{equation}
r^{4}a'^{2}=qQ^{2}\left(1-\frac{a'^{2}}{q\beta_{\mathrm{BI}}^{2}}\right),\qquad a'^{2}=\frac{qQ^{2}}{r^{4}+Q^{2}/\beta_{\mathrm{BI}}^{2}}.
\label{eq:gauss-algebra}
\end{equation}
The original unsquared equation then fixes the sign of $a'$, yielding
\begin{equation}
\begin{aligned}
a'&=-\frac{\sqrt q\,Q}{\sqrt{r^{4}+Q^{2}/\beta_{\mathrm{BI}}^{2}}},\qquad \mathscr S(r)=\sqrt{1+\frac{Q^{2}}{\beta_{\mathrm{BI}}^{2}r^{4}}},\qquad s=\frac1{\mathscr S}, \\ E(r)&\equiv\frac{F_{tr}}{\sqrt q}=\frac{Q}{\sqrt{r^{4}+Q^{2}/\beta_{\mathrm{BI}}^{2}}}.
\end{aligned}
\label{eq:electric-field}
\end{equation}

For nonzero charge, $|E|$ approaches $\beta_{\mathrm{BI}}$ as $r\to0$. The potential with $a(\infty)=0$ follows by setting $x=1/r$ in its radial integral:
\begin{equation}
\begin{aligned}
a(r)&=\sqrt q\,Q\int_{0}^{1/r}\frac{\mathrm{d}x}{\sqrt{1+Q^{2}x^{4}/\beta_{\mathrm{BI}}^{2}}}=\frac{\sqrt q\,Q}{r}\mathcal H(r), \\ \mathcal H(r)&={}_{2}F_{1}\!\left(\frac14,\frac12;\frac54;-\frac{Q^{2}}{\beta_{\mathrm{BI}}^{2}r^{4}}\right).
\end{aligned}
\label{eq:electric-potential}
\end{equation}

Before substituting the electric solution, Eq.~\eqref{eq:stress-tensors} gives
\begin{equation}
-T^{t}{}_{t\,\mathrm{BI}}=\frac{a'^{2}}{qs}-\beta_{\mathrm{BI}}^{2}(1-s),\qquad T^{r}{}_{r\,\mathrm{BI}}=T^{t}{}_{t\,\mathrm{BI}},\qquad T^{\theta}{}_{\theta\,\mathrm{BI}}=\beta_{\mathrm{BI}}^{2}(1-s).
\label{eq:electric-stress-components}
\end{equation}
Using $a'^{2}/q=\beta_{\mathrm{BI}}^{2}(1-s^{2})$ and $s=\mathscr S^{-1}$ then yields the energy density and transverse pressure:
\begin{equation}
\begin{aligned}
T^{\mu}{}_{\nu\,\mathrm{BI}}&=\operatorname{diag}\!\left(-\rho,-\rho,p_{\perp},p_{\perp}\right), \qquad \rho&=\beta_{\mathrm{BI}}^{2}(\mathscr S-1),\qquad p_{\perp}=\beta_{\mathrm{BI}}^{2}\left(1-\frac1{\mathscr S}\right).
\end{aligned}
\label{eq:rho-pressure}
\end{equation}

The radial pressure is $p_{r}=-\rho$, and electromagnetic energy conservation assumes the simple form
\begin{equation}
\rho'+\frac{2}{r}\left(\rho+p_{\perp}\right)=0.
\label{eq:matter-conservation}
\end{equation}

This identity will relate the temporal and angular gravitational equations. It also highlights a direct check of the nonlinear matter source independently of the metric integration.

Let us consider that $q_{0}=1+\xi b^{2}$, and take $q_{0}>0$. For $u=u_{0}$, substituting the norm and Eq.~\eqref{eq:lambda-curvature} into the metric equation gives the two scalar residuals
\begin{equation}
\begin{aligned}
\mathcal M(r)&=q_{0}\left(f+rf'\right)-q+\xi u_{0}^{2}+\kappa q r^{2}\rho, \\ \mathcal P(r)&=q_{0}\left(\frac{f''}{2}+\frac{f'}r\right)-\kappa q p_{\perp}.
\end{aligned}
\label{eq:radial-residuals}
\end{equation}

Writing $\mathcal E_{\mu\nu}$ for the left side of Eq.~\eqref{eq:metric-eom} after moving all terms to that side, its nontrivial components are
\begin{equation}
\mathcal E_{tt}=-\frac{f\mathcal M}{qr^{2}},\qquad \mathcal E_{rr}=\frac{\mathcal M}{r^{2}f},\qquad \mathcal E_{\theta\theta}=\frac{r^{2}\mathcal P}{q},\qquad \mathcal E_{\varphi\varphi}=\sin^{2}\theta\,\mathcal E_{\theta\theta}.
\label{eq:component-map}
\end{equation}

The mixed components vanish after the bumblebee equation is imposed. To see how the angular equation follows from the temporal equation, differentiate the first line of Eq.~\eqref{eq:radial-residuals}:
\begin{equation}
\mathcal M'=q_{0}\left(2f'+rf''\right)+\kappa q\left(2r\rho+r^{2}\rho'\right)=2r\left[q_{0}\left(\frac{f''}{2}+\frac{f'}r\right)-\kappa q p_{\perp}\right].
\label{eq:residual-derivative}
\end{equation}
Equation~\eqref{eq:matter-conservation} supplies the last equality and establishes $\mathcal M'=2r\mathcal P$.

In this case, the angular equation follows from the integrated temporal equation and the electromagnetic equation, while supplies an independent substitution check. Combining $\mathcal P=0$ with Eq.~\eqref{eq:lambda-curvature} determines the multiplier without fixing it in advance:
\begin{equation}
\lambda(r)=-\frac{\xi}{q_{0}}p_{\perp}(r)=-\frac{\xi\beta_{\mathrm{BI}}^{2}}{q_{0}}\left(1-\frac1{\mathscr S(r)}\right),\qquad \lambda'(r)=\frac{2\xi Q^{2}}{q_{0}r^{5}\mathscr S(r)^{3}}.
\label{eq:lambda-pressure}
\end{equation}

The charged geometry selects a radial multiplier even though the norm is constant and the potential vanishes on the solution. For $Q\xi\ne0$, a constant multiplier would remove this particular charged configuration. Its asymptotic value is zero, as appropriate to the present calculation with a vanishing cosmological constant.

The equation $\mathcal M=0$ is a first order equation for $rf$,
\begin{equation}
\frac{\mathrm{d}}{\mathrm{d}r}(rf)=\frac{q-\xi u_{0}^{2}}{q_{0}}-\frac{\kappa q}{q_{0}}r^{2}\rho(r).
\label{eq:master-ode}
\end{equation}

The integrating factor for Eq.~\eqref{eq:master-ode}, written as $f'+f/r=(q-\xi u_{0}^{2})/(q_{0}r)-\kappa q r\rho/q_{0}$, is $r$. To evaluate its matter integral, introduce $c=Q^{2}/\beta_{\mathrm{BI}}^{2}$. Since $r>0$, the source can be written as
\begin{equation}
r^{2}\rho(r)=\beta_{\mathrm{BI}}^{2}\left(\sqrt{r^{4}+c}-r^{2}\right),\qquad \int r^{2}\rho(r)\,\mathrm dr=\beta_{\mathrm{BI}}^{2}\left[\int\sqrt{r^{4}+c}\,\mathrm dr-\frac{r^{3}}3\right].
\label{eq:density-primitive-reduction}
\end{equation}
Writing $I_{c}=\int\sqrt{r^{4}+c}\,\mathrm dr$, integration by parts first gives $I_{c}=r\sqrt{r^{4}+c}-2\int r^{4}/\sqrt{r^{4}+c}\,\mathrm dr$. Substituting $r^{4}=(r^{4}+c)-c$ into the last integrand isolates $3I_{c}$ and gives
\begin{equation}
\int\sqrt{r^{4}+c}\,\mathrm{d}r=\frac{r\sqrt{r^{4}+c}}3+\frac{2c}{3}\int\frac{\mathrm{d}r}{\sqrt{r^{4}+c}},\qquad \int\frac{\mathrm{d}r}{\sqrt{r^{4}+c}}=-\frac{\mathcal H(r)}r.
\label{eq:integration-by-parts}
\end{equation}

The additive constant in this primitive can be chosen so that
\begin{equation}
\int r^{2}\rho(r)\,\mathrm dr=\frac{\beta_{\mathrm{BI}}^{2}r^{3}}3\left(\mathscr S(r)-1\right)-\frac{2Q^{2}}{3r}\mathcal H(r).
\label{eq:density-primitive}
\end{equation}
The same hypergeometric function that determines the electric potential integrates the energy density. At large radius, $\rho=Q^{2}/(2r^{4})+\mathcal O(r^{-8})$, so the integral from any $r>0$ to infinity converges. A useful definite integral is
\begin{equation}
\begin{aligned}
\mathcal J(r)&\equiv\int_{r}^{\infty}x^{2}\rho(x)\,\mathrm{d}x=\frac{\beta_{\mathrm{BI}}^{2}r^{3}}3\left(1-\mathscr S(r)\right)+\frac{2Q^{2}}{3r}\mathcal H(r), \qquad \mathcal J'(r)&=-r^{2}\rho(r),\qquad \mathcal J(\infty)=0.
\end{aligned}
\label{eq:tail-integral}
\end{equation}

The common electromagnetic correction to the lapse reads
\begin{equation}
\begin{aligned}
\Delta_{\mathrm{BI}}(r)={}&\frac{\kappa\beta_{\mathrm{BI}}^{2}r^{2}}3\left(1-\sqrt{1+\frac{Q^{2}}{\beta_{\mathrm{BI}}^{2}r^{4}}}\right) +\frac{2\kappa Q^{2}}{3r^{2}}\,{}_{2}F_{1}\!\left(\frac14,\frac12;\frac54;-\frac{Q^{2}}{\beta_{\mathrm{BI}}^{2}r^{4}}\right)=\frac{\kappa\mathcal J(r)}r,\qquad f_{\mathrm S}(r)=1-\frac{2m}{r}.
\end{aligned}
\label{eq:comparison-functions}
\end{equation}

Although the first term in the explicit expression is negative, the integral representation makes $\mathcal J(r)>0$ manifest for $Q\ne0$. Integrating Eq.~\eqref{eq:master-ode}, while retaining its integration constant, gives
\begin{equation}
rf(r)=\frac{q-\xi u_{0}^{2}}{q_{0}}r+\frac{\kappa q}{q_{0}}\mathcal J(r)+c_{m}.
\label{eq:metric-first-integral}
\end{equation}

\subsection{Purely radial vacuum configuration}
\label{subsec:radial-vev}

We begin with $u_{0}=u_{1}=0$, so that $B=v_{R}(r)\mathrm dr$ and the norm gives $v_{R}^{2} = q \,b^{2}/f_{R}$. The temporal bumblebee equation is then identically satisfied, while its radial equation determines $\lambda_{R}$ through Eq.~\eqref{eq:lambda-curvature}. The temporal and angular gravitational equations become
\begin{equation}
\begin{aligned}
f_{R}'+\frac{f_{R}}r&=\frac{q}{q_{0}r}-\frac{\kappa q}{q_{0}}r\rho(r), \qquad \frac{\mathrm d}{\mathrm dr}\left(r^{2}f_{R}'\right)&=\frac{2\kappa q}{q_{0}}r^{2}p_{\perp}(r).
\end{aligned}
\label{eq:radial-metric-odes}
\end{equation}
Multiplying the first equation by $r$ and using $\mathcal J'=-r^{2}\rho$ displays both the total derivative and its first integral:
\begin{equation}
\begin{aligned}
\frac{\mathrm d}{\mathrm dr}\left(rf_{R}\right)&=\frac{q}{q_{0}}\left(1-\kappa r^{2}\rho\right),  \qquad rf_{R}(r)&=\frac{q}{q_{0}}\left[r+\kappa\mathcal J(r)\right]+c_{m}.
\end{aligned}
\label{eq:radial-integration}
\end{equation}

Dividing the first integral by $r$ gives $f_{R}=q/q_{0}+c_{m}/r+\kappa q\mathcal J/(q_{0}r)$. Since $\mathcal J(\infty)=0$, its asymptotic constant is $q/q_{0}$, and the normalization $f_{R}(\infty)=1$ fixes $q=q_{0}=1+\xi b^{2}$. The angular equation is then $(r^{2}f_{R}')'=2\kappa r^{2}p_{\perp}$; together with Eq.~\eqref{eq:lambda-curvature}, it gives $\lambda_{R}=-\xi p_{\perp}/q_{0}$. Writing $c_{m}=-2m$, the complete line element is
\begin{equation}
\mathrm d s_{R}^{2}=-f_{R}(r)\mathrm d t^{2}+\frac{1+\xi b^{2}}{f_{R}(r)}\mathrm d r^{2}+r^{2}\mathrm d\theta^{2}+r^{2}\sin^{2}\theta\,\mathrm d\varphi^{2},
\label{eq:radial-line-element}
\end{equation}
where
\begin{equation}
f_{R}(r)=1-\frac{2m}{r}+\kappa\left[\frac{\beta_{\mathrm{BI}}^{2}r^{2}}3\left(1-\mathscr S(r)\right)+\frac{2Q^{2}}{3r^{2}}\mathcal H(r)\right]=1-\frac{2m}{r}+\Delta_{\mathrm{BI}}(r).
\label{eq:radial-lapse}
\end{equation}

In the coordinate order $(t,r,\theta,\varphi)$, the bumblebee expectation value, electromagnetic potential, and multiplier are explicitly
\begin{equation}
\begin{aligned}
B_{\mu}^{(R)}&=\left(0,\,\sigma b\sqrt{\frac{1+\xi b^{2}}{f_{R}(r)}},\,0,\,0\right), \\ A_{\mu}^{(R)}&=\left(\frac{\sqrt{1+\xi b^{2}}\,Q\mathcal H(r)}r,\,0,\,0,\,0\right), \\ \lambda_{R}(r)&=-\frac{\xi\beta_{\mathrm{BI}}^{2}}{1+\xi b^{2}}\left(1-\frac1{\mathscr S(r)}\right).
\end{aligned}
\label{eq:radial-fields}
\end{equation}

The antisymmetric bumblebee tensor vanishes, $H_{\mu\nu}^{(R)}=0$, while the independent nonzero electromagnetic component is $(F_{R})_{tr}=\sqrt{1+\xi b^{2}}\,Q/\sqrt{r^{4}+Q^{2}/\beta_{\mathrm{BI}}^{2}}$. Even with $H_{\mu\nu}=0$, the tensor $\nabla_{\mu}B_{\nu}$ and the nonminimal curvature interaction keep their gravitational role. The parameter $\xi$ enters the radial metric coefficient and the multiplier, whereas the lapse correction in Eq.~\eqref{eq:radial-lapse} has the Einstein--Born--Infeld form. The neutral limit gives the radial Casana geometry and the zero cosmological constant limit of Maluf and Neves~\cite{Casana2018,MalufNeves2021}. The behavior of the radial vector at the horizon will be addressed in Subsec.~\ref{subsec:horizon}.

\subsection{Temporal and radial VEVs with constant temporal amplitude}
\label{subsec:constant-temporal-vev}

We now retain $u_{0}\ne0$ and set $u_{1}=0$, so that $B=u_{0}\mathrm dt+v_{T}(r)\mathrm dr$. Its radial component obeys $v_{T}^{2}=q(b^{2}f_{T}+u_{0}^{2})/f_{T}^{2}$, and the constant temporal component gives $H_{\mu\nu}=0$. After the radial bumblebee equation determines $\lambda_{T}$, the two gravitational equations are
\begin{equation}
\begin{aligned}
f_{T}'+\frac{f_{T}}r&=\frac{q-\xi u_{0}^{2}}{q_{0}r}-\frac{\kappa q}{q_{0}}r\rho(r), \qquad \frac{\mathrm d}{\mathrm dr}\left(r^{2}f_{T}'\right)&=\frac{2\kappa q}{q_{0}}r^{2}p_{\perp}(r).
\end{aligned}
\label{eq:temporal-metric-odes}
\end{equation}
The constant temporal amplitude changes the constant source in the first equation. Multiplication by $r$ and one integration give
\begin{equation}
rf_{T}(r)=\frac{q-\xi u_{0}^{2}}{q_{0}}r+\frac{\kappa q}{q_{0}}\mathcal J(r)+c_{m},\qquad f_{T}(\infty)=\frac{q-\xi u_{0}^{2}}{q_{0}}.
\label{eq:temporal-metric-integration}
\end{equation}
We normalize the asymptotic Killing time by requiring $f_{T}(\infty)=1$ and define $c_{m}=-2m$. These choices imply
$q=q_{T}\equiv q_{0}+\xi u_{0}^{2}=1+\xi\left(b^{2}+u_{0}^{2}\right)>0.$

The full line element is
\begin{equation}
\mathrm d s_{T}^{2}=-f_{T}(r)\mathrm d t^{2}+\frac{1+\xi(b^{2}+u_{0}^{2})}{f_{T}(r)}\mathrm d r^{2}+r^{2}\mathrm d\theta^{2}+r^{2}\sin^{2}\theta\,\mathrm d\varphi^{2},
\label{eq:temporal-line-element}
\end{equation}
with $f_{T}(r)\equiv f_{\mathrm{BI}}(r)$ given explicitly by
\begin{equation}
\begin{aligned}
f_{T}(r)&=1-\frac{2m}{r}+\frac{\kappa q_{T}}{q_{0}r}\mathcal J(r) =1-\frac{2m}{r}+\frac{\kappa\, q_{T}}{q_{0}}\left[\frac{\beta_{\mathrm{BI}}^{2}r^{2}}3\left(1-\mathscr S(r)\right)+\frac{2Q^{2}}{3r^{2}}\mathcal H(r)\right].
\end{aligned}
\label{eq:metric-solution}
\end{equation}

The accompanying fields, with every covector component displayed, are
\begin{equation}
\begin{aligned}
B_{\mu}^{(T)}&=\left(u_{0},\,\frac{\sigma\sqrt{q_{T}}\sqrt{b^{2}f_{T}(r)+u_{0}^{2}}}{f_{T}(r)},\,0,\,0\right), \\ A_{\mu}^{(T)}&=\left(\frac{\sqrt{q_{T}}\,Q\mathcal H(r)}r,\,0,\,0,\,0\right), \\ \lambda_{T}(r)&=-\frac{\xi\beta_{\mathrm{BI}}^{2}}{1+\xi b^{2}}\left(1-\frac1{\mathscr S(r)}\right).
\end{aligned}
\label{eq:fields-solution}
\end{equation}
Here, $H_{\mu\nu}^{(T)}=0$ and $(F_{T})_{tr}=\sqrt{q_{T}}\,Q/\sqrt{r^{4}+Q^{2}/\beta_{\mathrm{BI}}^{2}}$. Equations~\eqref{eq:metric-solution} and \eqref{eq:fields-solution} give the charged solution at arbitrary $\xi$ in the stated domain. The constant temporal expectation value enters both the radial metric coefficient and the electromagnetic contribution to the lapse through $q_{T}/q_{0}$. Setting $u_{0}=0$ in Eqs.~\eqref{eq:temporal-line-element}--\eqref{eq:fields-solution} reproduces Eqs.~\eqref{eq:radial-line-element}--\eqref{eq:radial-fields}, including the radial VEV and the electric potential.

\subsection{Temporal and radial VEVs with varying temporal amplitude}
\label{subsec:varying-temporal-vev}

We next retain both integration constants in $u(r)=u_{0}+u_{1}/r$, with $u_{1}\ne0$, so that $B=\mathcal U(r)\mathrm dt+v_{\star}(r)\mathrm dr$ and $H_{tr}=u_{1}/r^{2}$. The temporal and radial Ricci projections remain equal, so Eq.~\eqref{eq:lambda-curvature} still determines the multiplier. In the gravitational equations, the temporal combinations of the bumblebee field are
\begin{equation}
u^{2}+2ruu'=u_{0}^{2}-\frac{u_{1}^{2}}{r^{2}},\qquad r^{2}u'^{2}=\frac{u_{1}^{2}}{r^{2}}.
\label{eq:temporal-profile-identities}
\end{equation}
The terms proportional to $u_{0}u_{1}$ cancel in the first expression. Combining the nonminimal curvature contribution with the bumblebee kinetic term changes the reduced residuals by
\begin{equation}
\mathcal M_{u_{1}}=\mathcal M+\frac{(\kappa-2\xi)u_{1}^{2}}{2r^{2}},\qquad \mathcal P_{u_{1}}=\mathcal P-\frac{(\kappa-2\xi)u_{1}^{2}}{2r^{4}}.
\label{eq:coulomb-residuals}
\end{equation}

Equivalently, before choosing the coupling, the radial differential equations read
\begin{equation}
\begin{aligned}
q_{0}\frac{\mathrm d}{\mathrm dr}(rf)&=q-\xi u_{0}^{2}-\kappa q r^{2}\rho-\frac{(\kappa-2\xi)u_{1}^{2}}{2r^{2}}, \qquad q_{0}\frac{\mathrm d}{\mathrm dr}\left(r^{2}f'\right)&=2\kappa q r^{2}p_{\perp}+\frac{(\kappa-2\xi)u_{1}^{2}}{r^{2}}.
\end{aligned}
\label{eq:varying-metric-odes}
\end{equation}
To retain the charged lapse already obtained for a constant temporal component, the additional terms in these two equations must vanish. For that lapse, their common condition is
\begin{equation}
(\kappa-2\xi)u_{1}^{2}=0.
\label{eq:temporal-amplitude-condition}
\end{equation}
In this case, $u_{1}\ne0$ selects $\xi=\kappa/2$ within the family displayed here. This condition identifies the coupling for which the nonzero bumblebee field strength leaves this metric unchanged; it does not restrict geometries with an additional radial contribution to the lapse. It is useful to display the parameters after making this substitution:
\begin{equation}
\begin{aligned}
\xi&=\frac{\kappa}{2},\qquad q_{0\star}=1+\frac{\kappa b^{2}}2,\qquad q_{\star}=1+\frac{\kappa}{2}\left(b^{2}+u_{0}^{2}\right), \qquad \mathcal U(r)&=u_{0}+\frac{u_{1}}r,\qquad u_{1}\ne0.
\end{aligned}
\label{eq:special-parameters}
\end{equation}

Using $q_{\star}-\kappa u_{0}^{2}/2=q_{0\star}$, the normalized temporal equation and its first integral take the form
\begin{equation}
\frac{\mathrm d}{\mathrm dr}(rf_{\star})=1-\frac{\kappa q_{\star}}{q_{0\star}}r^{2}\rho,\qquad rf_{\star}(r)=r+\frac{\kappa q_{\star}}{q_{0\star}}\mathcal J(r)+c_{m}.
\label{eq:special-metric-integration}
\end{equation}
The angular equation gives $f_{\star}''/2+f_{\star}'/r=\kappa q_{\star}p_{\perp}/q_{0\star}$ and $\lambda_{\star}=-\kappa p_{\perp}/(2q_{0\star})$. With $c_{m}=-2m$, the complete spacetime metric for this configuration is
\begin{equation}
\mathrm d s_{\star}^{2}=-f_{\star}(r)\mathrm d t^{2}+\frac{1+\frac{\kappa}{2}(b^{2}+u_{0}^{2})}{f_{\star}(r)}\mathrm d r^{2}+r^{2}\mathrm d\theta^{2}+r^{2}\sin^{2}\theta\,\mathrm d\varphi^{2},
\label{eq:special-line-element}
\end{equation}
where
\begin{equation}
f_{\star}(r)=1-\frac{2m}{r}+\frac{\kappa q_{\star}}{q_{0\star}}\left[\frac{\beta_{\mathrm{BI}}^{2}r^{2}}3\left(1-\mathscr S(r)\right)+\frac{2Q^{2}}{3r^{2}}\mathcal H(r)\right].
\label{eq:special-lapse}
\end{equation}

The bumblebee VEV now contains the independent amplitude $u_{1}$, and the full set of accompanying fields is
\begin{equation}
\begin{aligned}
B_{\mu}^{(\star)}&=\left(u_{0}+\frac{u_{1}}r,\,\frac{\sigma\sqrt{q_{\star}}\sqrt{b^{2}f_{\star}(r)+(u_{0}+u_{1}/r)^{2}}}{f_{\star}(r)},\,0,\,0\right), \\ A_{\mu}^{(\star)}&=\left(\frac{\sqrt{q_{\star}}\,Q\mathcal H(r)}r,\,0,\,0,\,0\right), \\ \lambda_{\star}(r)&=-\frac{\kappa\beta_{\mathrm{BI}}^{2}}{2+\kappa b^{2}}\left(1-\frac1{\mathscr S(r)}\right).
\end{aligned}
\label{eq:nonclosed-solution}
\end{equation}

Its independent field strength components and the bumblebee invariant are
\begin{equation}
\begin{aligned}
(H_{\star})_{tr}&=\frac{u_{1}}{r^{2}},\qquad (F_{\star})_{tr}=\frac{\sqrt{q_{\star}}\,Q}{\sqrt{r^{4}+Q^{2}/\beta_{\mathrm{BI}}^{2}}}, \qquad (H_{\star})_{\mu\nu}(H_{\star})^{\mu\nu}&=-\frac{2u_{1}^{2}}{q_{\star}r^{4}}.
\end{aligned}
\label{eq:special-strengths}
\end{equation}

The constants $Q$ and $u_{1}$ remain independent: the first controls the nonlinear electromagnetic source, whereas the second describes a bumblebee field strength whose metric contribution cancels at this coupling. Equation~\eqref{eq:coulomb-residuals} establishes this statement for the displayed metric without imposing a restriction on every possible solution of the theory. The neutral counterpart and the special role of this coupling have been examined in Refs.~\cite{ZhuLi2026,LiuWuWeiLiu2026}. In the common notation used below, this configuration has $q_{T}=q_{\star}$ and $f_{\mathrm{BI}}=f_{T}=f_{\star}$, with $q_{0}=q_{0\star}$.


\subsection{Field equations and future horizon}
\label{subsec:horizon}

For the shared discussion, the notation $(f_{\mathrm{BI}},q_{T},u)$ represents $(f_{R},q_{0},0)$ for the purely radial configuration, $(f_{T},q_{T},u_{0})$ for the constant temporal component, and $(f_{\star},q_{\star},\mathcal U)$ for the varying temporal component.

The field equations can now be checked directly from the integrated metric. Differentiating its matter contribution and using $\mathcal J'=-r^{2}\rho$ gives
\begin{equation}
\begin{aligned}
q_{0}(f_{\mathrm{BI}}+rf_{\mathrm{BI}}')-q_{T}+\xi u_{0}^{2}=\kappa q_{T}\mathcal J'=-\kappa q_{T}r^{2}\rho, \,\,\, q_{0}\left(\frac{f_{\mathrm{BI}}''}{2}+\frac{f_{\mathrm{BI}}'}r\right)&=\frac{\kappa q_{T}}{2r}\mathcal J''=-\kappa q_{T}\left(\rho+\frac r2\rho'\right)=\kappa q_{T}p_{\perp}.
\end{aligned}
\label{eq:integrated-metric-check}
\end{equation}
The last equality follows from Eq.~\eqref{eq:matter-conservation}. These relations set $\mathcal M=\mathcal P=0$ and verify the temporal, radial, and angular metric components through Eq.~\eqref{eq:component-map}. For the varying temporal component, Eq.~\eqref{eq:temporal-amplitude-condition} removes the additional terms in Eq.~\eqref{eq:coulomb-residuals}. Equation~\eqref{eq:lambda-curvature} verifies the radial bumblebee equation, while $(r^{2}u')'=0$ verifies its temporal component. The norm is recovered directly as $B_{\mu}B^{\mu}=-u^{2}/f_{\mathrm{BI}}+(b^{2}f_{\mathrm{BI}}+u^{2})/f_{\mathrm{BI}}=b^{2}$, for either sign of $\sigma$. Finally, differentiating the electric potential recovers Eq.~\eqref{eq:electric-field} and its Gauss first integral. In the accompanying Mathematica calculation, these steps are supplemented by substitution into all sixteen metric components, four bumblebee components, the multiplier equation, and four electromagnetic components, for both signs of $\sigma$ and for the three configurations displayed above. The tensor variations and component calculation employ the xAct framework and xTras~\cite{Nutma2014}.

The temporal component also permits a direct examination of the future horizon. Let $r_{h}>0$ be a simple outer zero of $f_{\mathrm{BI}}$, with $f_{\mathrm{BI}}>0$ for $r>r_{h}$. The horizon condition and the temporal metric equation give
\begin{equation}
2m=r_{h}+\frac{\kappa q_{T}}{q_{0}}\mathcal J(r_{h}),\qquad f_{\mathrm{BI}}'(r_{h})=\frac1{r_{h}}\left[1-\frac{\kappa q_{T}}{q_{0}}r_{h}^{2}\rho(r_{h})\right].
\label{eq:horizon-equations}
\end{equation}
For a simple outer horizon the second expression is positive. To examine the fields there, introduce an ingoing coordinate $w$ by
\begin{equation}
\mathrm{d}w=\mathrm{d}t+\frac{\sqrt{q_{T}}}{f_{\mathrm{BI}}}\,\mathrm{d}r,\qquad \mathrm{d}s^{2}=-f_{\mathrm{BI}}\mathrm{d}w^{2}+2\sqrt{q_{T}}\,\mathrm{d}w\,\mathrm{d}r+r^{2}\mathrm{d}\Omega^{2}.
\label{eq:ingoing-metric}
\end{equation}

The bumblebee one form becomes
\begin{equation}
B=u\,\mathrm{d}w+\sqrt{q_{T}}\,\frac{\sigma\sqrt{u^{2}+b^{2}f_{\mathrm{BI}}}-u}{f_{\mathrm{BI}}}\,\mathrm{d}r.
\label{eq:ingoing-vector}
\end{equation}

If $u(r_{h})\ne0$, choose $\sigma=\operatorname{sgn}u(r_{h})$ in a neighborhood of the horizon. Rationalizing its radial component gives
\begin{equation}
B_{r}^{(w,r)}=\frac{\sigma\sqrt{q_{T}}\,b^{2}}{\sqrt{u^{2}+b^{2}f_{\mathrm{BI}}}+|u|},\qquad \lim_{r\to r_{h}}B_{r}^{(w,r)}=\frac{\sqrt{q_{T}}\,b^{2}}{2u(r_{h})}.
\label{eq:horizon-vector}
\end{equation}

Both the metric and the bumblebee field  extend smoothly across the future horizon under these conditions. A regular electromagnetic potential in the same chart is $A=a(r)\mathrm{d}w$, which has the field strength of Eq.~\eqref{eq:electric-field}. The strictly radial configuration $u=0$ has a divergent bumblebee one form at a simple horizon.

The configuration with $u_{1}\ne0$ also admits a simple sufficient condition for reality at every positive radius. Since $\mathcal J\ge0$, Eq.~\eqref{eq:metric-solution} implies $f_{\mathrm{BI}}\ge1-2m/r$, and we obtain
\begin{equation}
r^{2}\left[u^{2}+b^{2}f_{\mathrm{BI}}\right]\ge u_{1}^{2}+2\left(u_{0}u_{1}-b^{2}m\right)r+\left(u_{0}^{2}+b^{2}\right)r^{2}.
\label{eq:reality-bound}
\end{equation}

For $m>0$, $u_{0}>0$, $u_{1}>0$, and $u_{0}u_{1}\ge b^{2}m$, the right side is strictly positive for $r>0$. With $\sigma=+1$, the vector is real throughout this domain and smooth at a simple future event horizon. As an explicit example in rescaled units, the notebook uses $\kappa=b=m=1$, $\xi=1/2$, $u_{0}=0.3$, $u_{1}=4$, $Q=0.6$, and $\beta_{\mathrm{BI}}=0.8$. It gives $q_{T}=1.545$, an outer horizon at $r_{h}\simeq1.90276970$, $f'_{\mathrm{BI}}(r_{h})\simeq0.49892004$, and $B_{r}^{(w,r)}(r_{h})\simeq0.25871721$. This example satisfies Eq.~\eqref{eq:reality-bound} at every positive radius.

Regularity at the horizon and bounded electric field are distinct from regularity at the center. For the present metric, the curvature invariants can be evaluated directly as
\begin{equation}
R=-\frac{f''}{q_{T}}-\frac{4f'}{q_{T}r}+\frac{2}{r^{2}}\left(1-\frac{f}{q_{T}}\right),\qquad R_{\mu\nu\alpha\rho}R^{\mu\nu\alpha\rho}=\frac{f''^{2}}{q_{T}^{2}}+\frac{4f'^{2}}{q_{T}^{2}r^{2}}+\frac4{r^{4}}\left(1-\frac{f}{q_{T}}\right)^{2}.
\label{eq:curvature-invariants}
\end{equation}

They remain finite at a smooth positive radius horizon, while the center is generically singular. At large radius the leading geometry is $\mathrm{d}s^{2}\simeq-\mathrm{d}t^{2}+q_{T}\mathrm{d}r^{2}+r^{2}\mathrm{d}\Omega^{2}$. A sphere at proper radial distance $\ell\simeq\sqrt{q_{T}}r$ has area $4\pi\ell^{2}/q_{T}$. The asymptotic geometry is therefore conical when $q_{T}\ne1$, and $m$ denotes the normalized coefficient of the $1/r$ term.

\begin{figure}[!htbp]
\centering
\includegraphics[width=0.49\linewidth]{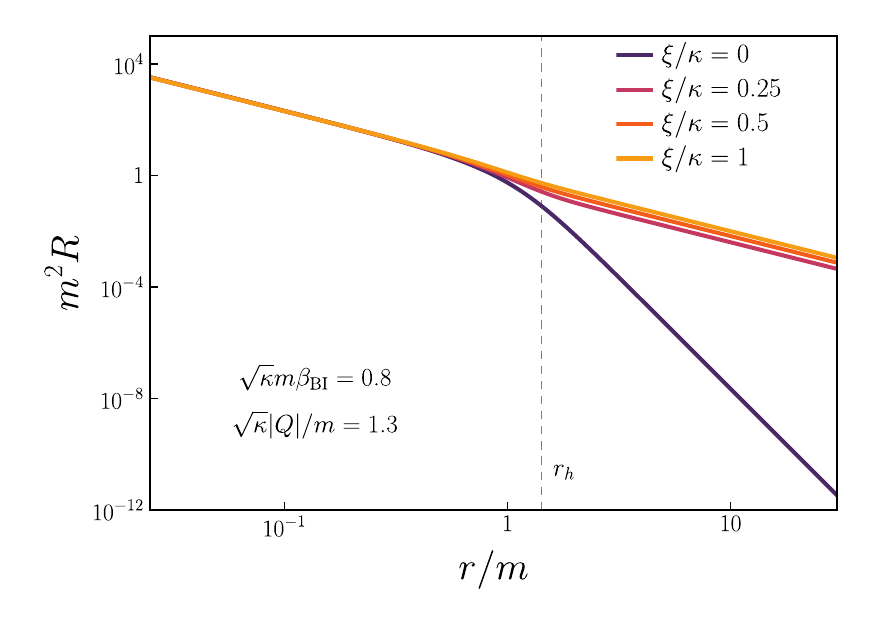}\hfill\includegraphics[width=0.49\linewidth]{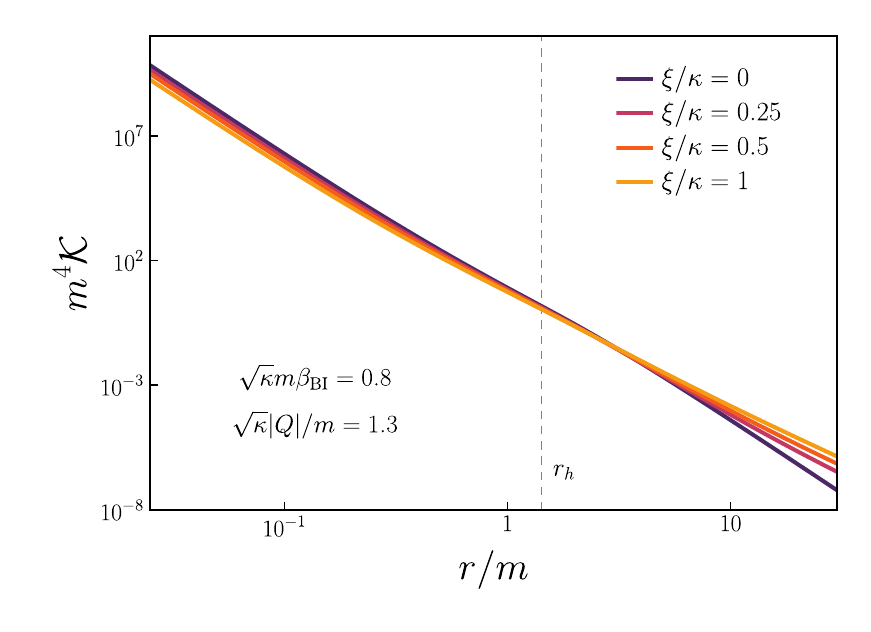}
\caption{Ricci scalar \(m^2R\) (left) and Kretschmann invariant \(m^4\mathcal K\) (right) for the radial configuration, with \(\kappa b^2=1\), \(\sqrt{\kappa}|Q|/m=1.3\), \(\sqrt{\kappa}\,m\beta_{\mathrm{BI}}=0.8\), and \(u_0=u_1=0\). The curves vary \(\xi/\kappa=0,0.25,0.5,1\); the zero coupling curve supplies the Einstein Born Infeld reference. Both axes are logarithmic. The vertical dashed line marks the common outer horizon, \(r_h/m\simeq1.422369\). The horizon radius is independent of the radial bumblebee coupling, while both curvature invariants retain its contribution.}
\label{fig:curvature_radial}
\end{figure}

\begin{figure}[!htbp]
\centering
\includegraphics[width=0.49\linewidth]{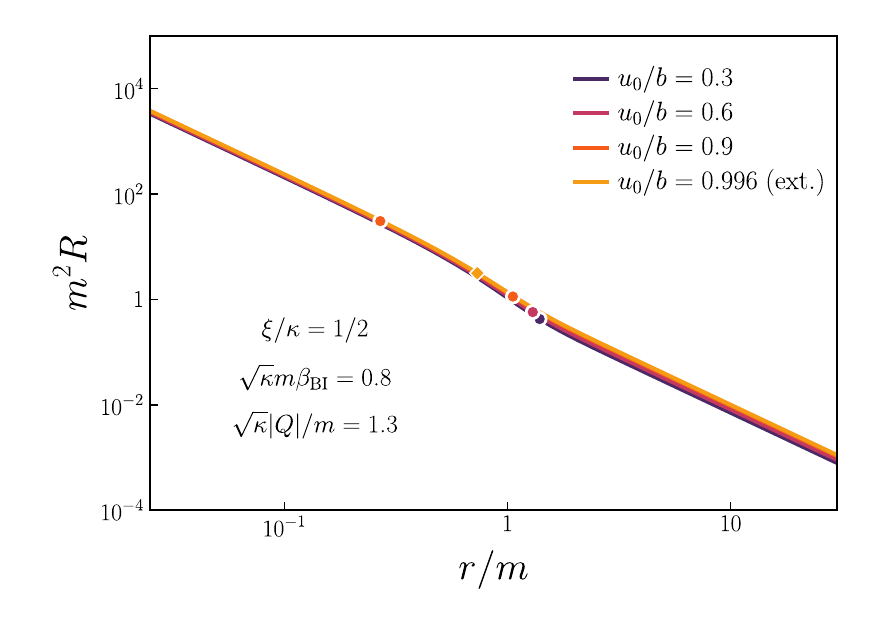}\hfill\includegraphics[width=0.49\linewidth]{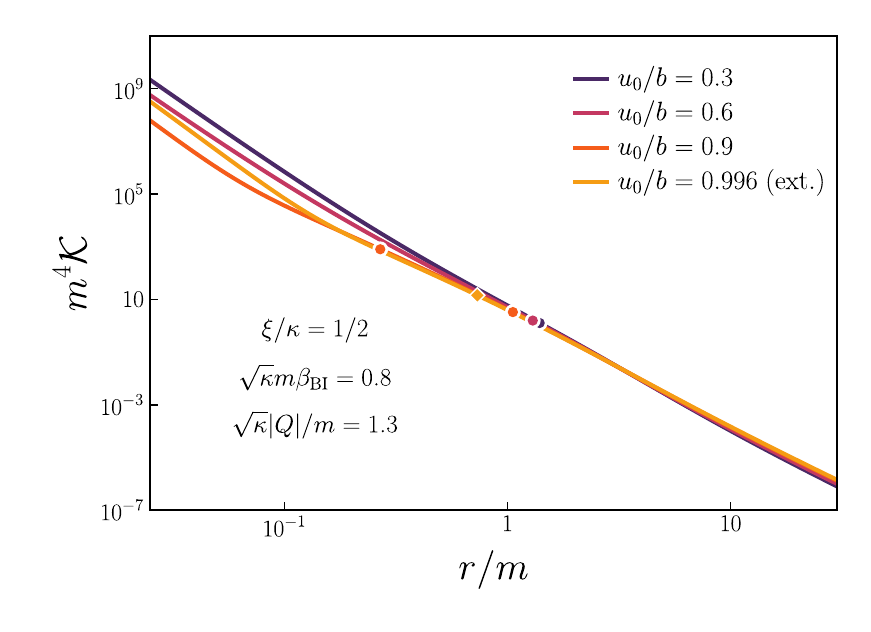}
\caption{Ricci scalar (left) and Kretschmann invariant (right) for the constant temporal configuration at \(\kappa b^2=1\), \(\xi/\kappa=1/2\), \(\sqrt{\kappa}|Q|/m=1.3\), \(\sqrt{\kappa}\,m\beta_{\mathrm{BI}}=0.8\), and \(u_1=0\). The temporal amplitude takes \(u_0/b=0.3,0.6,0.9,0.995953\), with the last value evaluated at its exact extremal limit. The first two curves have one positive horizon, the third has two, and the fourth has one degenerate horizon. Circles mark simple horizons and diamonds mark the degenerate horizon. The logarithmic axes and normalization follow Fig.~\ref{fig:curvature_radial}.}
\label{fig:curvature_constant_temporal}
\end{figure}

\begin{figure}[!htbp]
\centering
\includegraphics[width=0.49\linewidth]{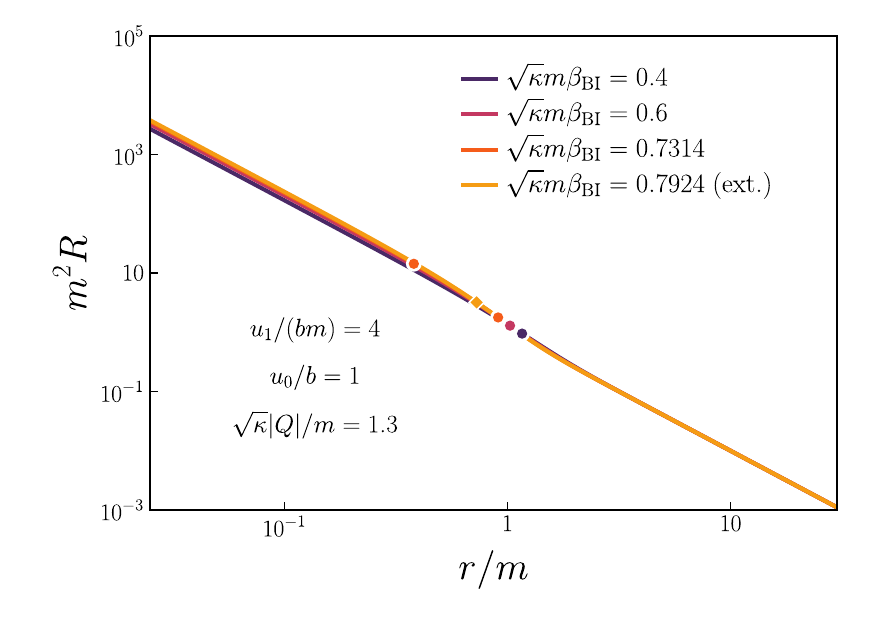}\hfill\includegraphics[width=0.49\linewidth]{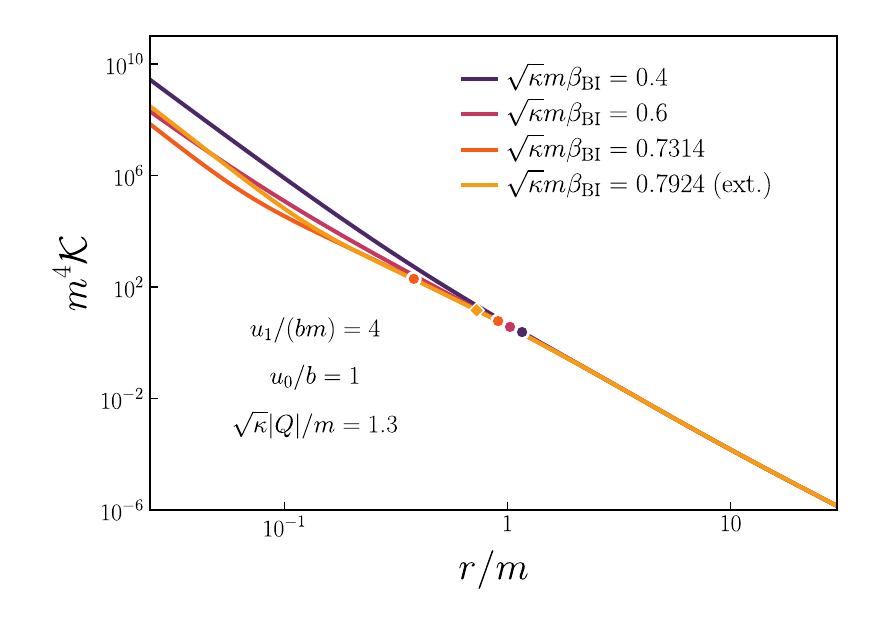}
\caption{Ricci scalar (left) and Kretschmann invariant (right) for the varying temporal configuration at \(\kappa b^2=1\), \(\xi/\kappa=1/2\), \(\sqrt{\kappa}|Q|/m=1.3\), \(u_0/b=1\), and \(u_1/(bm)=4\). The curves vary \(\sqrt{\kappa}\,m\beta_{\mathrm{BI}}=0.4,0.6,0.731370,0.792423\), with the last value evaluated at its exact extremal limit. The first two curves have one positive horizon, the third has two, and the fourth has one degenerate horizon. The markers and logarithmic axes follow Fig.~\ref{fig:curvature_constant_temporal}. The chosen field amplitudes satisfy the sufficient global reality condition, while \(u_1\) leaves these metric invariants unchanged.}
\label{fig:curvature_varying_temporal}
\end{figure}

In Figs.~\ref{fig:curvature_radial}, \ref{fig:curvature_constant_temporal}, and \ref{fig:curvature_varying_temporal}, we present the Ricci scalar and the Kretschmann invariant, \(\mathcal K=R_{\mu\nu\alpha\beta}R^{\mu\nu\alpha\beta}\), for the three bumblebee configurations. The curves are evaluated with the exact lapse and the expressions above. For the radial configuration, varying the coupling changes the curvature even though the lapse and its horizon positions retain their values. Moreover, the temporal amplitude and the nonlinear electromagnetic scale produce the distinct profiles displayed in the remaining panels. Both invariants remain finite at the indicated metric horizons, while the growth toward small radius identifies the central curvature singularity. At large radius, the Einstein Born Infeld reference in Fig.~\ref{fig:curvature_radial} decays faster than the configurations with a constant radial deformation. On the other hand, the constant and varying temporal configurations give identical invariants at matched metric parameters and \(\xi=\kappa/2\), independently of \(u_1\).

\subsection{Limits and comparison with earlier geometries}
\label{subsec:limits}

The familiar limits make the electromagnetic and Lorentz symmetry breaking contributions explicit. Expanding the exact solution at large $\beta_{\mathrm{BI}}$, or at large $r$ with the other parameters fixed, gives
\begin{equation}
\begin{aligned}
f_{\mathrm{BI}}={}&1-\frac{2m}{r}+\frac{\kappa q_{T}Q^{2}}{2q_{0}r^{2}}-\frac{\kappa q_{T}Q^{4}}{40q_{0}\beta_{\mathrm{BI}}^{2}r^{6}}+\frac{\kappa q_{T}Q^{6}}{144q_{0}\beta_{\mathrm{BI}}^{4}r^{10}} +\mathcal O\!\left(\frac{\kappa q_{T}Q^{8}}{q_{0}\beta_{\mathrm{BI}}^{6}r^{14}}\right).
\end{aligned}
\label{eq:weak-nonlinearity}
\end{equation}

The Maxwell term is followed by the first two nonlinear corrections with the normalization of Eq.~\eqref{eq:maxwell-lagrangian}. In this limit, $a=\sqrt{q_{T}}Q/r$ and $\lambda=-\xi Q^{2}/(2q_{0}r^{4})$. At $\xi=0$, one has $q_{0}=q_{T}=1$ and $\lambda=0$, recovering the Einstein--Born--Infeld geometry~\cite{CaiPangWang2004}. This limit applies to the arbitrary coupling solution with $u_{1}=0$.

For $Q=0$, the matter integral vanishes, and $f=1-2m/r$ with $\lambda=0$. Taking $u_{0}=u_{1}=0$ gives the radial Casana solution~\cite{Casana2018}, which is also the zero cosmological constant limit of Maluf and Neves~\cite{MalufNeves2021}. With a constant nonzero temporal component, the neutral metric has $q=q_{T}$ and lies within the closed bumblebee constructions discussed in Ref.~\cite{Ovcharenko2026}. The term in Eq.~\eqref{eq:metric-solution} is the nonlinear electromagnetic deformation of these neutral geometries.

Our earlier \textit{metric--affine} vacuum solutions provide a complementary geometric comparison~\cite{AraujoFilho2023,AraujoFilho2026}. Their independent connection is governed by an auxiliary metric, and the curvature interaction uses the traceless tensor $B^{\mu}B^{\nu}-B^{2}g^{\mu\nu}/4$. In physical areal coordinates and normalized time, the metrics studied there take the form
\begin{equation}
\mathrm{d}s_{\mathrm A}^{2}=-F_{\mathrm A}(R)\mathrm{d}t_{\infty}^{2}+\frac{\mathcal D}{F_{\mathrm A}(R)}\mathrm{d}R^{2}+R^{2}\mathrm{d}\Omega^{2},\qquad F_{\mathrm A}(R)=1-\frac{2\mathcal M_{\mathrm A}}R.
\label{eq:affine-comparison}
\end{equation}

Using the notation $x_{\mathrm A}=\xi_{\mathrm A}b_{\mathrm A}^{2}$ and $y_{\mathrm A}=\xi_{\mathrm A}b_{0,\mathrm A}^{2}$, their deformation is
\begin{equation}
\mathcal D=\frac{1+3x_{\mathrm A}/4}{(1-x_{\mathrm A}/4)\left[1-y_{\mathrm A}\sqrt{(1-x_{\mathrm A}/4)/(1+3x_{\mathrm A}/4)}\right]}.
\label{eq:affine-deformation}
\end{equation}

The radial solution of Ref.~\cite{AraujoFilho2023} corresponds to $y_{\mathrm A}=0$. Their neutral physical geometries agree with the present neutral metric when $\mathcal D=q_{T}$ and $\mathcal M_{\mathrm A}=m$. A nonzero $Q$ supplies the additional radial function $\mathcal J/r$ in Eq.~\eqref{eq:metric-solution}, beyond a constant conical deformation. The multiplier profile and, for Eq.~\eqref{eq:nonclosed-solution}, the nonvanishing invariant $H_{\mu\nu}H^{\mu\nu}$ further distinguish the underlying fields.

The relation to existing bumblebee solutions with nonlinear electrodynamics is equally specific. Li and L\"u~\cite{LiLu2026} construct charged geometries using an auxiliary electromagnetic scalar and direct couplings between matter and the bumblebee field, with the norm potential evaluated at $U=U_{Y}=0$. Here, the linear norm constraint leaves $V_{X}=\lambda(r)/2$, and Eq.~\eqref{eq:lambda-pressure} determines this response from the electromagnetic pressure. The resulting solution combines minimal Born--Infeld matter, temporal and radial bumblebee components, and an explicit future horizon extension within a single set of field equations.


\subsection{Horizon structure and parameter domains}
\label{subsec:horizon_parameter_domains}

The lapse functions obtained above permit a direct comparison of the horizons supported by the three bumblebee configurations. We use the exact electromagnetic contribution throughout and count its positive radius zeros. For the parameter plots, the dimensionless bare charge, nonlinear scale, and temporal amplitude are written as
\begin{equation}
\widehat Q=\frac{\sqrt{\kappa}|Q|}{m},\qquad \widehat\beta=\sqrt{\kappa}\,m\beta_{\mathrm{BI}},\qquad \widehat u_0=\frac{u_0}{b}.
\label{eq:horizon_scaled_parameters}
\end{equation}
All displayed configurations have \(\kappa b^2=1\) and \(\xi/\kappa=1/2\). The radial, constant temporal, and varying temporal cases are examined through different parameter slices, so that each figure resolves a dependence present in its lapse. Moreover, the horizon domains are determined from the exact minimum of \(r f_{\mathrm{BI}}\), which also locates a degenerate zero.

To obtain this criterion, it is convenient to introduce
\begin{equation}
\mathcal A=\frac{\kappa q_T}{q_0},\qquad Z(r)=r f_{\mathrm{BI}}(r),\qquad Z'(r)=1-\mathcal A r^2\rho(r),\qquad Z''(r)=2\mathcal A r p_{\perp}(r)>0\quad(Q\neq0).
\label{eq:horizon_convexity}
\end{equation}
The positivity of the transverse pressure gives a strictly convex function. In addition, the finite electromagnetic energy integral at the origin determines the mass scale
\begin{equation}
m_0=\frac{\mathcal A\,\Gamma(1/4)^2}{12\sqrt{\pi}}\sqrt{\beta_{\mathrm{BI}}}\,|Q|^{3/2},\qquad Z(0^{+})=2(m_0-m),\qquad \chi=\mathcal A\beta_{\mathrm{BI}}|Q|.
\label{eq:horizon_central_mass}
\end{equation}
For \(\chi\leq1\), \(Z\) increases throughout the positive radial interval. The geometry has one positive simple horizon for \(m>m_0\), whereas \(m\leq m_0\) belongs to the region with zero positive horizons. For \(\chi>1\), the unique minimum occurs at
\begin{equation}
r_{\mathrm e}^{\,2}=\frac{\mathcal A Q^2}{2}-\frac{1}{2\mathcal A\beta_{\mathrm{BI}}^2},\qquad m_{\mathrm e}=\frac{r_{\mathrm e}+\mathcal A J(r_{\mathrm e})}{2}<m_0.
\label{eq:horizon_extremal_mass}
\end{equation}
In this regime, the number of distinct positive horizon radii is
\begin{equation}
N_h=\begin{cases}0,&m<m_{\mathrm e},\\1\ \text{degenerate},&m=m_{\mathrm e},\\2\ \text{simple},&m_{\mathrm e}<m<m_0,\\1\ \text{simple},&m\geq m_0.\end{cases}
\label{eq:horizon_classification}
\end{equation}
The extremal curve therefore describes the merger of two horizons at a positive radius. On the other hand, at \(m=m_0\) the inner radius reaches the origin. When \(\chi>1\), the outer horizon remains positive at this transition; when \(\chi\leq1\), the transition marks the loss of the sole positive horizon. These two boundaries are distinguished in the parameter maps.

\begin{figure}[tbp]
\centering
\includegraphics[width=0.49\linewidth]{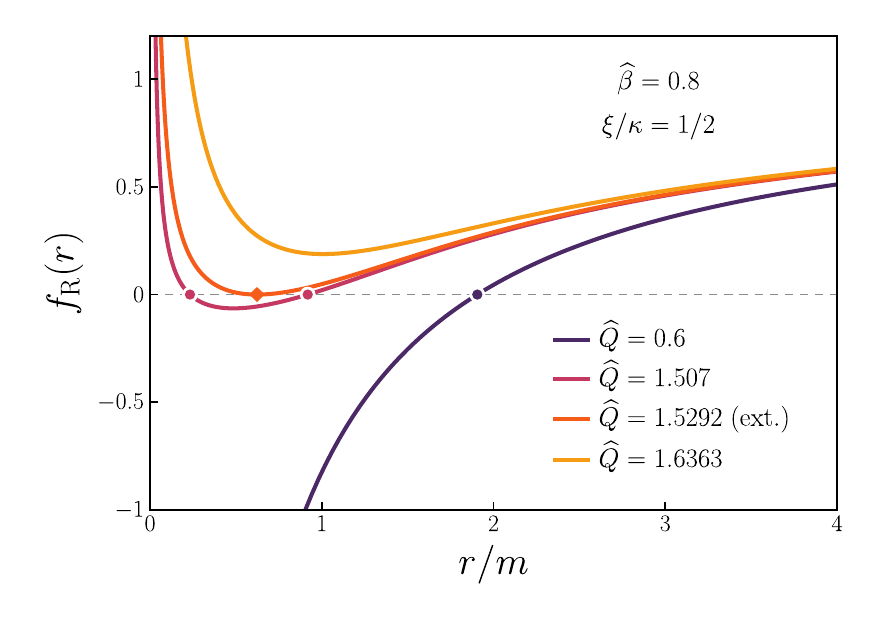}\hfill\includegraphics[width=0.49\linewidth]{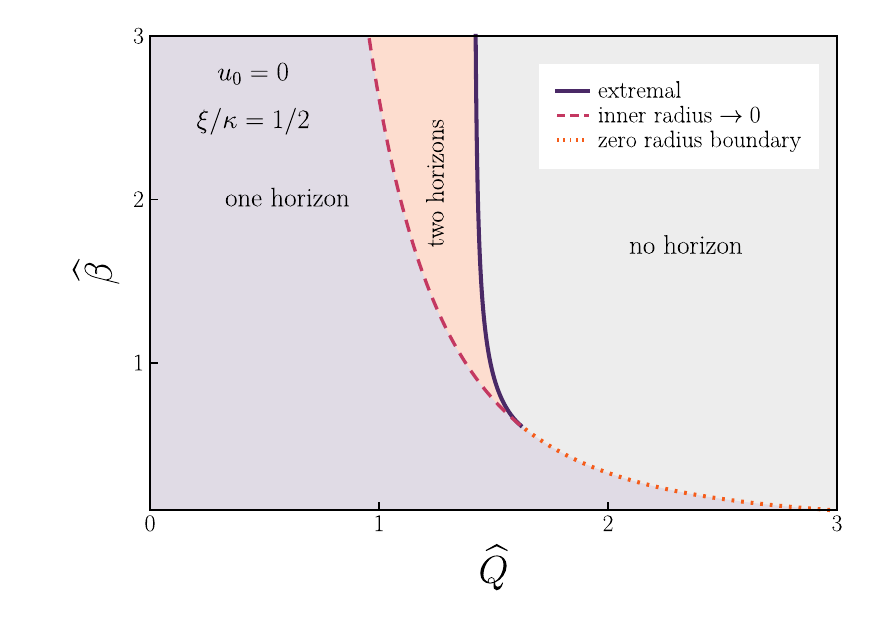}
\caption{Lapse of the purely radial configuration for selected bare charges at \(\widehat\beta=0.8\) (left), and its metric horizon domains in the \((\widehat Q,\widehat\beta)\) plane (right). The coupling combinations are \(\kappa b^2=1\) and \(\xi/\kappa=1/2\). Colored circles mark simple positive roots, while the diamond marks a degenerate root. The maps distinguish one simple horizon, two simple horizons, and zero positive horizons. The boundary labels identify finite radius extremality and the central transition separately.}
\label{fig:horizon_radial}
\end{figure}

Figure~\ref{fig:horizon_radial} shows the radial solution as the bare charge varies at fixed nonlinear scale. Since its lapse retains the Einstein--Born--Infeld form, the bumblebee radial deformation preserves these horizon radii at fixed \(m,Q,\beta_{\mathrm{BI}}\). For \(\widehat\beta=0.8\), the central transition occurs at \(\widehat Q\simeq1.484679\), while the two horizons merge at \(\widehat Q\simeq1.529240\) and \(r_{\mathrm e}/m\simeq0.622926\). In other words, the charge values between these transitions support two positive horizons.

\begin{figure}[tbp]
\centering
\includegraphics[width=0.49\linewidth]{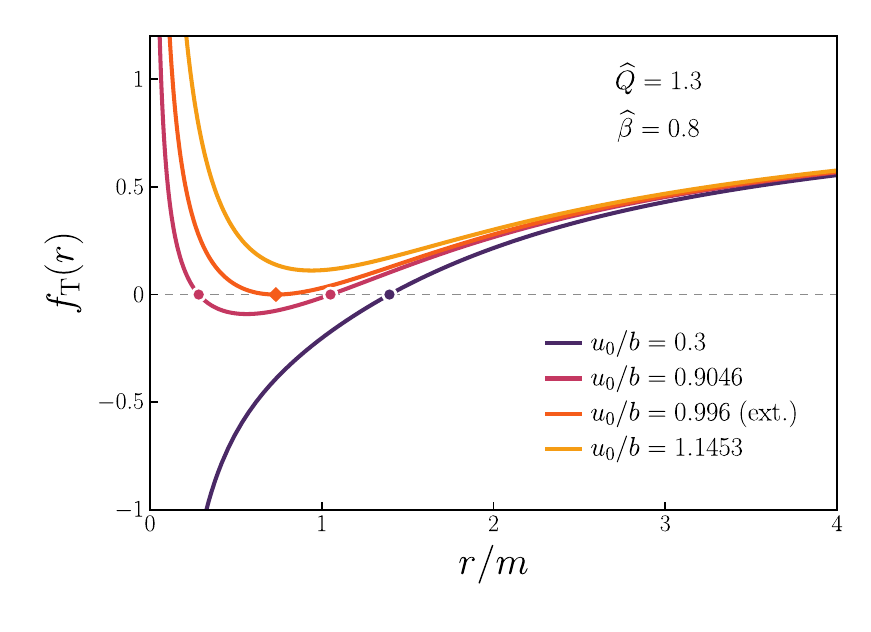}\hfill\includegraphics[width=0.49\linewidth]{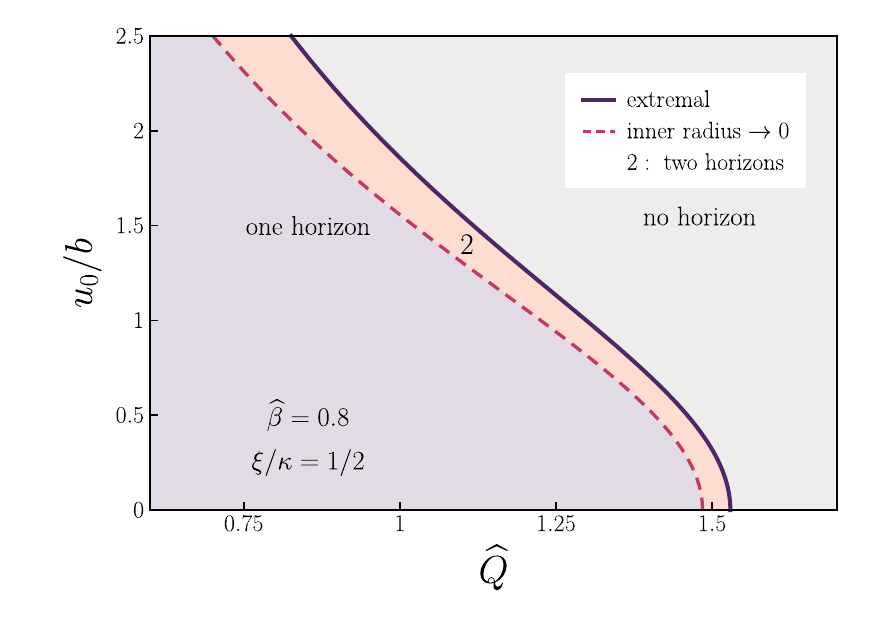}
\caption{Lapse of the constant temporal configuration for selected \(\widehat u_0\) at \(\widehat Q=1.3\) and \(\widehat\beta=0.8\) (left), and metric horizon domains in the \((\widehat Q,\widehat u_0)\) plane at the same nonlinear scale (right). The coupling combinations and the horizon markers follow Fig.~\ref{fig:horizon_radial}. }
\label{fig:horizon_constant_temporal}
\end{figure}

The constant temporal amplitude also changes the electromagnetic contribution to the lapse. For positive \(\xi\), increasing \(\widehat u_0\) raises \(f_T\) at a prescribed radius. Figure~\ref{fig:horizon_constant_temporal} follows this effect at \(\widehat Q=1.3\) and \(\widehat\beta=0.8\). The inner horizon appears from the origin at \(\widehat u_0\simeq0.813307\), and the two horizons merge at \(\widehat u_0\simeq0.995953\), with \(r_{\mathrm e}/m\simeq0.732986\). The corresponding parameter map shows how the temporal amplitude changes the charge interval supporting a metric black hole.

\begin{figure}[tbp]
\centering
\includegraphics[width=0.49\linewidth]{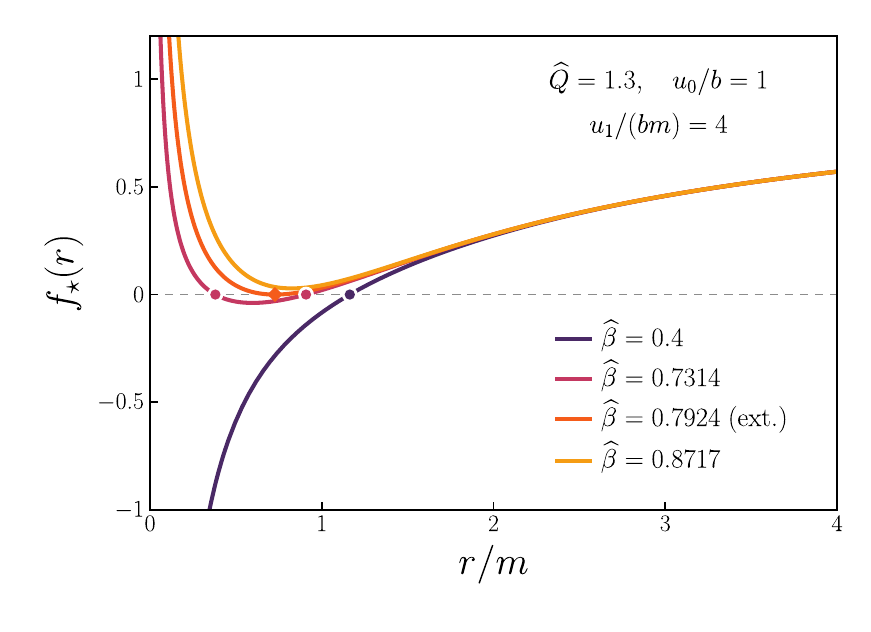}\hfill\includegraphics[width=0.49\linewidth]{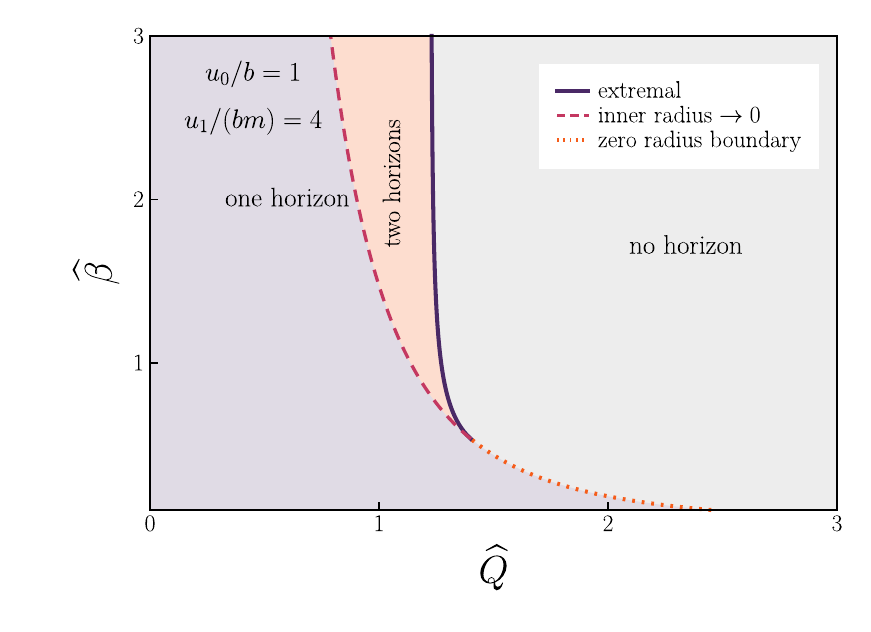}
\caption{Lapse of the varying temporal configuration for selected \(\widehat\beta\) at \(\widehat Q=1.3\) (left), and metric horizon domains in the \((\widehat Q,\widehat\beta)\) plane (right). Both panels use \(\widehat u_0=1\), \(\kappa b^2=1\), and \(\xi/\kappa=1/2\). The independent amplitude satisfies \(u_0u_1=4b^2m\). Its value leaves the metric horizon map unchanged. The marker conventions follow Fig.~\ref{fig:horizon_radial}.}
\label{fig:horizon_varying_temporal}
\end{figure}

For the varying temporal configuration, Figure~\ref{fig:horizon_varying_temporal} instead varies the nonlinear scale at \(\widehat Q=1.3\) and \(\widehat u_0=1\). Its metric coincides with the constant temporal solution at matched parameters and \(\xi=\kappa/2\), so every allowed \(u_1\) gives the same horizon locations. The displayed nonzero \(u_1\) is chosen so that \(u_0u_1=4b^2m\), satisfying the sufficient reality condition established above. The central transition occurs at \(\widehat\beta\simeq0.670317\); the horizons merge at \(\widehat\beta\simeq0.792423\), with \(r_{\mathrm e}/m\simeq0.727647\). At fixed charge, \(\partial\rho/\partial\beta_{\mathrm{BI}}=\beta_{\mathrm{BI}}(S-1)^2/S>0\), so increasing the nonlinear scale raises the lapse and moves the two horizons toward their merger. In summary, the figures separate the effects of electromagnetic nonlinearity from those of the temporal amplitude.

\begin{figure}[tbp]
\centering
\includegraphics[width=0.55\linewidth]{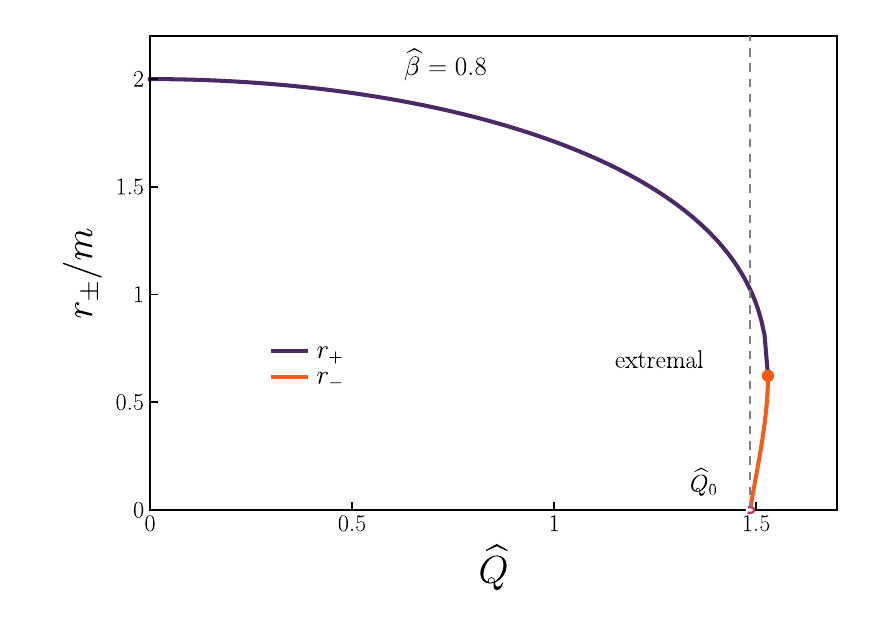}
\caption{Positive horizon radii of the radial configuration as functions of \(\widehat Q\) at \(\widehat\beta=0.8\). The inner branch begins at zero radius at the central transition and joins the outer branch at the positive extremal radius. The zero radius endpoint is shown as a limiting point and is excluded from the positive horizon count. The coupling combinations follow Fig.~\ref{fig:horizon_radial}.}
\label{fig:horizon_branches}
\end{figure}

The movement of the two radii is displayed directly in Fig.~\ref{fig:horizon_branches}. Along a simple horizon, differentiation at fixed \(m,\mathcal A,\beta_{\mathrm{BI}}\) gives
\begin{equation}
\frac{\partial r_h}{\partial |Q|}=-\frac{\mathcal A |Q|}{1-\mathcal A r_h^2\rho(r_h)}\int_{r_h}^{\infty}\frac{\mathrm{d}x}{x^2S(x)}.
\label{eq:horizon_motion}
\end{equation}
The denominator is positive at the outer horizon and negative at the inner horizon. Thereby, increasing the charge contracts the outer radius and expands the inner one toward their common extremal value. The region maps naturally classify the metric horizons; the accompanying vector fields keep the horizon regularity properties derived in the preceding discussion.

\section{Geodesic motion}
\label{sec:geodesic_motion}

The three charged solutions obtained above share a radial structure that permits a common treatment of particle motion. We use the notation of the preceding section and consider a simple outer horizon, with $f_{\mathrm{BI}}>0$ in the exterior. The analysis applies to neutral massive particles and to null probes minimally coupled to the spacetime metric. The electromagnetic characteristics of the Born--Infeld field are considered in the subsequent optical analysis \cite{Breton2002Geodesics,Novello2000NLED}. Our purpose here is to determine the capture threshold, the circular and eccentric timelike orbits, and the timescales that distinguish the three bumblebee configurations.

\subsection{Conserved quantities and radial dynamics}
\label{subsec:geodesic_dynamics}

For an affine parameter $\lambda$, the geodesic Lagrangian in the equatorial plane is
\begin{equation}
2\mathcal{L}=-f_{\mathrm{BI}}\dot{t}^{\,2}+\frac{q_T}{f_{\mathrm{BI}}}\dot{r}^{\,2}+r^2\dot{\phi}^{\,2}=-\varepsilon,\qquad E=f_{\mathrm{BI}}\dot{t},\qquad L=r^2\dot{\phi}.
\label{eq:geo_integrals}
\end{equation}
Here, a dot denotes $\mathrm{d}/\mathrm{d}\lambda$, while $\varepsilon=0$ and $\varepsilon=1$ describe null and timelike motion, respectively. For a massive particle, $\lambda$ is its proper time and the conserved quantities are expressed per unit rest mass. We choose $E>0$ and orient the orbital plane so that $L>0$ for the nonradial trajectories. Spherical symmetry preserves the initial conditions $\theta=\pi/2$ and $\dot{\theta}=0$. Substitution of the conserved quantities into the normalization condition gives
\begin{equation}
q_T\dot{r}^{\,2}=\mathcal{R}_{\varepsilon}(r),\qquad \mathcal{R}_{\varepsilon}(r)=E^2-V_{\varepsilon}(r),\qquad V_{\varepsilon}(r)=f_{\mathrm{BI}}(r)\left(\varepsilon+\frac{L^2}{r^2}\right).
\label{eq:geo_radial}
\end{equation}
The allowed intervals satisfy $\mathcal{R}_{\varepsilon}\geq0$, and their endpoints determine the radial turning points. In addition, an exterior maximum of $V_{\varepsilon}$ supplies a potential barrier, while a minimum supports stable oscillations. These conditions give the same effective potential description used for Einstein--Born--Infeld trajectories and neutral motion in the Reissner--Nordstr\"om geometry \cite{Breton2002Geodesics,Pugliese2011Circular}.

\begin{figure}[tbp]
\centering
\includegraphics[width=0.49\linewidth]{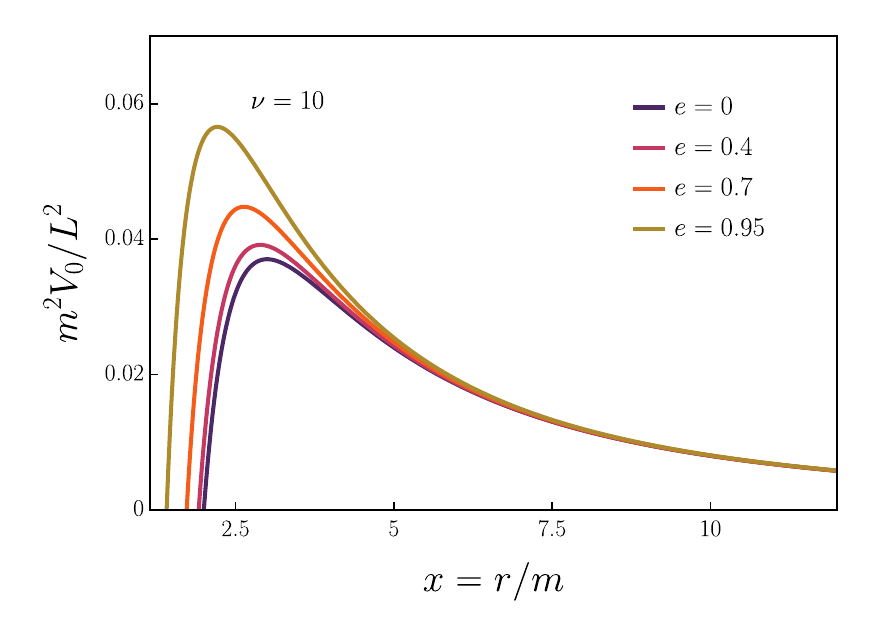}\hfill\includegraphics[width=0.49\linewidth]{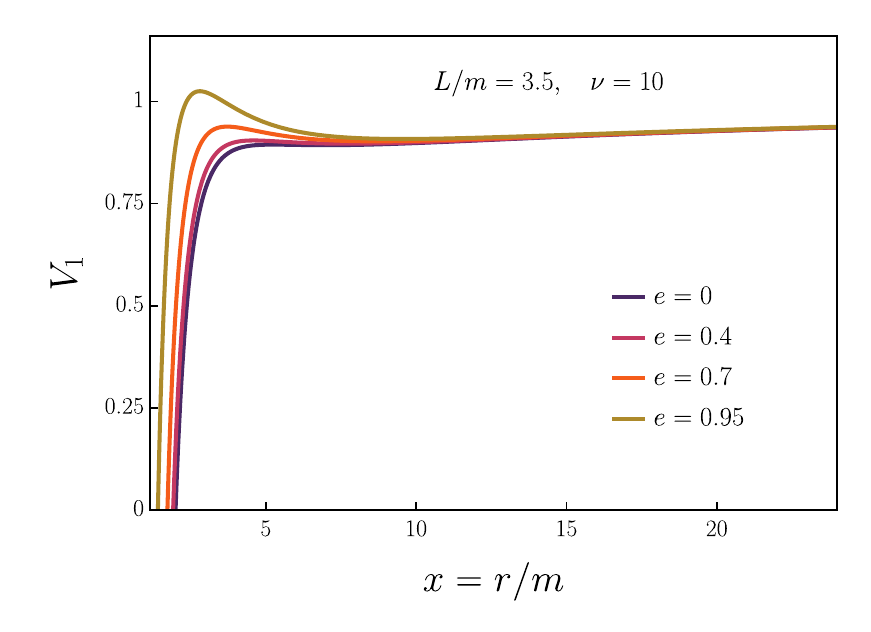}
\caption{Metric null potential $m^2V_0/L^2$ (left) and timelike potential $V_1$ for $L/m=3.5$ (right), evaluated with the exact charged lapse. The charged curves have $\nu=10$, while the neutral curve has $e=\nu=0$. Every displayed geometry possesses a simple outer horizon. Their independence of an additional variation of $q_T$ assumes that $m,e,\nu$ are held fixed.}
\label{fig:geo_potentials}
\end{figure}

For the numerical comparisons, it is convenient to introduce the dimensionless combinations
\begin{equation}
x=\frac{r}{m},\qquad e^2=\frac{\kappa q_TQ^2}{2q_0m^2},\qquad \nu=\frac{Q^2}{\beta_{\mathrm{BI}}^2m^4}.
\label{eq:geo_dimensionless}
\end{equation}
The parameter $e$ measures the charge coefficient in the Maxwell limit, whereas $\nu/x^4$ measures the electromagnetic nonlinearity at radius $x$. The neutral geometry corresponds to $e=\nu=0$. At fixed nonzero charge, $\nu=0$ denotes the Maxwell limit. Figure~\ref{fig:geo_potentials} displays the exact potentials for several values of $e$. At the displayed parameters, increasing the charge raises the null barrier and shifts its maximum inward. The timelike potential exhibits both a maximum and a minimum, thereby separating capture from the bound trajectories considered below.

An evolution system that passes smoothly through the turning points follows directly from Eq.~\eqref{eq:geo_radial}:
\begin{equation}
\dot{r}=v_r,\qquad \dot{v}_r=-\frac{V_{\varepsilon}'(r)}{2q_T},\qquad \dot{\phi}=\frac{L}{r^2}.
\label{eq:geo_evolution}
\end{equation}
We integrate this system with the exact lapse and monitor Eq.~\eqref{eq:geo_radial} along each trajectory. The radial and angular equations remain regular at a simple horizon, which shows the endpoint for the exterior capture curves. For $L\neq0$, the inverse radius $w=1/r$ also gives
\begin{equation}
q_T\frac{\mathrm{d}^2w}{\mathrm{d}\phi^2}=-\frac{1}{2}\frac{\mathrm{d}}{\mathrm{d}w}\left[f_{\mathrm{BI}}(1/w)\left(\frac{\varepsilon}{L^2}+w^2\right)\right].
\label{eq:geo_orbit}
\end{equation}
In this regard, at fixed lapse and conserved quantities, the radial deformation changes the angular accumulation through $\phi/\sqrt{q_T}$. More precisely, if $r_1$ denotes the solution with $q_T=1$, the same $E,L$, and the same initial radial orientation, the aligned trajectories obey
\begin{equation}
r_{q_T}(\phi)=r_1\!\left(\frac{\phi-\phi_0}{\sqrt{q_T}}\right),\qquad r_1(0)=r_0.
\label{eq:geo_angular_scaling}
\end{equation}
This relation isolates the contribution of the radial metric coefficient. A comparison at fixed $m,Q,\beta_{\mathrm{BI}}$ also takes into account the change of the lapse produced by a temporal bumblebee amplitude.

\subsection{Null trajectories, scattering, and capture}
\label{subsec:geodesic_null}

For a metric null probe, the ratio $d_{\gamma}=L/E$ is invariant under an affine rescaling, and the radial equation becomes
\begin{equation}
q_T\dot{r}^{\,2}=E^2\left[1-d_{\gamma}^{\,2}W_g(r)\right],\qquad W_g(r)=\frac{f_{\mathrm{BI}}(r)}{r^2}.
\label{eq:geo_null_radial}
\end{equation}
The exterior capture orbit is selected by the global maximum of $W_g$. Denoting its radius by $r_{\mathrm{ph},g}$, we have \cite{Claudel2001Photon}
\begin{equation}
W_g'(r_{\mathrm{ph},g})=0,\qquad W_g''(r_{\mathrm{ph},g})<0,\qquad d_{\mathrm{c},g}^{\,2}=\frac{1}{W_g(r_{\mathrm{ph},g})}.
\label{eq:geo_null_critical}
\end{equation}
An incident ray with $d_{\gamma}<d_{\mathrm{c},g}$ crosses the barrier and reaches the horizon. For $d_{\gamma}>d_{\mathrm{c},g}$, the outer turning point satisfies $d_{\gamma}^{\,2}W_g(r_{\min})=1$, and the ray returns to the distant region. At the threshold, the limiting trajectory approaches the unstable circular orbit. These statements concern rays incident from outside the outermost relevant barrier; notice that the global maximum prescription keeps the appropriate capture boundary when further stationary points occur.

\begin{figure}[tbp]
\centering
\includegraphics[width=0.49\linewidth]{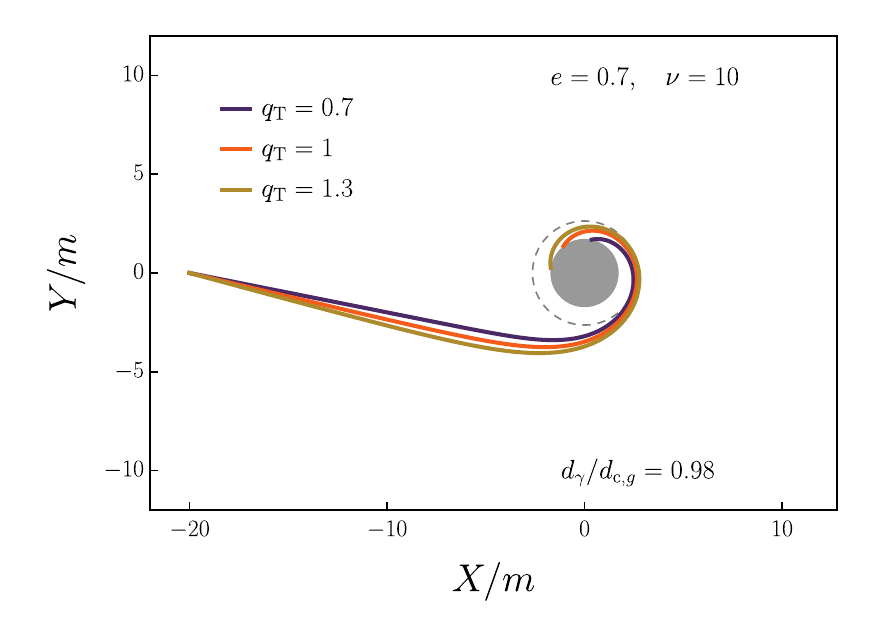}\hfill\includegraphics[width=0.49\linewidth]{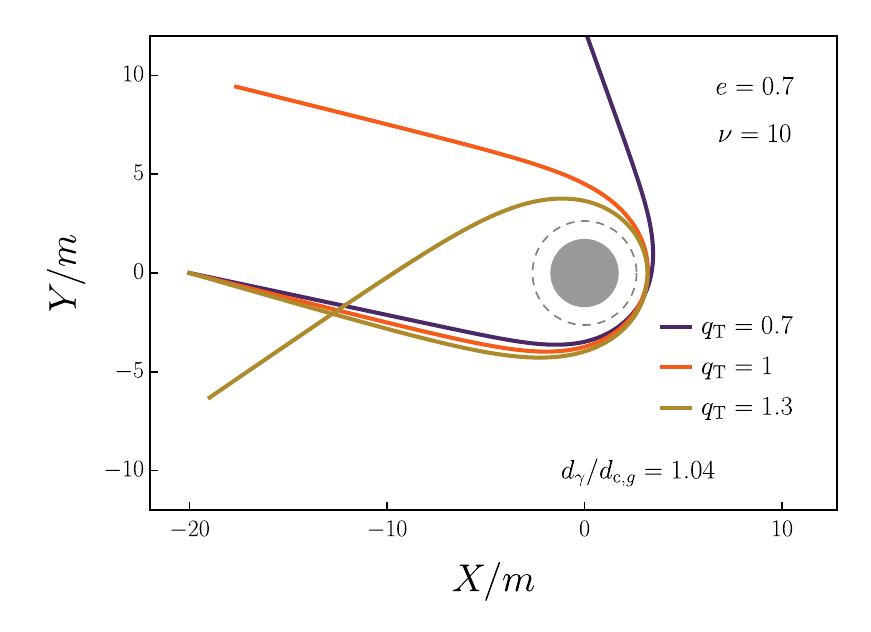}
\caption{Captured (left) and scattered (right) metric null trajectories for $e=0.7$, $\nu=10$, and $q_T=0.7,1,1.3$. The launch data are $E=1$, $r_0=20m$, and $\phi_0=\pi$, with $d_{\gamma}/d_{\mathrm{c},g}=0.98$ and $1.04$, respectively. The gray disk indicates the outer horizon and the dashed circle the metric photon orbit. All curves in each panel share the same lapse and local incident direction.}
\label{fig:geo_null_paths}
\end{figure}

To compare the same incident ray across different radial deformations, we specify $E$ and $d_{\gamma}$, instead of an independent coordinate radial velocity. At a common initial radius, the ingoing data and the local angle $\psi_0$ measured from the inward radial direction by a static observer satisfy
\begin{equation}
\dot{r}_0=-\frac{E}{\sqrt{q_T}}\sqrt{1-\frac{d_{\gamma}^{\,2}f_{\mathrm{BI}}(r_0)}{r_0^2}},\qquad \dot{\phi}_0=\frac{Ed_{\gamma}}{r_0^2},\qquad \sin\psi_0=\frac{d_{\gamma}\sqrt{f_{\mathrm{BI}}(r_0)}}{r_0}.
\label{eq:geo_null_launch}
\end{equation}
In this manner, fixing the lapse and $d_{\gamma}$ also fixes the local launch direction. Figure~\ref{fig:geo_null_paths} shows capture and scattering at common values of $d_{\gamma}/d_{\mathrm{c},g}$. The larger radial deformation increases the angular winding in both families, while the capture classification remains fixed. The displayed coordinates $X=r\cos\phi$ and $Y=r\sin\phi$ represent areal radius and azimuth; proper radial lengths follow from the metric.

\subsection{Angular winding and the instability of critical motion}
\label{subsec:geodesic_critical}

For a scattered metric ray with source and receiver at large radius, the total change of azimuth is
\begin{equation}
\Delta\phi_g(d_{\gamma})=2\sqrt{q_T}\int_{r_{\min}}^{\infty}\frac{d_{\gamma}\,\mathrm{d}r}{r^2\sqrt{1-d_{\gamma}^{\,2}W_g(r)}}.
\label{eq:geo_null_angle}
\end{equation}
The asymptotic radial deformation requires a specified angular reference. Setting the lapse to unity in this integral gives $\Delta\phi_{\mathrm{cone}}=\pi\sqrt{q_T}$. In this manner, the excess over a coordinate angle $\pi$ and the additional bending relative to the same conical background are related by
\begin{equation}
\alpha_g=\Delta\phi_g-\pi=\pi\left(\sqrt{q_T}-1\right)+\alpha_g^{\mathrm{cone}},\qquad \alpha_g^{\mathrm{cone}}=\Delta\phi_g-\pi\sqrt{q_T}.
\label{eq:geo_angle_reference}
\end{equation}
This separation keeps the asymptotic geometry in the interpretation of the trajectory \cite{Gibbons2008GaussBonnet}. The finite distance angles and the weak field observables are considered in the Solar System section. Here the exact integral also determines the repeated azimuthal revolutions near the capture threshold.

Let $f_c=f_{\mathrm{BI}}(r_{\mathrm{ph},g})$, with the same subscript denoting evaluation of its derivatives. A perturbation of the critical radius grows as $\delta r\propto\exp(\lambda_{t,g}t)$, where \cite{Cardoso2009Lyapunov}
\begin{equation}
\lambda_{t,g}^{\,2}=-\frac{f_cr_{\mathrm{ph},g}^{\,2}}{2q_T}W_g''(r_{\mathrm{ph},g})=\frac{f_c\left(2f_c-r_{\mathrm{ph},g}^{\,2}f_c''\right)}{2q_Tr_{\mathrm{ph},g}^{\,2}},\qquad \Omega_{\mathrm{ph},g}=\frac{\sqrt{f_c}}{r_{\mathrm{ph},g}}=\frac{1}{d_{\mathrm{c},g}}.
\label{eq:geo_null_lyapunov}
\end{equation}
The exponent is expressed in the asymptotically normalized Killing time already adopted for the solutions. At fixed lapse, $\lambda_{t,g}$ scales as $q_T^{-1/2}$, whereas $\Omega_{\mathrm{ph},g}$ retains its value. In other words, the radial deformation increases the time spent near the circular orbit while preserving its orbital frequency.

Moreover, expansion of Eq.~\eqref{eq:geo_null_angle} around a nondegenerate exterior maximum yields the logarithmic behavior \cite{Bozza2002Strong}
\begin{equation}
\Delta\phi_g=-\overline{a}_g\ln\left(\frac{d_{\gamma}}{d_{\mathrm{c},g}}-1\right)+\overline{b}_{\phi,g}+o(1),\qquad \overline{a}_g=\sqrt{\frac{2q_T}{2f_c-r_{\mathrm{ph},g}^{\,2}f_c''}}=\frac{\Omega_{\mathrm{ph},g}}{\lambda_{t,g}}.
\label{eq:geo_strong_angle}
\end{equation}
The finite coefficient $\overline{b}_{\phi,g}$ follows by subtracting the displayed logarithm from the exact angular integral and taking $d_{\gamma}\to d_{\mathrm{c},g}^{+}$. In this regard, the same local instability that controls departure from the photon orbit fixes the divergent angular winding. The neutral limit gives $\overline{a}_g=\sqrt{q_T}$ and $\lambda_{t,g}=1/(3\sqrt{3q_T}\,m)$, which naturally shows a direct check against the neutral bumblebee exterior \cite{Casana2018Bumblebee}.

\subsection{Circular timelike orbits and marginal stability}
\label{subsec:geodesic_circular}

Circular massive trajectories satisfy $\mathcal{R}_1=\mathcal{R}_1'=0$. Solving these two conditions gives
\begin{equation}
E_c^2=\frac{2f_{\mathrm{BI}}^2}{D},\qquad L_c^2=\frac{r^3f_{\mathrm{BI}}'}{D},\qquad D(r)=2f_{\mathrm{BI}}-rf_{\mathrm{BI}}',\qquad \Omega_{\phi}^2=\left(\frac{\mathrm{d}\phi}{\mathrm{d}t}\right)^2=\frac{f_{\mathrm{BI}}'}{2r}.
\label{eq:geo_circular_integrals}
\end{equation}
The orbit lies in the static exterior and obeys $D>0$ and $f_{\mathrm{BI}}'>0$. Its stability follows from the second derivative of the potential at fixed $L$. Let us define
\begin{equation}
\mathcal{S}(r)=f_{\mathrm{BI}}f_{\mathrm{BI}}''-2\left(f_{\mathrm{BI}}'\right)^2+\frac{3f_{\mathrm{BI}}f_{\mathrm{BI}}'}{r},\qquad V_1''(r_c)=\frac{2\mathcal{S}(r_c)}{D(r_c)},
\label{eq:geo_stability_function}
\end{equation}
we obtain stable circular motion for $\mathcal{S}>0$. The innermost stable circular orbit is the inner boundary of the stable segment connected to large radius. It satisfies $\mathcal{S}(r_{\mathrm{ISCO}})=0$ together with the circular orbit conditions \cite{Pugliese2011Circular}. It is important to mention that this prescription selects the relevant orbit throughout the numerical analysis.

The matter equations derived above also provide convenient exact derivatives of the lapse,
\begin{equation}
f_{\mathrm{BI}}'=\frac{1-f_{\mathrm{BI}}}{r}-\frac{\kappa q_T}{q_0}r\rho,\qquad f_{\mathrm{BI}}''=-\frac{2(1-f_{\mathrm{BI}})}{r^2}+\frac{2\kappa q_TQ^2}{q_0r^4S}.
\label{eq:geo_lapse_derivatives}
\end{equation}
Notice that these identities show the complete Born--Infeld dependence and avoid numerical differentiation of the hypergeometric function. In addition, the marginally bound circular orbit has $E_c=1$ and is determined by
\begin{equation}
2f_{\mathrm{BI}}^2-2f_{\mathrm{BI}}+rf_{\mathrm{BI}}'=0.
\label{eq:geo_marginal_bound}
\end{equation}
Its unstable exterior case gives the capture threshold for neutral particles released with unit specific energy. For a prescribed $E>1$, the corresponding threshold follows from $E_c(r_u)=E$, $L_{\mathrm{crit}}=L_c(r_u)$, and $d_{\mathrm{mass}}=L_{\mathrm{crit}}/\sqrt{E^2-1}$, where $r_u$ belongs to the unstable case accessible from infinity.

\begin{figure}[tbp]
\centering
\includegraphics[width=0.49\linewidth]{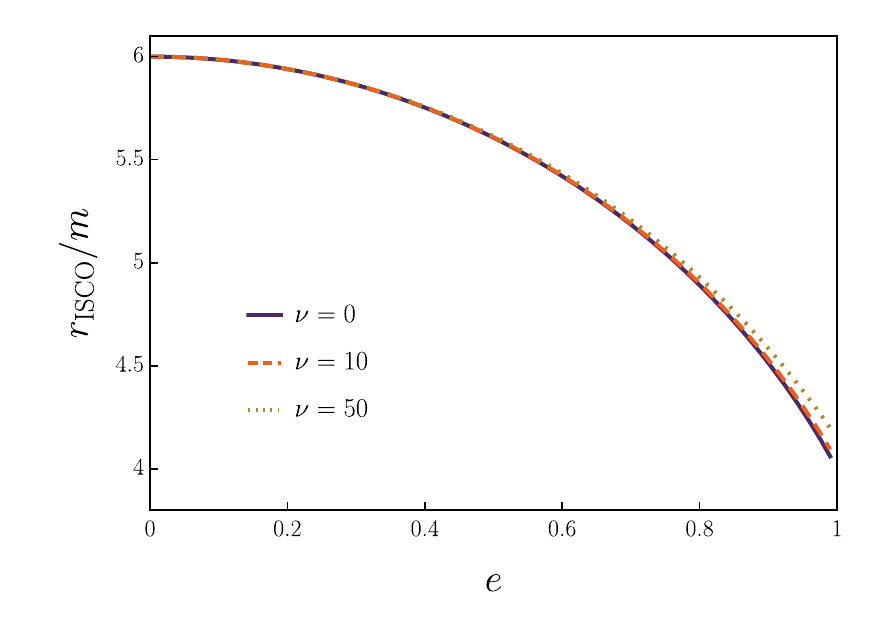}\hfill\includegraphics[width=0.49\linewidth]{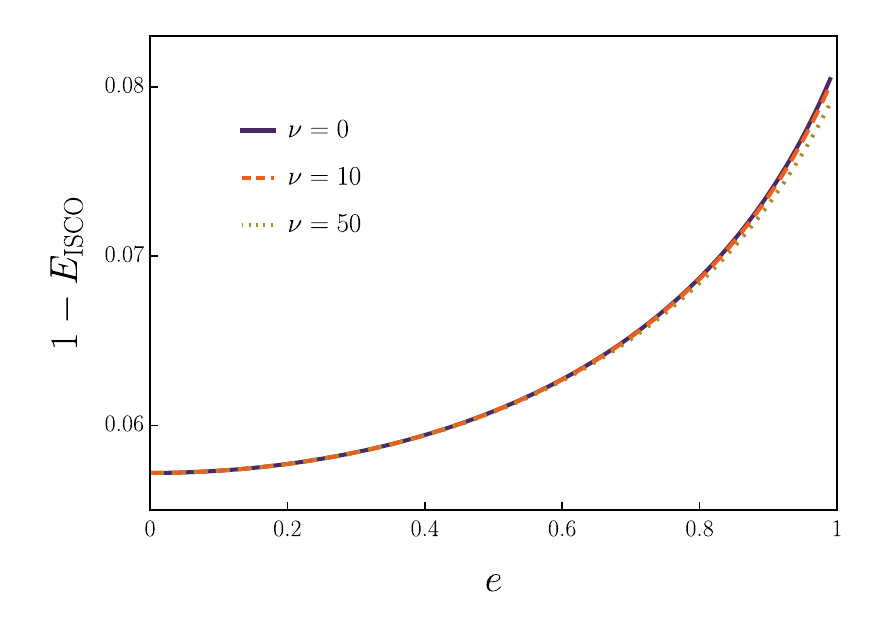}
\caption{Innermost stable circular orbit (left) and specific binding energy $1-E_{\mathrm{ISCO}}$ (right) as functions of the geometric charge for $\nu=0,10,50$. The curves use the exact lapse and the stable case connected to infinity. The physical neutral endpoint has $e=\nu=0$; the charged curves approach the same neutral metric continuously. At fixed $m,e,\nu$, both quantities keep their values under a variation of the radial coefficient $q_T$.}
\label{fig:geo_isco}
\end{figure}

The Maxwell limit supplies an analytic comparison. Writing $x_0$ for its outer physical ISCO root, the condition becomes
\begin{equation}
x_0^3-6x_0^2+9e^2x_0-4e^4=0,\qquad 0\leq e\leq1.
\label{eq:geo_rn_isco}
\end{equation}
It connects $x_0=6$ in the neutral geometry to $x_0=4$ at the extremal Maxwell charge. For $\nu/x_0^4\ll1$, the leading nonlinear displacement is
\begin{equation}
\frac{r_{\mathrm{ISCO}}}{m}=x_0+\frac{e^2\nu\left(12x_0^2+x_0-12e^2\right)}{60x_0^4\left(x_0^2-4x_0+3e^2\right)}+O(\nu^2).
\label{eq:geo_isco_bi_shift}
\end{equation}
The displayed correction is positive on the Maxwell black hole case with $0<e\leq1$. At fixed geometric charge, the leading Born--Infeld contribution moves the marginally stable orbit outward. Figure~\ref{fig:geo_isco} extends this result with the exact lapse and also displays the specific binding energy $\epsilon_{\mathrm{bind}}=1-E_{\mathrm{ISCO}}$. In other words, this quantity gives the geodesic energy available between a particle at rest in the distant region and the marginally stable circular orbit. Over the displayed range, increasing the nonlinear parameter at fixed charge reduces this binding energy.

\begin{figure}[tbp]
\centering
\includegraphics[width=0.49\linewidth]{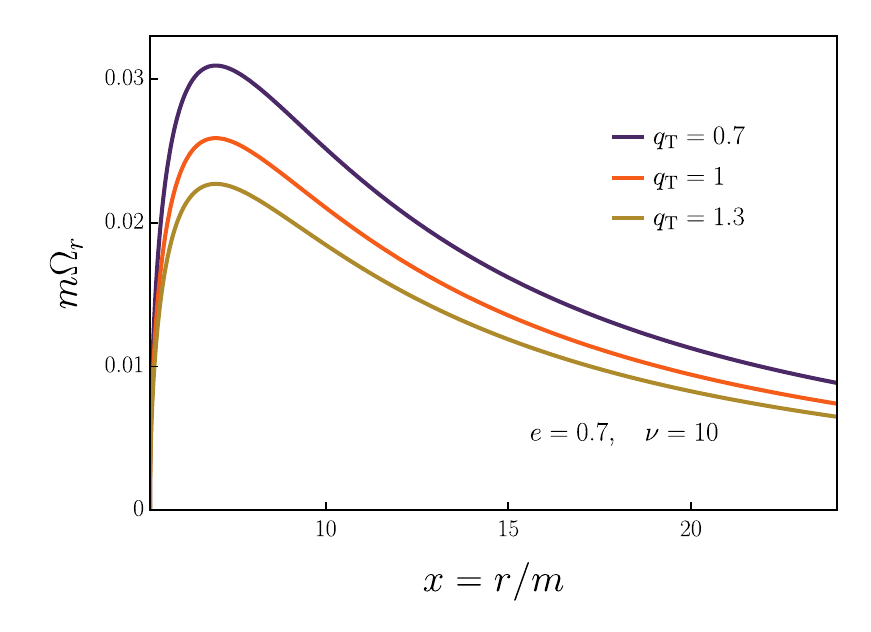}\hfill\includegraphics[width=0.49\linewidth]{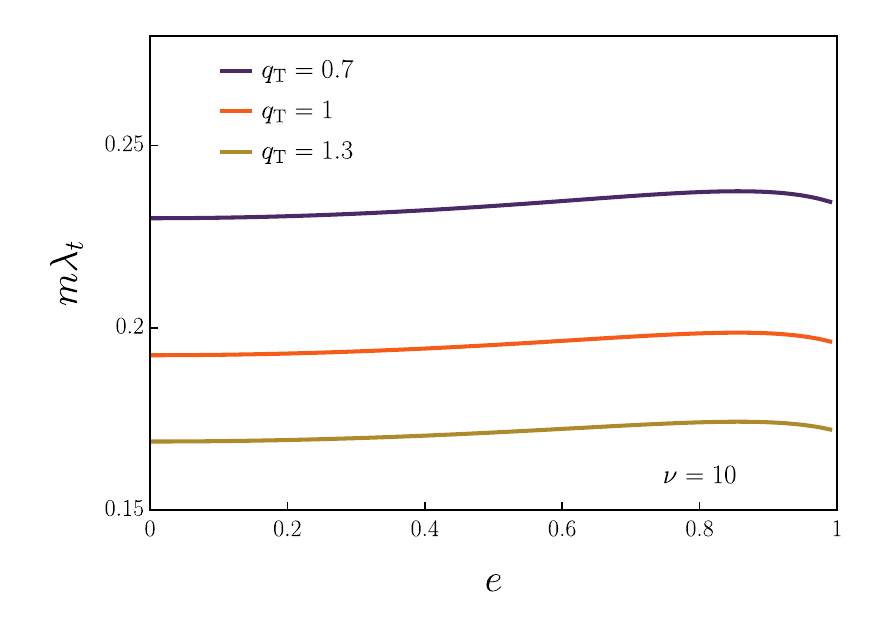}
\caption{Radial epicyclic frequency of stable timelike circular orbits for $e=0.7$ (left) and metric photon instability exponent as a function of charge (right), with $\nu=10$ and $q_T=0.7,1,1.3$. The neutral point uses $e=\nu=0$. The frequencies refer to the normalized Killing time.}
\label{fig:geo_timescales}
\end{figure}

Small radial oscillations about a stable circular orbit obey $\mathrm{d}^2\delta r/\mathrm{d}t^2+\Omega_r^2\delta r=0$, with
\begin{equation}
\Omega_r^2=\frac{\mathcal{S}}{2q_T},\qquad \frac{\Omega_r^2}{\Omega_{\phi}^2}=\frac{rf_{\mathrm{BI}}f_{\mathrm{BI}}''/f_{\mathrm{BI}}'-2rf_{\mathrm{BI}}'+3f_{\mathrm{BI}}}{q_T}.
\label{eq:geo_epicycle}
\end{equation}
On the other hand, the azimuthal frequency depends on the lapse alone. Figure~\ref{fig:geo_timescales} compares the radial oscillation frequency with the null instability exponent. Both scale as $q_T^{-1/2}$ at fixed lapse, and $\Omega_r$ vanishes at the ISCO. For the neutral solution these expressions reduce to
\begin{equation}
\Omega_{\phi}^2=\frac{m}{r^3},\qquad \Omega_r^2=\frac{m(r-6m)}{q_Tr^4},\qquad \Delta\varpi_{\mathrm{circ}}=2\pi\left[\sqrt{\frac{q_T}{1-6m/r}}-1\right].
\label{eq:geo_neutral_frequencies}
\end{equation}
The last relation gives the advance of the periapsis in the limit of small eccentricity. 

\subsection{Eccentric bound trajectories and orbital frequencies}
\label{subsec:geodesic_bound}

For a bound timelike trajectory, let $r_p$ and $r_a$ be the periastron and apastron, with $r_h<r_p<r_a$. The conditions $\mathcal{R}_1(r_p)=\mathcal{R}_1(r_a)=0$ fix its conserved quantities as
\begin{equation}
L^2=\frac{r_p^2r_a^2(f_a-f_p)}{r_a^2f_p-r_p^2f_a},\qquad E^2=\frac{f_pf_a(r_a^2-r_p^2)}{r_a^2f_p-r_p^2f_a},\qquad f_p=f_{\mathrm{BI}}(r_p),\qquad f_a=f_{\mathrm{BI}}(r_a).
\label{eq:geo_bound_integrals}
\end{equation}
We select positive $L^2$, $0<E^2<1$, and $\mathcal{R}_1>0$ between the two simple turning points. These requirements identify the exterior bound interval of the exact charged geometry. The elapsed Killing time and accumulated azimuth over a radial period are
\begin{equation}
T_r=2\sqrt{q_T}\int_{r_p}^{r_a}\frac{E\,\mathrm{d}r}{f_{\mathrm{BI}}(r)\sqrt{\mathcal{R}_1(r)}},\qquad \Phi_r=2\sqrt{q_T}\int_{r_p}^{r_a}\frac{L\,\mathrm{d}r}{r^2\sqrt{\mathcal{R}_1(r)}}.
\label{eq:geo_bound_periods}
\end{equation}
Following the organization of bound motion in terms of fundamental frequencies \cite{Schmidt2002Celestial,Fujita2009Bound}, we obtain
\begin{equation}
\Omega_r^{\mathrm{bound}}=\frac{2\pi}{T_r},\qquad \Omega_{\phi}^{\mathrm{bound}}=\frac{\Phi_r}{T_r},\qquad \Delta\varpi=\Phi_r-2\pi.
\label{eq:geo_bound_frequencies}
\end{equation}
At fixed lapse and turning radii, both $T_r$ and $\Phi_r$ scale as $\sqrt{q_T}$. As a consequence,, the mean azimuthal frequency maintains its value, while the radial frequency scales as $q_T^{-1/2}$. A closed orbit occurs when $\Phi_r/(2\pi)$ is rational; it is worth highlithing that the continuous changes of the parameters instead generate the familiar precessing patterns \cite{Levin2008Periodic}.

\begin{figure}[tbp]
\centering
\includegraphics[width=0.49\linewidth]{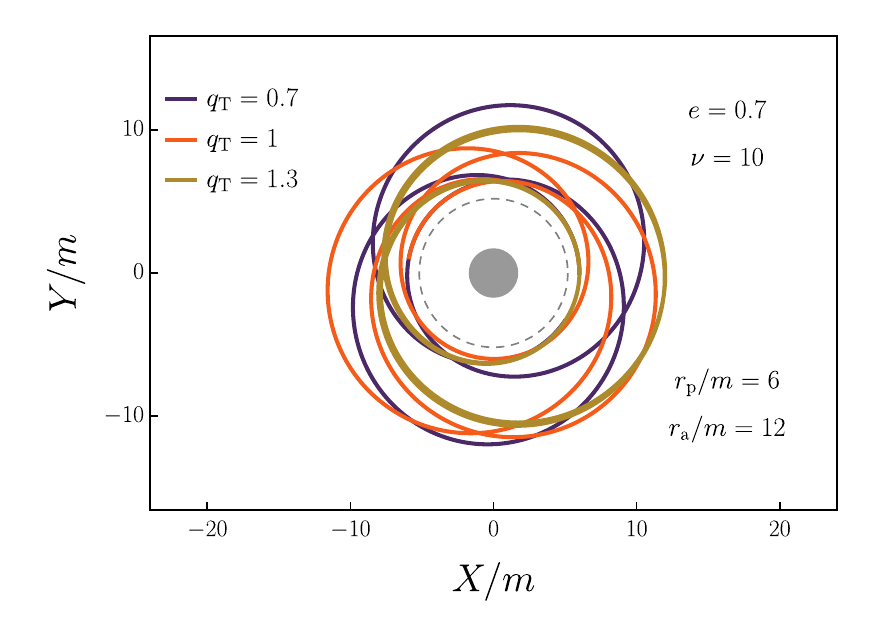}\hfill\includegraphics[width=0.49\linewidth]{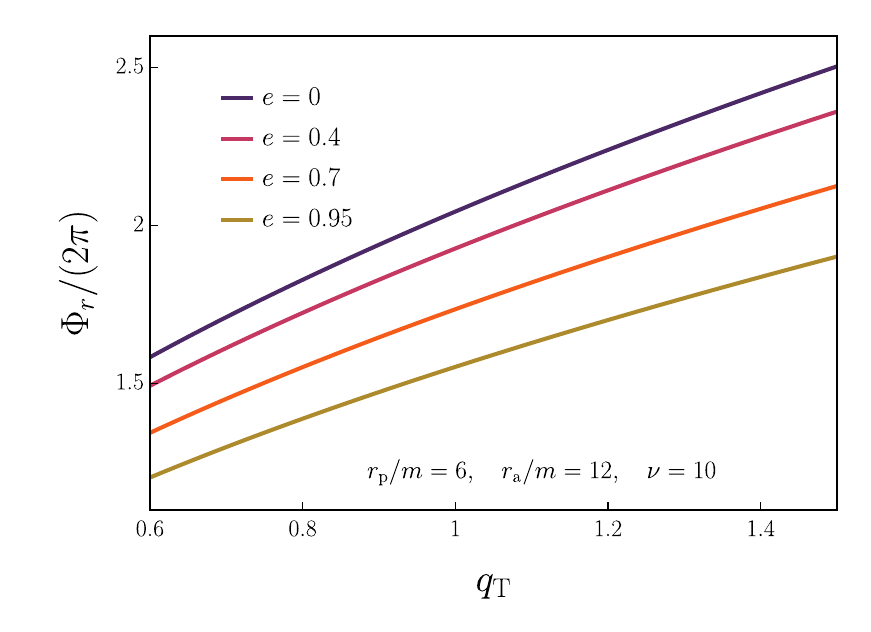}
\caption{Bound timelike trajectories with $r_p=6m$ and $r_a=12m$ (left), and accumulated azimuth $\Phi_r/(2\pi)$ per radial period (right). The trajectories have $e=0.7$, $\nu=10$, and $q_T=0.7,1,1.3$; the angle curves compare several charges at the same turning radii, with $\nu=10$ for charged configurations and $e=\nu=0$ for the neutral one. The gray disk marks the outer horizon and the dashed circle the ISCO. Each trajectory is drawn over two radial periods.}
\label{fig:geo_bound_paths}
\end{figure}

For numerical quadrature, we use the radial parametrization
\begin{equation}
r(\chi)=\frac{mp}{1+\epsilon_{\mathrm{orb}}\cos\chi},\qquad p=\frac{2r_pr_a}{m(r_p+r_a)},\qquad \epsilon_{\mathrm{orb}}=\frac{r_a-r_p}{r_a+r_p}.
\label{eq:geo_anomaly}
\end{equation}
The zeros of $\mathrm{d}r/\mathrm{d}\chi$ cancel the square root endpoint behavior of the integrals in Eq.~\eqref{eq:geo_bound_periods}. This permits an evaluation at finite electromagnetic nonlinearity with the turning radii held fixed. Figure~\ref{fig:geo_bound_paths} illustrates the resulting orbits and the accumulated angle. The dependence on $q_T$ follows the exact angular scaling, whereas the change with charge reflects the charged lapse.

The boundary between eccentric bound motion and plunge is obtained when the inner turning point becomes a double root,
\begin{equation}
\mathcal{R}_1(r_p)=0,\qquad \mathcal{R}_1'(r_p)=0,\qquad \mathcal{R}_1(r_a)=0.
\label{eq:geo_separatrix}
\end{equation}
As this boundary is approached, the particle spends an increasing fraction of its radial period near the unstable circular orbit and completes repeated azimuthal revolutions there. Besides connecting the bound trajectories to capture, these conditions determine the charged separatrix directly from the exact lapse. The neutral limit provides an analytic check:
\begin{equation}
E^2=\frac{(p-2)^2-4\epsilon_{\mathrm{orb}}^2}{p(p-3-\epsilon_{\mathrm{orb}}^2)},\qquad \frac{L^2}{m^2}=\frac{p^2}{p-3-\epsilon_{\mathrm{orb}}^2},\qquad \frac{\mathrm{d}\phi}{\mathrm{d}\chi}=\sqrt{\frac{q_Tp}{p-6-2\epsilon_{\mathrm{orb}}\cos\chi}},\qquad p>6+2\epsilon_{\mathrm{orb}}.
\label{eq:geo_neutral_bound}
\end{equation}
Its separatrix is $p=6+2\epsilon_{\mathrm{orb}}$. For the charged solutions, Eq.~\eqref{eq:geo_separatrix} supplies the corresponding boundary and retains all nonlinear electromagnetic terms. The exact periapsis observable in Eq.~\eqref{eq:geo_bound_frequencies} also gives rise to the starting point for the weak field estimates discussed later.


\subsection{Comparison of the three bumblebee configurations}
\label{subsec:geodesic_configurations}

The common equations above cover each solution through its lapse and radial coefficient. For the purely radial configuration, the lapse retains the Einstein Born--Infeld form at fixed $m,Q,\beta_{\mathrm{BI}}$. Varying the bumblebee coupling preserves its turning radii, capture impact parameter, circular specific energy, and ISCO radius. Its accumulated angles and radial timescales show the dependence on $q_0$ described above.

For a constant temporal amplitude, the electromagnetic contribution to the lapse also depends on the bumblebee parameters. At fixed $m,Q,\beta_{\mathrm{BI}}$, the circular orbit locations and the capture threshold As a consequence, change together with the radial timescales. In addition, the solution with a varying temporal amplitude at $\xi=\kappa/2$ has the same metric as the constant temporal solution with matched remaining parameters. Their complete geodesic dynamics then satisfies
\begin{equation}
\left.\mathcal{R}_{\varepsilon}^{(\star)}\right|_{\xi=\kappa/2}=\left.\mathcal{R}_{\varepsilon}^{(T)}\right|_{\xi=\kappa/2},\qquad \frac{\partial\mathcal{R}_{\varepsilon}^{(\star)}}{\partial u_1}=0,\qquad \frac{\partial T_r^{(\star)}}{\partial u_1}=\frac{\partial\Phi_r^{(\star)}}{\partial u_1}=0.
\label{eq:geo_temporal_degeneracy}
\end{equation}
The derivatives hold at fixed remaining field parameters and prescribed conserved quantities or turning radii. In other words, the independent bumblebee field strength of the varying temporal configuration accompanies the same neutral test particle trajectories.

Table~\ref{tab:geo_configurations} evaluates the characteristic timelike quantities for the example used in the preceding section. The two temporal rows coincide, as required by Eq.~\eqref{eq:geo_temporal_degeneracy}. Notice that by considering all together, the circular orbit data, eccentric angular accumulation, and critical instability times give us complementary information: the first depends on the lapse, while the latter two also resolve the constant radial deformation.

\begin{table}[tbp]
\centering
\caption{Timelike orbital quantities for $\kappa=b=m=1$, $\xi=1/2$, $Q=0.6$, and $\beta_{\mathrm{BI}}=0.8$. The temporal configurations use $u_0=0.3$, with $u_1=4$ in the varying case. The periastron and apastron for the last column are $6m$ and $12m$. }
\label{tab:geo_configurations}
\begin{tabular}{lcccc}
\toprule
Configuration & $r_{\mathrm{ISCO}}/m$ & $E_{\mathrm{ISCO}}$ & $L_{\mathrm{ISCO}}/m$ & $\Phi_r/(2\pi)$ \\
\midrule
Radial & 5.720842 & 0.940268 & 3.374442 & 2.344315 \\
Constant temporal & 5.712159 & 0.940186 & 3.371651 & 2.374801 \\
Varying temporal & 5.712159 & 0.940186 & 3.371651 & 2.374801 \\
\bottomrule
\end{tabular}
\end{table}


\section{Photon spheres and shadows}
\label{sec:photon-shadows}

The charged solutions obtained above allow us to examine how the bumblebee configuration and nonlinear electrodynamics affect light propagation. Circular photon orbits determine the capture threshold, whose projection onto the observer's sky defines the shadow boundary. In this sense, the photon sphere is characterized by an areal radius, whereas the shadow is characterized by a critical impact parameter. We consider the three bumblebee configurations, specify the electromagnetic propagation law, and compare the resulting radii with the Event Horizon Telescope observations of Sgr~A$^*$ and M87$^*$~\cite{sh:Perlick2022,sh:EHTSgr2022VI,sh:Kocherlakota2021}.


\subsection{Photon propagation and circular orbits}
\label{subsec:shadow-optics}

We write the common metric and its normalized lapse as
\begin{align}
\mathrm{d}s^2&=-f(r)\mathrm{d}t^2+\frac{q_T}{f(r)}\mathrm{d}r^2+r^2\mathrm{d}\Omega^2,\notag\\ f(r)&=1-\frac{2m}{r}+\eta\Delta_{\mathrm{BI}}(r),\qquad \eta=\frac{q_T}{q_0},\qquad q_0=1+\xi b^2>0,
\label{eq:sh-metric}
\end{align}
where $q_T>0$, $m>0$, and $\Delta_{\mathrm{BI}}=\kappa\mathcal J/r$ has the normalization established above. Here, $r_h$ denotes a simple outer horizon, with $f>0$ for $r>r_h$. We consider the solutions with vanishing cosmological constant derived in this manuscript and normalize $f(\infty)=1$, thereby fixing the Killing time entering the conserved photon energy.

It is important to mention that a separate Maxwell probe follows the null cone of $g_{\mu\nu}$, while disturbances of the nonlinear electromagnetic field follow the characteristics of its Lagrangian~\cite{sh:Novello2000,sh:Gibbons2001,sh:Abalos2015}. If the electric Lagrangian $\mathcal L(\mathcal I)=\beta_{\mathrm{BI}}^2[1-\sqrt{1+\mathcal I/(2\beta_{\mathrm{BI}}^2)}]$ defines the complete theory, its polarizations comprise the ordinary metric cone and an extraordinary cone determined by
\begin{equation}
\mathcal G^{\mu\nu}k_\mu k_\nu=0,\qquad \mathcal G^{\mu\nu}=\mathcal L_{\mathcal I}g^{\mu\nu}-4\mathcal L_{\mathcal I\mathcal I}F^\mu{}_{\alpha}F^{\alpha\nu},\qquad \frac{\mathcal L_{\mathcal I\mathcal I}}{\mathcal L_{\mathcal I}}=-\frac{1}{4\beta_{\mathrm{BI}}^2s^2}.
\label{eq:sh-characteristic}
\end{equation}
On the electric background, $s^2=1-E^2/\beta_{\mathrm{BI}}^2=(1+Q^2/\beta_{\mathrm{BI}}^2r^4)^{-1}$. Up to a conformal factor, a covariant representative of the extraordinary optical metric reads
\begin{equation}
\mathrm{d}\hat s^2=-f(r)\mathrm{d}t^2+\frac{q_T}{f(r)}\mathrm{d}r^2+C_{\mathrm{BI}}(r)\mathrm{d}\Omega^2,\qquad C_{\mathrm{BI}}(r)=r^2w(r),\qquad w(r)=1+\frac{Q^2}{\beta_{\mathrm{BI}}^2r^4}.
\label{eq:sh-effective}
\end{equation}
On the other hand, the complete determinant form of Born--Infeld theory also admits the electric restriction employed above. With $\mathcal K=F_{\mu\nu}\widetilde F^{\mu\nu}$, it reads
\begin{equation}
\mathcal L_{\mathrm{det}}=\beta_{\mathrm{BI}}^2\left[1-\sqrt{1+\frac{\mathcal I}{2\beta_{\mathrm{BI}}^2}-\frac{\mathcal K^2}{16\beta_{\mathrm{BI}}^4}}\right].
\label{eq:sh-determinant}
\end{equation}
Its background equations coincide with those used here because $\mathcal K=0$, while its second variation places both polarizations on the cone in Eq.~\eqref{eq:sh-effective}. In this manner, the full Born--Infeld completion gives a single optical shadow; the literal $\mathcal L(\mathcal I)$ theory instead permits polarization dependent capture boundaries~\cite{sh:Gibbons2001,sh:Abalos2015}. Both interpretations share the same background solutions.

We denote the two cases by $a=g,\mathrm{BI}$, with $C_g=r^2$ and $C_{\mathrm{BI}}=r^2w$. In the equatorial plane, the null Hamiltonian associated with the corresponding optical representative gives
\begin{equation}
E_\gamma=f\dot t,\qquad L_\gamma=C_a\dot\varphi,\qquad d=\frac{L_\gamma}{E_\gamma},\qquad q_T\dot r^2=E_\gamma^2-L_\gamma^2W_a(r),\qquad W_a(r)=\frac{f(r)}{C_a(r)}.
\label{eq:sh-radial}
\end{equation}
The affine parameter in this equation belongs to the chosen conformal representative; the trajectory and $d$ are conformally invariant. A circular photon orbit satisfies
\begin{equation}
W_a'(r_{\mathrm{ph},a})=0,\qquad W_a''(r_{\mathrm{ph},a})<0,\qquad r_{\mathrm{ph},a}>r_h,\qquad d_{c,a}^2=\frac{C_a(r_{\mathrm{ph},a})}{f(r_{\mathrm{ph},a})}.
\label{eq:sh-circular}
\end{equation}
For distant illumination, the capture boundary is fixed by the global exterior maximum of $W_a$, including when several stationary radii occur. The numerical analysis selects this maximum and verifies its instability.

The metric photon sphere admits a particularly simple exact equation. Using $\mathcal J'=-r^2\rho$ in $rf'-2f=0$, we obtain
\begin{equation}
r_{\mathrm{ph},g}-3m+\frac{\kappa\eta}{2}\left[r_{\mathrm{ph},g}^3\rho(r_{\mathrm{ph},g})+3\mathcal J(r_{\mathrm{ph},g})\right]=0.
\label{eq:sh-photon-integral}
\end{equation}
The terms proportional to $\beta_{\mathrm{BI}}^2r^3(\mathcal S-1)$ cancel between $r^3\rho$ and $3\mathcal J$. In this regard, Eq.~\eqref{eq:sh-photon-integral} reduces to
\begin{equation}
r_{\mathrm{ph},g}^2-3mr_{\mathrm{ph},g}+\kappa\eta Q^2\mathcal H(r_{\mathrm{ph},g})=0,\qquad \mathcal H(r)={}_2F_1\!\left(\frac14,\frac12;\frac54;-\frac{Q^2}{\beta_{\mathrm{BI}}^2r^4}\right).
\label{eq:sh-photon-exact}
\end{equation}
For the Born--Infeld cone, the angular coefficient in Eq.~\eqref{eq:sh-effective} contributes an additional term. The corresponding exact condition is
\begin{equation}
rf'(r)-2f(r)+\frac{4Q^2f(r)}{\beta_{\mathrm{BI}}^2r^4+Q^2}=0,\qquad r=r_{\mathrm{ph},\mathrm{BI}}.
\label{eq:sh-photon-BI}
\end{equation}
Equations~\eqref{eq:sh-photon-exact} and~\eqref{eq:sh-photon-BI} maintain the complete electromagnetic nonlinearity and apply to each charged bumblebee configuration through its value of $\eta$.


\subsection{Shadow radii and the three bumblebee configurations}
\label{subsec:shadow-casees}

The observer's sky connects the circular orbit with the angular shadow. For a static observer at $r_o>r_{\mathrm{ph},g}$, the physical tetrad gives~\cite{sh:Perlick2022}
\begin{equation}
\sin^2\alpha_g=\frac{d_{c,g}^2f(r_o)}{r_o^2}.
\label{eq:sh-angle-g}
\end{equation}
For a Born--Infeld ray, its group direction is determined by the tangent to the optical null trajectory. Measuring that direction with the physical metric tetrad gives
\begin{equation}
\tan^2\alpha_{\mathrm{BI}}=\frac{r_o^2d_{c,\mathrm{BI}}^2f(r_o)}{C_{\mathrm{BI}}(r_o)^2\left[1-d_{c,\mathrm{BI}}^2f(r_o)/C_{\mathrm{BI}}(r_o)\right]}.
\label{eq:sh-angle-BI}
\end{equation}
This expression approaches Eq.~\eqref{eq:sh-angle-g} when $w\to1$. At large radius, $f\to1$ and $w\to1$, so the shadow radii on the observer's transverse screen become
\begin{equation}
R_{\mathrm{sh},g}=\frac{r_{\mathrm{ph},g}}{\sqrt{f(r_{\mathrm{ph},g})}},\qquad R_{\mathrm{sh},\mathrm{BI}}=\frac{r_{\mathrm{ph},\mathrm{BI}}\sqrt{w(r_{\mathrm{ph},\mathrm{BI}})}}{\sqrt{f(r_{\mathrm{ph},\mathrm{BI}})}},\qquad \theta_{\mathrm{sh},a}=\frac{2R_{\mathrm{sh},a}}{D_A}.
\label{eq:sh-radii}
\end{equation}
At this stage, $\theta_{\mathrm{sh},a}$ is the angular diameter and $D_A$ the angular diameter distance for the adopted source geometry. This convention accounts for the asymptotically conical radial geometry. It is worth pointing out that no additional factor $\sqrt{q_T}$ enters the local angle when expressed through the areal transverse distance. Spherical symmetry fixes the circular boundary $\alpha^2+\beta_{\mathrm{sky}}^2=R_{\mathrm{sh},a}^2$ for every parameter value considered.

The substitution rules for the three solutions can be collected as
\begin{equation}
\begin{array}{c|c|c} 
\text{Bumblebee configuration} & q_T & \eta \\ 
\hline
\hline
u_0=u_1=0 & 1+\xi b^2 & 1 \\ u_0\ne0,\ u_1=0 & 1+\xi(b^2+u_0^2) & 1+\dfrac{\xi u_0^2}{1+\xi b^2} \\ u_1\ne0,\ \xi=\kappa/2 & 1+\dfrac{\kappa}{2}(b^2+u_0^2) & 1+\dfrac{\kappa u_0^2}{2+\kappa b^2} \end{array}.
\label{eq:sh-casees}
\end{equation}
For the purely radial configuration, both optical radii are independent of $\xi b^2$ at fixed $(m,Q,\beta_{\mathrm{BI}})$: the deformation enters the radial metric coefficient, while the lapse and electric invariant retain their Einstein Born--Infeld forms. For a constant temporal component, the shadow depends on $\xi$, $b$, and $u_0$ through $\eta$. The varying temporal component gives the same radii at $\xi=\kappa/2$ for every permitted $u_1$; the radial sign $\sigma$ is likewise absent. In this regard, the three families satisfy exact optical relations.

The neutral limit yields a further common result,
\begin{equation}
Q=0:\qquad r_h=2m,\qquad r_{\mathrm{ph},g}=r_{\mathrm{ph},\mathrm{BI}}=3m,\qquad R_{\mathrm{sh},g}=R_{\mathrm{sh},\mathrm{BI}}=3\sqrt3\,m.
\label{eq:sh-neutral}
\end{equation}
The neutral Casana limit and the neutral \textit{metric--affine} comparison geometries share these radii in physical areal coordinates and normalized time, with $m$ replaced by their mass coefficient. In this sense, shadow size leaves their constant radial deformation undetermined. On the other hand, the radial coefficient affects the departure rate of neighboring rays from a circular orbit. In Killing time, this rate is
\begin{equation}
\lambda_{t,a}^2=-\frac{f(r_{\mathrm{ph},a})C_a(r_{\mathrm{ph},a})}{2q_T}W_a''(r_{\mathrm{ph},a}),\qquad \Omega_{\mathrm{ph},a}=\sqrt{\frac{f(r_{\mathrm{ph},a})}{C_a(r_{\mathrm{ph},a})}}=\frac{1}{R_{\mathrm{sh},a}}.
\label{eq:sh-instability}
\end{equation}
This relation identifies a possible complementary role for time dependent lensing observables while preserving the exact size degeneracy, as discussed in \cite{al2026shadow}.


\subsection{Dependence on charge, nonlinearity, and the temporal amplitude}
\label{subsec:shadow-parameters}

For the numerical analysis, it is convenient to introduce
\begin{equation}
x=\frac rm,\qquad e^2=\frac{\kappa\eta Q^2}{2m^2},\qquad \nu=\frac{Q^2}{\beta_{\mathrm{BI}}^2m^4},\qquad h(x,\nu)={}_2F_1\!\left(\frac14,\frac12;\frac54;-\frac{\nu}{x^4}\right).
\label{eq:sh-dimensionless}
\end{equation}
The dimensionless charge $e$ includes the bumblebee factor and the electromagnetic normalization of the action. In these variables, the exact lapse can be written in a form that also avoids subtraction of nearly equal square roots,
\begin{equation}
f(x)=1-\frac2x+\frac{2e^2}{3x^2}\left[2h(x,\nu)-\frac{1}{1+\sqrt{1+\nu/x^4}}\right],\qquad x_{\mathrm{ph},g}^2-3x_{\mathrm{ph},g}+2e^2h(x_{\mathrm{ph},g},\nu)=0.
\label{eq:sh-numerical}
\end{equation}
All figures use this exact lapse. For $Q\ne0$, $e$ and $\nu$ are independent coordinates; the physical neutral point is $e=\nu=0$. Before evaluating a shadow, we require a simple outer horizon. For $\nu>0$, $\mathcal M(x)=x f(x)+2$ has its minimum at $x_{\min}=\sqrt{e^2-\nu/(4e^2)}$ when $2e^2>\sqrt\nu$; otherwise its infimum is approached as $x\to0$, with value $\Gamma(1/4)^2e^2/(3\sqrt\pi\,\nu^{1/4})$. A simple outer horizon exists when this minimum or infimum is below $2$, which naturally restricts the analysis to the assumed capture geometry.

\begin{figure}[t]
\centering
\includegraphics[width=.49\linewidth]{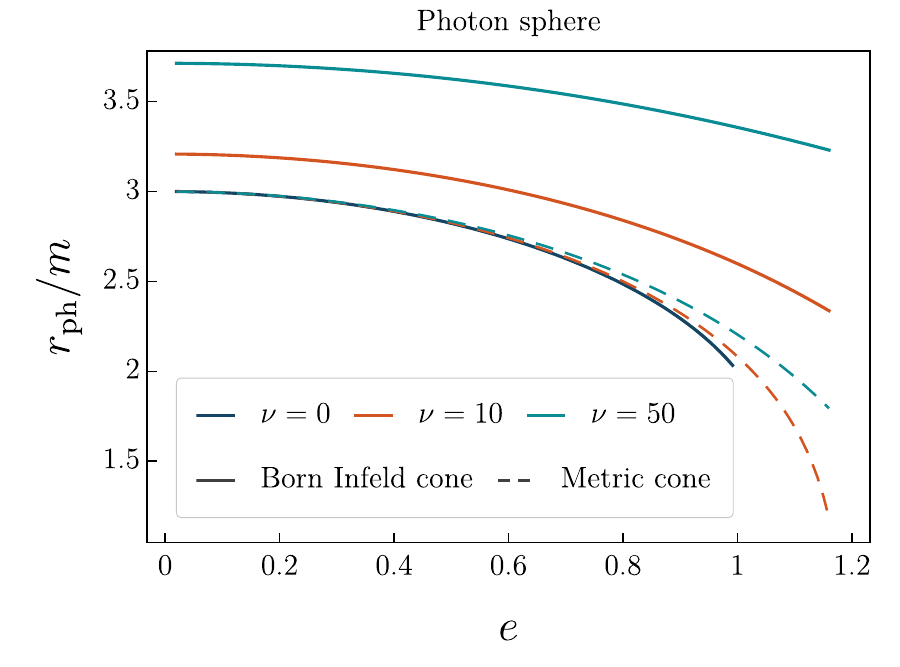}\hfill
\includegraphics[width=.49\linewidth]{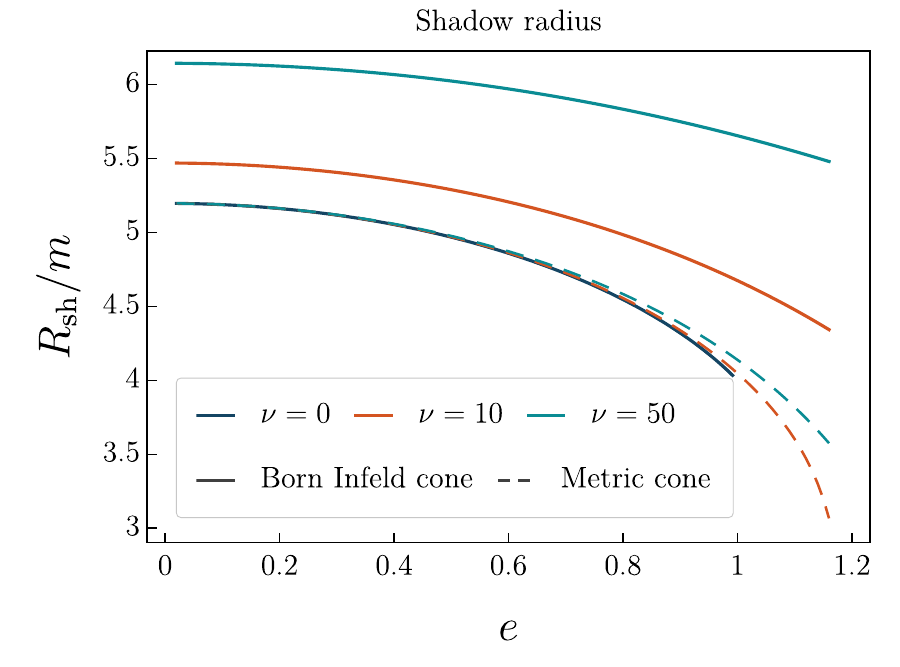}
\caption{Exact photon sphere radii (left) and shadow radii (right) as functions of the dimensionless geometric charge $e$, for $\nu=0,10,50$. Solid curves refer to the Born--Infeld optical cone and dashed curves to the metric cone; the two coincide at $\nu=0$. }
\label{fig:sh-charge}
\end{figure}

\begin{figure}[t]
\centering
\includegraphics[width=.6\linewidth]{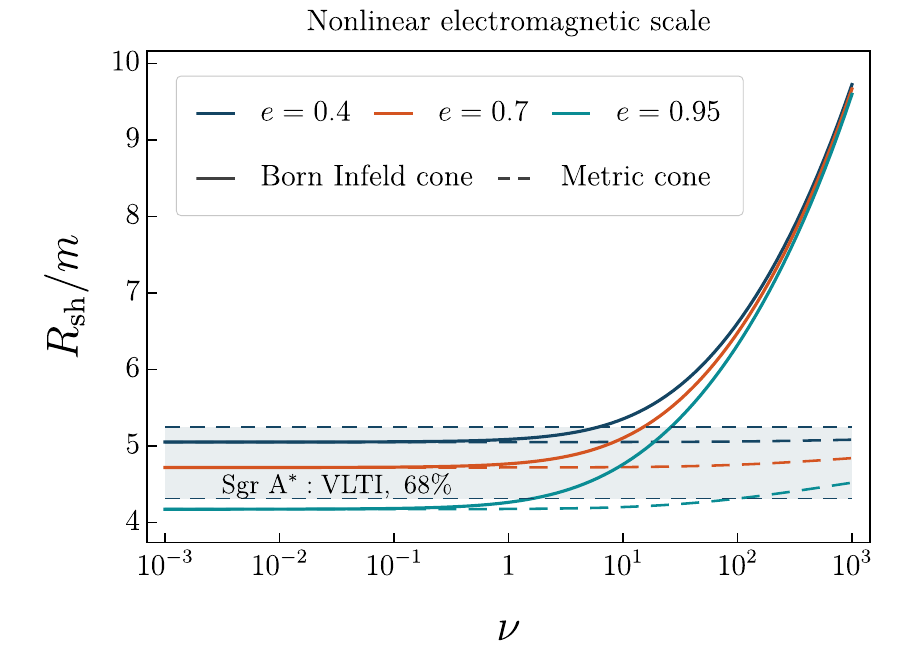}
\caption{Shadow radii as functions of the nonlinear parameter $\nu$ for $e=0.4,0.7,0.95$. Solid and dashed curves denote the Born--Infeld and metric cones, respectively. The shaded interval is the conditional $68\%$ shadow size interval inferred for Sgr~A$^*$ using the VLTI mass to distance prior. }
\label{fig:sh-nonlinear}
\end{figure}

The Maxwell limit supplies analytic expressions for comparison,
\begin{equation}
f_0(x)=1-\frac2x+\frac{e^2}{x^2},\qquad x_0=\frac{3+\sqrt{9-8e^2}}2,\qquad \frac{R_0}{m}=\sqrt{\frac{2x_0^3}{x_0-1}},\qquad 0\le e\le1.
\label{eq:sh-RN}
\end{equation}
In this limit, notice that both cones coincide. Increasing the charge reduces the photon sphere radius from $3m$ to $2m$ and the shadow radius from $3\sqrt3\,m$ to $4m$. At finite $\nu$, the leading nonlinear correction to the lapse is
\begin{equation}
f(x)=f_0(x)-\frac{e^2\nu}{20x^6}+O(\nu^2),\qquad \frac{\nu}{x^4}\ll1.
\label{eq:sh-weakBI}
\end{equation}
Writing $f_{0p}=f_0(x_0)$, the circular orbits and their projected radii In this regard, they become
\begin{align}
x_{\mathrm{ph},g}&=x_0+\frac{e^2\nu}{5x_0^4(2x_0-3)}+O(\nu^2), & x_{\mathrm{ph},\mathrm{BI}}&=x_0+\frac{\nu}{2x_0-3}\left(\frac{e^2}{5x_0^4}+\frac{2f_{0p}}{x_0^2}\right)+O(\nu^2), \\ \frac{R_{\mathrm{sh},g}}{R_0}&=1+\frac{e^2\nu}{40x_0^6f_{0p}}+O(\nu^2), & \frac{R_{\mathrm{sh},\mathrm{BI}}}{R_0}&=1+\frac{e^2\nu}{40x_0^6f_{0p}}+\frac{\nu}{2x_0^4}+O(\nu^2).
\label{eq:sh-expansions}
\end{align}
Both corrections enlarge the shadow at fixed $e$: the nonlinear energy density changes the lapse, while the Born--Infeld cone adds an angular contribution. Stationarity of the unperturbed impact parameter allows the leading correction to the shadow radius to be evaluated at $x_0$. In this manner, Eqs.~\eqref{eq:sh-expansions} explain the weakly nonlinear behavior, while the exact curves in Figs.~\ref{fig:sh-charge} and~\ref{fig:sh-nonlinear} extend it to finite $\nu$.

\begin{figure}[t]
\centering
\includegraphics[width=.49\linewidth]{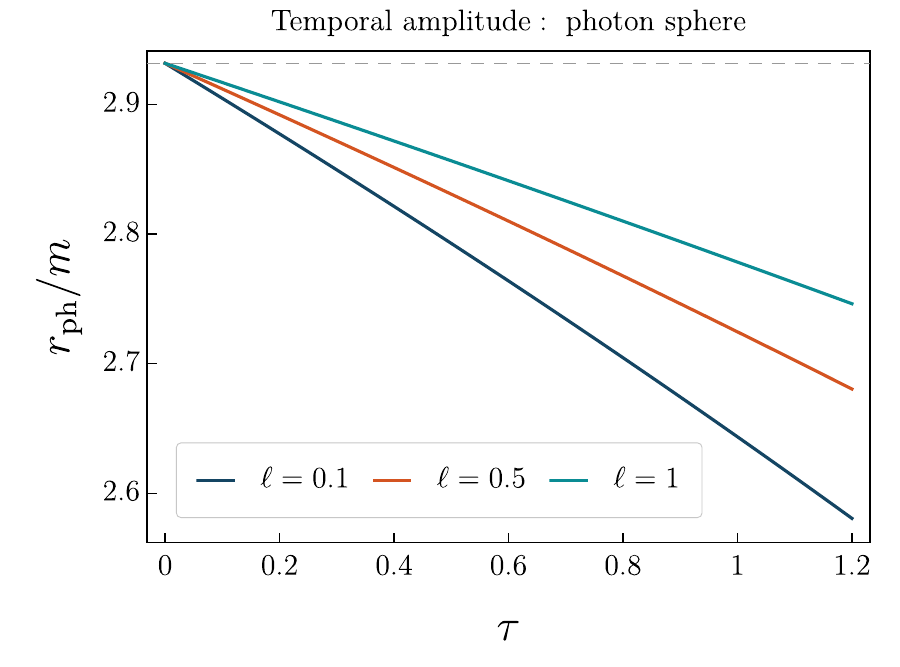}\hfill
\includegraphics[width=.49\linewidth]{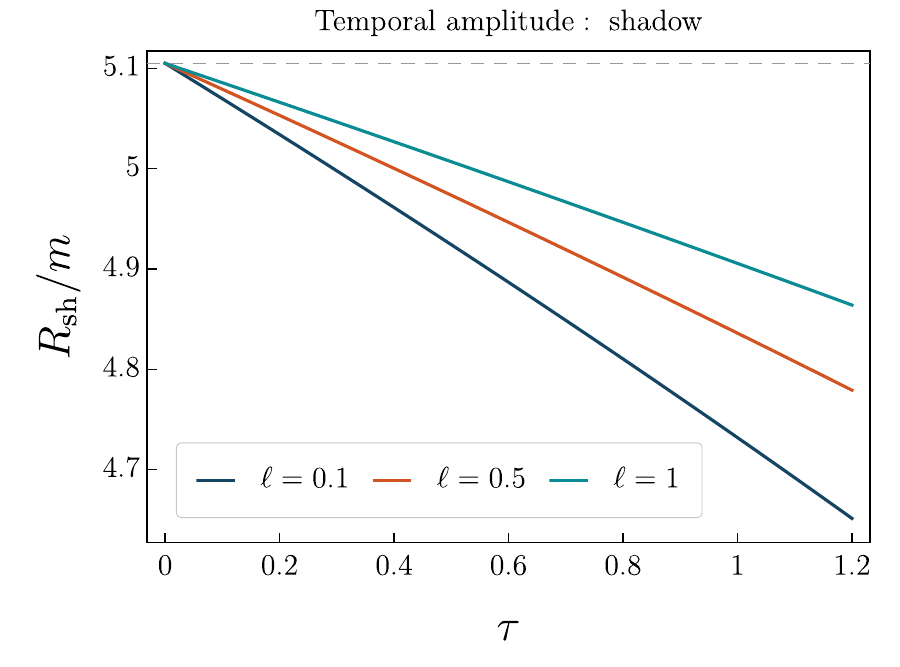}
\caption{Born--Infeld photon sphere radius (left) and shadow radius (right) as functions of $\tau=\xi u_0^2$, for $\ell=\xi b^2=0.1,0.5,1$. The bare charge is fixed at $\sqrt{\kappa Q^2/(2m^2)}=0.7$ and $\nu=10$, so $e^2=0.49[1+\tau/(1+\ell)]$ along each curve. The horizontal dashed line marks the purely radial value $\tau=0$. }
\label{fig:sh-bumblebee}
\end{figure}

\begin{figure}[t]
\centering
\includegraphics[width=.49\linewidth]{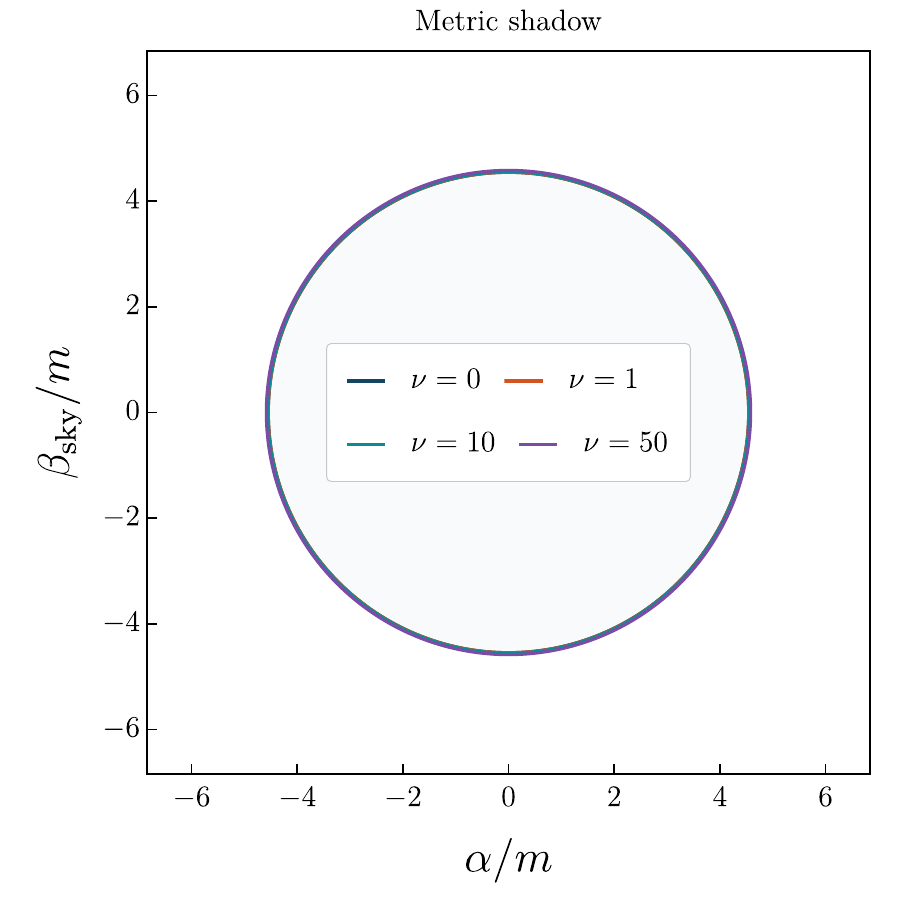}\hfill
\includegraphics[width=.49\linewidth]{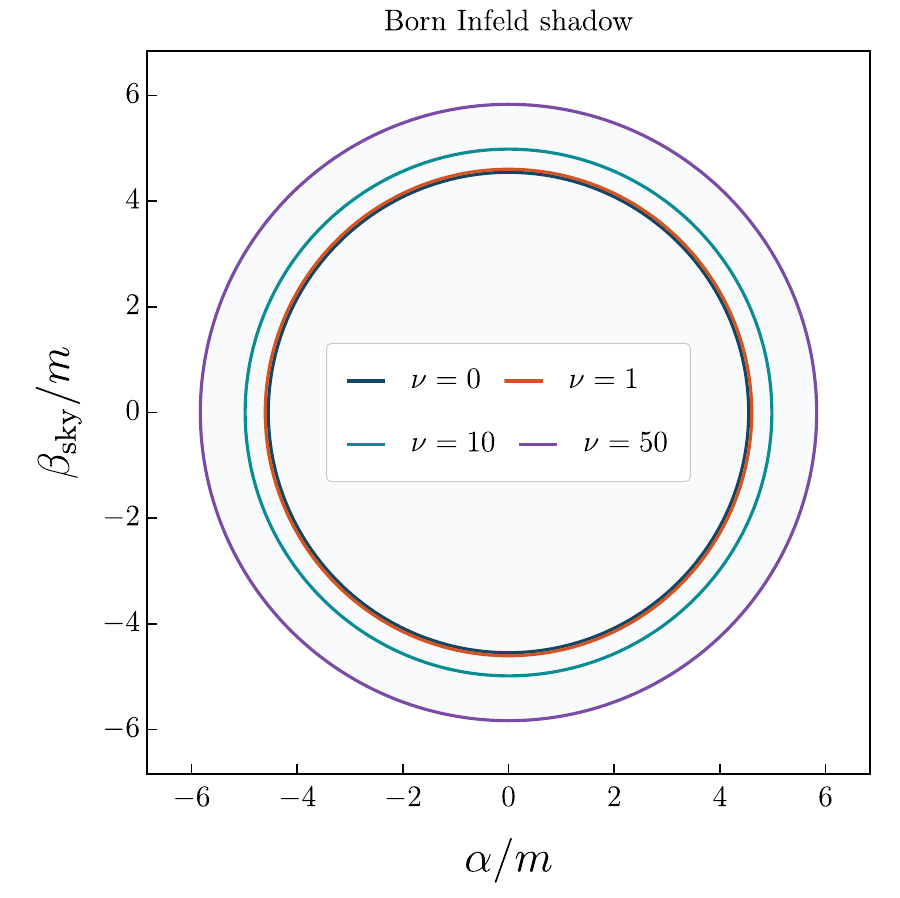}
\caption{Circular capture boundaries for $e=0.8$ and $\nu=0,1,10,50$, using the metric cone (left) and the Born--Infeld cone (right). The screen coordinates are normalized by $m$. }
\label{fig:sh-outlines}
\end{figure}

The bumblebee dependence follows directly from $e^2=\eta\kappa Q^2/(2m^2)$. At fixed $(m,Q,\beta_{\mathrm{BI}})$, stationarity of the critical impact parameter gives, for either cone,
\begin{equation}
\left.\frac{\partial\ln R_{\mathrm{sh},a}}{\partial\eta}\right|_{m,Q,\beta_{\mathrm{BI}}}=-\frac{\Delta_{\mathrm{BI}}(r_{\mathrm{ph},a})}{2f(r_{\mathrm{ph},a})}<0\qquad (Q\ne0).
\label{eq:sh-eta-derivative}
\end{equation}
The positive integral representation of $\mathcal J$ fixes this sign: a larger $\eta$ reduces the shadow along a continuous capture segment. Increasing $u_0^2$ increases $\eta$ for $\xi>0$ and decreases it for $\xi<0$, within $q_0,q_T>0$. The parameters are related by
$\,\ell=\xi b^2$, $\tau=\xi u_0^2$, $\, \eta=1+\frac{\tau}{1+\ell}$.
Figure~\ref{fig:sh-bumblebee} illustrates the positive coupling case at fixed bare charge $\sqrt{\kappa Q^2/(2m^2)}=0.7$ and $\nu=10$. A larger temporal amplitude reduces the radii, while increasing $\ell$ at fixed positive $\tau$ moderates this change. Every curve recovers the purely radial value at $\tau=0$. Figure~\ref{fig:sh-outlines} shows the corresponding circular boundaries.

Table~\ref{tab:sh-example} gives the exact radii for the example above: $\kappa=b=m=1$, $\xi=1/2$, $Q=0.6$, and $\beta_{\mathrm{BI}}=0.8$. The radial scenario has $u_0=0$, while both temporal cases use $u_0=0.3$, with $u_1=4$ admitted by the varying it. Their identical metrics and electromagnetic invariants yield identical radii.

\begin{table}[t]
\caption{Exact exterior radii for the manuscript's example, with the radial configuration included at the same $(\kappa,b,m,\xi,Q,\beta_{\mathrm{BI}})$. The value $\nu=0.5625$ is common to all rows. The two photon radii are areal radii, and the two shadow radii are distant screen radii.}
\label{tab:sh-example}
\centering
\begin{tabular}{lccccc}
\toprule Configuration & $r_h/m$ & $r_{\mathrm{ph},g}/m$ & $r_{\mathrm{ph},\mathrm{BI}}/m$ & $R_{\mathrm{sh},g}/m$ & $R_{\mathrm{sh},\mathrm{BI}}/m$ \\ \midrule Radial & 1.905748 & 2.874880 & 2.890797 & 5.034449 & 5.054908 \\ Constant temporal & 1.902770 & 2.870951 & 2.886944 & 5.029401 & 5.049950 \\ Varying temporal & 1.902770 & 2.870951 & 2.886944 & 5.029401 & 5.049950 \\ \bottomrule
\end{tabular}
\end{table}


\subsection{Compatibility with \texorpdfstring{Sgr~A$^*$ and M87$^*$}{Sgr A* and M87*}}
\label{subsec:shadow-bounds}

We express the observational comparison through the fractional deviation of the shadow diameter from its Schwarzschild value,
\begin{equation}
\delta_a=\frac{R_{\mathrm{sh},a}}{3\sqrt3\,m}-1,\qquad \theta_{\mathrm{sh},a}=6\sqrt3\,\frac{m}{D_A}(1+\delta_a).
\label{eq:sh-delta}
\end{equation}
The Event Horizon Telescope analysis of Sgr~A$^*$ uses the relation between the observed emission ring and the geometric shadow using simulated images, and combines it with independent stellar measurements of the mass to distance ratio~\cite{sh:EHTSgr2022VI}. Its two fiducial $68\%$ results are
\begin{equation}
\delta_{\mathrm{Sgr},\mathrm{VLTI}}=-0.08^{+0.09}_{-0.09},\qquad \delta_{\mathrm{Sgr},\mathrm{Keck}}=-0.04^{+0.09}_{-0.10}.
\label{eq:sh-Sgr-data}
\end{equation}
The case regarded to M87$^*$ comparison is based on the stellar dynamical mass prior gives $\delta_{\mathrm{M87}}=-0.01\pm0.17$ at $68\%$~\cite{sh:EHTM872019VI,sh:Kocherlakota2021}. Retaining its central value, the resulting radius intervals are
\begin{equation}
\begin{array}{c|c|c} \text{Source and prior} & \delta\ \text{interval} & R_{\mathrm{sh}}/m\ \text{interval} \\ \hline \mathrm{Sgr~A^*,\ VLTI} & [-0.17,\ 0.01] & [4.31281,\ 5.24811] \\ \mathrm{Sgr~A^*,\ Keck} & [-0.14,\ 0.05] & [4.46869,\ 5.45596] \\ \mathrm{M87^*,\ stellar} & [-0.18,\ 0.16] & [4.26084,\ 6.02754] \end{array}.
\label{eq:sh-radius-bounds}
\end{equation}
It is important to mention that the Sgr~A$^*$ rows use alternative priors for the same EHT observations and are treated separately. For M87$^*$, the stellar dynamical prior supplies independent mass information; using the EHT inferred mass instead would reuse the shadow information.

\begin{figure}[t]
\centering
\includegraphics[width=.49\linewidth]{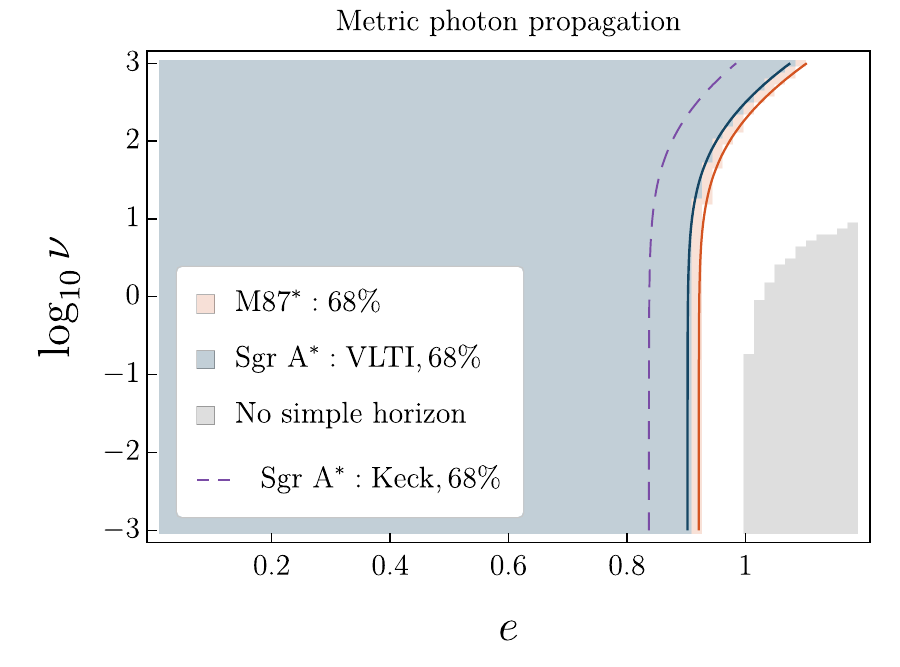}\hfill
\includegraphics[width=.49\linewidth]{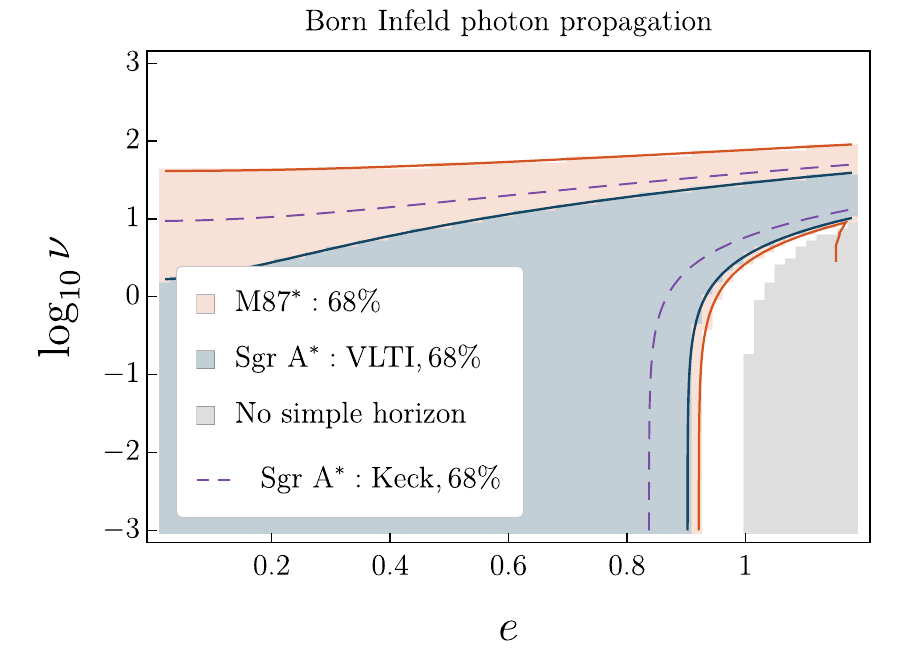}
\caption{Conditional parameter regions obtained from the calibrated $68\%$ shadow size intervals, using the metric cone (left) and the Born--Infeld cone (right). Orange boundaries enclose the M87$^*$ region based on its stellar dynamical prior; blue marks the Sgr~A$^*$ VLTI region, and dashed purple curves mark the alternative Keck boundaries. The blue region also satisfies the adopted M87$^*$ interval, while pale orange identifies additional M87$^*$ compatibility. Gray marks configurations without a simple outer horizon.}
\label{fig:sh-regions}
\end{figure}

For a specified propagation cone, let $\mathcal R_a(e,\nu)=R_{\mathrm{sh},a}/m$ denote the exact prediction. The conditional compatibility region is then
\begin{equation}
\mathcal A_{i,a}=\left\{(e,\nu):\ \text{simple outer horizon},\ \ R_{i,-}\le\mathcal R_a(e,\nu)\le R_{i,+}\right\},
\label{eq:sh-region}
\end{equation}
where $R_{i,\pm}$ are the endpoints in Eq.~\eqref{eq:sh-radius-bounds}. Figure~\ref{fig:sh-regions} translates these published intervals into parameter sensitivities for both cones under the adopted image. For the literal $\mathcal L(\mathcal I)$ theory, the relative polarized intensities also enter. In this regard, the maps give conditional compatibility regions at the stated $68\%$ level.

The Maxwell limit makes the charge constraints particularly transparent. Inverting Eq.~\eqref{eq:sh-RN} at the lower endpoints of Eq.~\eqref{eq:sh-radius-bounds} gives
\begin{equation}
\begin{aligned}
e&\lesssim0.90244\quad(\mathrm{Sgr~A^*,\ VLTI}),\\ e&\lesssim0.83733\quad(\mathrm{Sgr~A^*,\ Keck}),\\ e&\lesssim0.92149\quad(\mathrm{M87^*}),
\end{aligned}
\label{eq:sh-e-bounds}
\end{equation}
as conditional $68\%$ bounds. The corresponding metric photon sphere radii satisfy $r_{\mathrm{ph}}/m\gtrsim2.28816$, $2.42074$, and $2.24276$, respectively, with upper endpoint $3$. These intervals follow from the model relation between the orbit and its shadow. On the other hand, doubling the quoted errors admits the full Reissner--Nordstr\"om black hole interval $0\le e\le1$ in every size test. This specifies the statistical strength of Eq.~\eqref{eq:sh-e-bounds} and motivates retaining both parameters at finite nonlinearity.

In terms of the original parameters, the Maxwell bounds take the form
\begin{equation}
\frac{\kappa Q^2}{2m^2}\left(1+\frac{\xi u_0^2}{1+\xi b^2}\right)\le e_{\max,i}^2,\qquad \frac{\xi u_0^2}{1+\xi b^2}\le\frac{2m^2e_{\max,i}^2}{\kappa Q^2}-1\quad(Q\ne0).
\label{eq:sh-original-bounds}
\end{equation}
The parenthesis is unity for the radial case and is evaluated at $\xi=\kappa/2$ for the varying temporal part. Equation~\eqref{eq:sh-original-bounds} constrains charge together with the temporal amplitude. Also, in the neutral limit, shadow size leaves $\xi b^2$, $u_0$, and the nonlinear scale free. The constant $u_1$ remains absent at the special coupling.

Finite $\nu$ supplies conditional bounds on the Born--Infeld scale itself. Let us write 
\begin{equation}
\widehat\beta=m\beta_{\mathrm{BI}}\sqrt{\frac{\kappa\eta}{2}},\qquad \nu=\frac{e^2}{\widehat\beta^2},\qquad \beta_{\mathrm{BI}}=\frac{|Q|}{m^2\sqrt\nu},
\label{eq:sh-beta-hat}
\end{equation}
so that we can map any allowed $\nu$ interval at a specified charge directly into an allowed $\beta_{\mathrm{BI}}$ interval. For the Born--Infeld cone, the Sgr~A$^*$ VLTI interval gives the exact numerical examples
\begin{equation}
\begin{array}{c|c|c} e & \text{allowed }\nu\text{ interval} & \widehat\beta_{\min} \\ \hline 0.40 & [0,\ 6.06332] & 0.16244 \\ 0.70 & [0,\ 15.05440] & 0.18041 \\ 0.80 & [0,\ 19.11357] & 0.18299 \\ 0.95 & [1.59088,\ 26.17850] & 0.18567 \end{array}.
\label{eq:sh-beta-bounds}
\end{equation}
The endpoint $\nu=0$ denotes the Maxwell limit. For $e=0.95$, one also obtains $\widehat\beta\lesssim0.75319$: finite nonlinearity first brings the shadow into the observed interval and eventually above its upper edge. The optical contribution thereby admits charges above the Maxwell bound while restricting the nonlinear scale.

The M87$^*$ observations from 2017, 2018, and 2021 give a persistent emission ring diameter of $43.9\pm0.6\,\mu\mathrm{as}$~\cite{sh:EHTM872025}. Its use in Eq.~\eqref{eq:sh-delta} needs adjust between the emission ring and the critical curve, together with an independent prior on the mass to distance ratio. In this regard, we use Eq.~\eqref{eq:sh-radius-bounds} for the numerical bounds and regard the recent ring measurement as complementary input for a future image fit.


\subsection{Complementarity with stellar orbital data}
\label{subsec:shadow-stellar}

\begin{figure}[t]
\centering
\includegraphics[width=.6\linewidth]{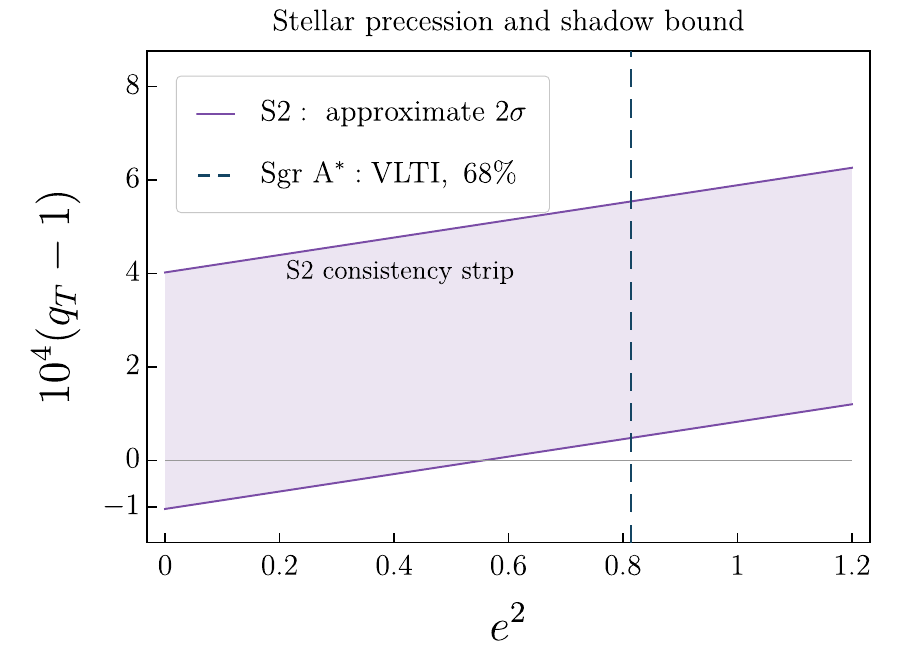}
\caption{Complementarity between the approximate orbital sensitivity inferred at two standard deviations from the combined GRAVITY stellar precession result and the conditional $68\%$ Sgr~A$^*$ VLTI shadow bound in the Maxwell limit. The purple band follows Eq.~\eqref{eq:sh-S2-bound}; the dashed blue vertical line marks $e^2=(0.90244)^2$, with the compatible Maxwell charges on its left.}
\label{fig:sh-S2}
\end{figure}

Stellar dynamics around Sgr~A$^*$ probes the radial deformation independently of the shadow size degeneracy. GRAVITY first detected the Schwarzschild precession of S2 and strengthened it through a combined analysis of S2, S29, S38, and S55~\cite{sh:GRAVITY2020,sh:GRAVITY2024}. The latter gives $f_{\mathrm{SP}}=1.133\pm0.113$ from MCMC, where $f_{\mathrm{SP}}$ multiplies the relativistic precession. In the weak field orbital regime, the metric predicts
\begin{equation}
\Delta\varpi=2\pi(\sqrt{q_T}-1)+\sqrt{q_T}\left(\frac{6\pi m}{p}-\frac{\pi e^2m}{p}\right)+\cdots,\qquad p=a(1-e_{\mathrm{orb}}^2).
\label{eq:sh-S2-precession}
\end{equation}
Here, $e_{\mathrm{orb}}$ is the orbital eccentricity, distinct from the charge $e$. Higher orbital orders and the nonlinear contribution are omitted; the latter's weak field expansion requires $\nu(m/r)^4\ll1$ along the orbit. To first order in $q_T-1$, Eq.~\eqref{eq:sh-S2-precession} gives
\begin{equation}
\frac{p}{6m}(q_T-1)-\frac{e^2}{6}\simeq f_{\mathrm{SP}}-1.
\label{eq:sh-S2-combination}
\end{equation}
Products of $q_T-1$ with relativistic orbital terms are omitted. Using the S2 reference precession $6\pi m/p=12.1'$ and the quoted MCMC uncertainty from the combined stellar analysis gives the illustrative interval of two standard deviations
\begin{equation}
-0.093\lesssim\frac{p}{6m}(q_T-1)-\frac{e^2}{6}\lesssim0.359,\qquad -1.04\times10^{-4}\lesssim q_T-1\lesssim4.02\times10^{-4}\quad(Q=0).
\label{eq:sh-S2-bound}
\end{equation}
This interval assigns the precession residual to the displayed metric terms. It is worth pointing out that a posterior for this geometry requires fitting the stellar mass distribution, astrometric frame, mass, distance, and orbital parameters. Figure~\ref{fig:sh-S2} instead illustrates the published precession sensitivity alongside the conditional Maxwell shadow bound. The orbital band probes $q_T$, while the shadow probes the charge combination in the lapse.

For the three solutions, $q_T-1=\xi b^2$ in the radial case, $q_T-1=\xi(b^2+u_0^2)$ for a constant temporal component, and $q_T-1=\kappa(b^2+u_0^2)/2$ for the varying component. In this manner, orbital motion gives the radial deformation with charge corrections, while the shadow probes the charged optical potential. Both ones keep the degeneracy in $u_1$ at $\xi=\kappa/2$. Combining Sgr~A$^*$ and M87$^*$ can test a common Born--Infeld scale once their charges and exterior bumblebee configurations are specified. For doing so, notice that their masses enter through $\widehat\beta_i=m_i\beta_{\mathrm{BI}}\sqrt{\kappa\eta_i/2}$, which connectes the analytic solutions with observations across different black hole masses indeed.


\section{Solar System estimates and bounds}
\label{sec:solar_bounds}

Solar System observations give rise to a useful setting to estimate the allowed magnitude of the parameters entering the geometries obtained above. We take into account the assumption that the solar exterior is described by the same spherical metric, with ordinary matter coupled minimally to the spacetime geometry. The resulting estimates apply to the parameters of this solar configuration; their application to another source follows upon specifying the corresponding exterior bumblebee field. Planetary motion and the propagation of electromagnetic signals offer complementary probes of the radial deformation and the charged contribution \cite{ss:Will2014,ss:Casana2018}. We use the notation introduced in the previous section and work in the regime $|q_T-1|\ll1$.


\subsection{Orbital motion and perihelion advance}
\label{subsec:ss_orbits}

For an electrically neutral test body in the equatorial plane, let $\mathcal E=f_{\mathrm{BI}}\dot t$ and $\mathcal L=r^2\dot\phi$ denote the conserved energy and angular momentum per unit rest mass, where a dot represents differentiation with respect to proper time. The radial equation is
\begin{equation}
q_T\dot r^{,2}=\mathcal E^2-f_{\mathrm{BI}}(r)\left(1+\frac{\mathcal L^2}{r^2}\right).
\label{eq:ss_radial}
\end{equation}

Introducing $\psi=\phi/\sqrt{q_T}$ and $w=1/r$ gives a convenient orbital equation,
\begin{equation}
\frac{\mathrm{d}^2w}{\mathrm{d}\psi^2} = -\frac{1}{2}\frac{\mathrm{d}}{\mathrm{d}w}\left[f_{\mathrm{BI}}(1/w)\left(\mathcal L^{-2} + w^2\right)\right].
\label{eq:ss_binet}
\end{equation}

The rescaled angle is useful for integrating the trajectory, while the physical azimuth $\phi$ retains its period $2\pi$. In this manner, the radial deformation contributes to the angle between successive perihelia even at the leading orbital order. To display the electromagnetic terms compactly, we denote the coefficients of $r^{-2}$ and $-r^{-6}$ in the previously obtained lapse expansion by
\begin{equation}
C_2=\frac{\kappa q_TQ^2}{2q_0},\qquad C_6=\frac{\kappa q_TQ^4}{40q_0\beta_{\mathrm{BI}}^2},
\label{eq:ss_coefficients}
\end{equation}
where $C_2$ has dimensions of length squared and $C_6$ has dimensions of length to the sixth power. Let $a$, $e$, and $p=a(1-e^2)$ be the semimajor axis, eccentricity, and semilatus rectum of the reference orbit. Substituting $w=(1+e\cos\psi)/p$ into the perturbing terms of Eq.~\eqref{eq:ss_binet} yields the perihelion advance per radial period,
\begin{equation}
\Delta\varpi=2\pi\left(\sqrt{q_T}-1\right)+\sqrt{q_T}\left[\frac{6\pi m}{p}-\frac{\pi C_2}{mp}+\frac{15\pi C_6}{mp^5}\left(1+\frac{3e^2}{2}+\frac{e^4}{8}\right)\right].
\label{eq:ss_precession}
\end{equation}

Equation~\eqref{eq:ss_precession} shows the leading contribution of each radial correction. Its expansion assumes $m/p\ll1$, $C_2/(mp)\ll1$, $C_6/(mp^5)\ll1$, and $Q^2/(\beta_{\mathrm{BI}}^2r_{\mathrm p}^4)\ll1$, where $r_{\mathrm p}=a(1-e)$. Corrections containing an additional factor of $m/p$, or a product of the small electromagnetic perturbations, are omitted. The electric contribution produces a retrograde shift, as in the Reissner--Nordstr\"om problem \cite{ss:Iorio2012}, while the leading Born--Infeld correction produces a prograde shift. Notice that the eccentricity dependence of the latter has been taken into account, which permits the same expression to be used for different planetary orbits.

After subtracting the Schwarzschild contribution and expanding to first order in $q_T-1$, the anomalous advance takes the form
\begin{equation}
\frac{\Delta\varpi_{\mathrm{anom}}}{\pi}=q_T-1-\frac{C_2}{mp}+\frac{15C_6}{mp^5}\left(1+\frac{3e^2}{2}+\frac{e^4}{8}\right).
\label{eq:ss_anomalous}
\end{equation}

Products of $q_T-1$ with the small orbital corrections are omitted in Eq.~\eqref{eq:ss_anomalous}. The three contributions have different dependences on $p$, which explicitly allow us to see the traces for observations of several planets to distinguish the deformation from the electromagnetic parameters.


\subsection{Estimates from planetary ephemerides}
\label{subsec:ss_ephemerides}

Published supplementary perihelion advances supply a direct scale for these estimates. The INPOP15a analysis gives the Mercury interval $0.0\pm3.1\,\mathrm{mas\,cy^{-1}}$ under its $35\%$ residual criterion \cite{ss:Fienga2016}. For Mars, EPM2011 gives $-0.020\pm0.037\,\mathrm{mas\,cy^{-1}}$ \cite{ss:Pitjeva2013}. Here ``$\mathrm{mas}$'' denotes milliarcseconds and ``$\mathrm{cy}$'' denotes a Julian century \cite{ss:FiengaMinazzoli2024}.

We also take into account the solar Lense--Thirring contribution. Its magnitude is approximately $2.0\,\mathrm{mas\,cy^{-1}}$ for Mercury and $0.03\,\mathrm{mas\,cy^{-1}}$ for Mars \cite{ss:Iorio2011}. This gives rise to its full magnitude in the estimate gives the illustrative envelopes
\begin{equation}
\mathcal B_{\mathrm{Mercury}}=5.1\,\mathrm{mas\,cy^{-1}},\qquad \mathcal B_{\mathrm{Mars}}=0.087\,\mathrm{mas\,cy^{-1}}.
\label{eq:ss_budgets}
\end{equation}

For Mercury we add the frame dragging allowance to the quoted interval, while for Mars we add it to the sum of the absolute central value and the quoted uncertainty. This prescription specifies the numerical allowance used below. The solar quadrupole and planetary perturbations enter through the fitted ephemeris model; their correlations with the new parameters can be included by fitting the modified equations of motion together with the observations \cite{ss:FiengaMinazzoli2024}.

For a planet of orbital period $P_i$, let $N_i$ be its number of radial revolutions per Julian century. Converting $\mathcal B_i$ to radians per century gives the inequality
\begin{equation}
\left|q_T-1-\frac{C_2}{mp_i}+\frac{15C_6}{mp_i^5}\left(1+\frac{3e_i^2}{2}+\frac{e_i^4}{8}\right)\right|\lesssim\frac{\mathcal B_i}{\pi N_i},\qquad N_i=\frac{36525\,\mathrm{d}}{P_i}.
\label{eq:ss_joint_inequality}
\end{equation}

In the neutral case, each planet gives an estimate of $|q_T-1|$. We use $m_\odot\simeq1.4766\times10^3\,\mathrm m$ and the orbital values $P_{\mathrm{Mercury}}=87.969\,\mathrm d$, $a_{\mathrm{Mercury}}=5.7909\times10^{10}\,\mathrm m$, $e_{\mathrm{Mercury}}=0.2056$, $P_{\mathrm{Mars}}=686.980\,\mathrm d$, $a_{\mathrm{Mars}}=2.27956\times10^{11}\,\mathrm m$, and $e_{\mathrm{Mars}}\simeq0.0934$ \cite{ss:NASAPlanets}. With $1\,\mathrm{mas}=4.8481368\times10^{-9}\,\mathrm{rad}$, Eq.~\eqref{eq:ss_joint_inequality} gives
\begin{equation}
|q_T-1|\lesssim1.9\times10^{-11}\quad\text{(Mercury)},\qquad |q_T-1|\lesssim2.6\times10^{-12}\quad\text{(Mars)},\qquad Q=0.
\label{eq:ss_neutral_bounds}
\end{equation}

In addition, if $q_T$ is fixed independently and the nonlinear term is negligible, the Mars allowance gives the charge sensitivity
\begin{equation}
C_2\lesssim8.5\times10^2\,\mathrm m^2,\qquad \sqrt{C_2}\lesssim2.9\times10^1\,\mathrm m,\qquad q_T=1.
\label{eq:ss_charge_bound}
\end{equation}

The quantity constrained here is the geometric charge coefficient in Eq.~\eqref{eq:ss_coefficients}, with the electromagnetic normalization of the present action. For a simultaneous estimate, Eq.~\eqref{eq:ss_joint_inequality} supplies the allowed combinations directly. In particular, the different semilatus recta of Mercury and Mars prevent a cancellation between the first two terms from occurring with the same amplitude in both orbits.


\subsection{Light deflection and radio time delay}
\label{subsec:ss_light}

The neutral configuration also permits a direct comparison with light propagation experiments. At $Q=0$, the Born--Infeld background field vanishes and electromagnetic disturbances follow the metric null cone. Let $d=\mathcal L_\gamma/\mathcal E_\gamma$ be the photon impact parameter. For the azimuthal separation between a source and a receiver, the null equations give
\begin{equation}
\Delta\phi_{\mathrm{RS}}=\sqrt{q_T}\sum_{i=\mathrm S,\mathrm R}\int_{r_0}^{r_i}\frac{d\,\mathrm{d}r}{r^2\sqrt{1-d^2f_{\mathrm{BI}}(r)/r^2}},\qquad d^2f_{\mathrm{BI}}(r_0)=r_0^2,
\label{eq:ss_light_azimuth}
\end{equation}
where $r_0$ is the radius of closest approach. A static observer measures a local photon angle satisfying $\sin\Psi_i=d\sqrt{f_{\mathrm{BI}}(r_i)}/r_i$. Taking the source angle on the ingoing part of the trajectory, the finite distance bending angle is $\alpha=\Psi_{\mathrm R}-\Psi_{\mathrm S}+\Delta\phi_{\mathrm{RS}}$ \cite{ss:Ishihara2016}. In the large distance limit,
\begin{equation}
\alpha_\infty=\pi\left(\sqrt{q_T}-1\right)+\frac{4m\sqrt{q_T}}{d}+O\!\left(\frac{m^2}{d^2}\right),\qquad Q=0.
\label{eq:ss_deflection}
\end{equation}
The first term records the conical contribution, while the second gives the mass dependent deflection. For comparison with astrometric observations, the finite source and receiver positions and the fitted source directions enter through Eq.~\eqref{eq:ss_light_azimuth}.

For a radio signal passing near the Sun, the radial time integral contains the same factor $\sqrt{q_T}$. At superior conjunction, its leading round trip logarithmic contribution is
\begin{equation}
\Delta T_{\log}=4m\sqrt{q_T}\ln\!\left(\frac{4r_{\mathrm E}r_{\mathrm R}}{d^2}\right),\qquad d\ll r_{\mathrm E},r_{\mathrm R},\qquad Q=0.
\label{eq:ss_shapiro}
\end{equation}

The time is expressed in length units; division by $c$ restores seconds. The complete travel time also includes the propagation time in the conical background and the endpoint clock conversion. Cassini measured $\gamma-1=(2.1\pm2.3)\times10^{-5}$ by using the frequency shift of the radio link \cite{ss:Bertotti2003}. Comparing the logarithmic coefficient in Eq.~\eqref{eq:ss_shapiro} with $2(1+\gamma)m$ gives, for independently specified mass and endpoint geometry,
\begin{equation}
\gamma_{\log}-1=2\left(\sqrt{q_T}-1\right)=q_T-1+O\!\left((q_T-1)^2\right).
\label{eq:ss_gamma_log}
\end{equation}

The Cassini result corresponds to a coefficient sensitivity $|\gamma_{\log}-1|\lesssim6.7\times10^{-5}$ when the absolute central value and twice the quoted uncertainty are added. Here, $\gamma_{\log}$ labels the amplitude of this particular logarithmic term. It is worth mentioning that the orbital estimates in Eq.~\eqref{eq:ss_neutral_bounds} already place this fractional logarithmic correction at the $10^{-12}$--$10^{-11}$ level.
\subsection{Bounds on the bumblebee and Born--Infeld parameters}
\label{subsec:ss_parameters}

The neutral Mars estimate translates directly into the combination of bumblebee parameters appearing in the metric,
\begin{equation}
|\xi|\left(b^2+u_0^2\right)\lesssim2.6\times10^{-12}.
\label{eq:ss_bumblebee_bound}
\end{equation}

For a purely radial expectation value, Eq.~\eqref{eq:ss_bumblebee_bound} reduces to $|\xi|b^2\lesssim2.6\times10^{-12}$. Moreover, the solution with a varying temporal amplitude at $\xi=\kappa/2$ obeys the corresponding estimate $\kappa(b^2+u_0^2)\lesssim5.2\times10^{-12}$. The amplitude $u_1$ remains free in these metric tests at that coupling, since it leaves the displayed metric unchanged.

For a specified nonzero charge, the orbital signal also probes the Born--Infeld scale. If the lower order deformation and charge contributions have been set independently, the remaining allowance gives
\begin{equation}
C_6\lesssim C_{6,\max}^{(i)}=\frac{mp_i^5}{15\left(1+3e_i^2/2+e_i^4/8\right)}\frac{\mathcal B_i}{\pi N_i},\qquad \beta_{\mathrm{BI}}\gtrsim\left(\frac{\kappa q_TQ^4}{40q_0C_{6,\max}^{(i)}}\right)^{1/2}.
\label{eq:ss_bi_bound}
\end{equation}

For Mercury, the adopted allowance gives $C_{6,\max}^{(\mathrm{Mercury})}\simeq9.2\times10^{44}\,\mathrm m^6$. Equation~\eqref{eq:ss_bi_bound} is a conditional sensitivity to the nonlinear coefficient, evaluated with $Q^2/(\beta_{\mathrm{BI}}^2r_{\mathrm p}^4)\ll1$. Its information on $\beta_{\mathrm{BI}}$ depends on the independently specified charge and on this expansion domain. In the neutral limit, the orbital and radio observables show their bounds on the bumblebee deformation for every value of $\beta_{\mathrm{BI}}$.

Furthermore, the exact lapse provides an orbital test at finite nonlinear scale through
\begin{equation}
\Delta\varpi_{\mathrm{exact}}=2\sqrt{q_T}\,\mathcal L\int_{r_{\mathrm p}}^{r_{\mathrm a}}\frac{\mathrm{d}r}{r^2\sqrt{\mathcal E^2-f_{\mathrm{BI}}(r)\left(1+\mathcal L^2/r^2\right)}}-2\pi.
\label{eq:ss_exact_precession}
\end{equation}

The constants $\mathcal E$ and $\mathcal L$ are fixed by the two radial turning points $r_{\mathrm p}$ and $r_{\mathrm a}$. Furthermore, notice that this expression incorporates the full Born--Infeld dependence in the same perihelion observable and connects the analytic estimates above with a numerical fit of the charged geometry.


\section{Conclusion}

In this work, we obtained exact static and spherically symmetric charged solutions in bumblebee gravity with minimally coupled Born--Infeld electrodynamics, a spacelike vector norm, and a vanishing cosmological constant. An independently varied Lagrange multiplier enforced the norm constraint and acquired a radial profile determined by the transverse electromagnetic pressure, although the potential vanished on shell. This formulation accommodated purely radial configurations and configurations with constant or varying temporal amplitudes. At $\xi=\kappa/2$, the varying amplitude introduced a nonzero bumblebee field strength without changing the metric obtained for the constant temporal configuration.

Using the exact lapse, we classified the parameter domains with zero, one, or two positive metric horizons. The strict convexity of $r f_{\mathrm{BI}}(r)$ distinguished extremality at a positive radius, where two horizons merged, from the central transition at which a horizon reached the origin. At fixed $(m,Q,\beta_{\mathrm{BI}})$, the purely radial deformation preserved the Einstein--Born--Infeld horizon locations, whereas the temporal amplitude modified the charged lapse and its horizon domain. The independent amplitude $u_1$ left this classification unchanged at $\xi=\kappa/2$.

The temporal configurations admitted smooth extensions across a simple future horizon when $u(r_h)\ne0$ and the radial sign matched the temporal orientation. For the varying temporal component, we also established a sufficient condition for the vector to remain real at every positive radius and verified an explicit charged example. The purely radial configuration retained a divergent bumblebee one form at a simple horizon. These properties distinguished regularity of the full field configuration from regularity of the metric alone. The electric field remained bounded, while the center was generically singular. The appropriate Maxwell, Einstein--Born--Infeld, and neutral bumblebee limits were recovered, and the asymptotic geometry was conical whenever $q_T\ne1$.

The geodesic analysis identified complementary sensitivities to the lapse and the constant radial deformation. At fixed case, capture thresholds, circular orbit radii, and circular specific energies and angular momenta were independent of an additional variation of $q_T$. For bound trajectories with fixed turning radii, the radial period and accumulated azimuth instead scaled as $\sqrt{q_T}$. Radial oscillation frequencies and the instability rates of metric photon orbits scaled as $q_T^{-1/2}$ in the normalized Killing time. We also determined the ISCO and the boundary between eccentric bound motion and plunge. At fixed geometric charge, the leading Born--Infeld correction shifted the ISCO outward on the physical Maxwell black hole case. Remarkably, what we have got was that the orbital precession and radial timescales supplied information about the deformation that circular radii alone could not provide.

For the optical analysis, the electromagnetic action determined the relevant propagation law. Both polarizations shared a single optical cone in the full determinant Born--Infeld theory, whereas the literal $\mathcal L(\mathcal I)$ theory admitted ordinary and extraordinary characteristics despite having the same electric background. At fixed geometric charge, the leading nonlinear corrections enlarged the photon sphere and shadow radii. At fixed $(m,Q,\beta_{\mathrm{BI}})$, increasing $\eta=q_T/q_0$ reduced the shadow radius along an admissible capture scenario. The purely radial deformation remained absent from the shadow size, and the neutral configurations shared $r_{\mathrm{ph}}=3m$ and $R_{\mathrm{sh}}=3\sqrt{3}\,m$. Another remark showed was the fact that, remarkably, the dependence of circular ray instability rates on the radial coefficient nevertheless suggested a complementary role for time dependent lensing observables.

Comparisons with the results for the shadow intervals for Sgr~A$^*$ and M87$^*$ yielded conditional constraints on the charge and nonlinear scale. In the Maxwell limit, the adopted $68\%$ intervals gave approximately $e\lesssim0.90$ and $e\lesssim0.84$ for the alternative VLTI and Keck priors of Sgr~A$^*$, respectively, and $e\lesssim0.92$ for M87$^*$, where $e^2=\kappa\eta Q^2/(2m^2)$. At finite nonlinearity, the compatibility regions restricted $\beta_{\mathrm{BI}}$ at specified charge and propagation law. Their interpretation depended on image calibration and independent mass information; doubling the adopted errors admitted the entire Maxwell black hole charge interval. The GRAVITY precession result gave complementary sensitivity to the radial deformation together with the charge contribution.

Under the assumption of a neutral solar exterior, the adopted Mercury and Mars perihelion allowances yielded illustrative sensitivities of $|q_T-1|\lesssim1.9\times10^{-11}$ and $|q_T-1|\lesssim2.6\times10^{-12}$, respectively. The latter translated into $|\xi|(b^2+u_0^2)\lesssim2.6\times10^{-12}$ for the solar configuration. For a specified nonzero charge, the remaining nonlinear orbital allowance supplied a conditional lower bound on $\beta_{\mathrm{BI}}$ when the lower order contributions were independently used and the expansion remained valid.

Future work could investigate gravitational lensing in the weak and strong deflection regimes, the effects of lensing on neutrino oscillations, matter accretion, quasinormal modes, and gravitational wave emission associated with periodic orbits in the black hole spacetimes derived here.


\section*{Acknowledgments}
\hspace{0.5cm} A. A. Araújo Filho is supported by Conselho Nacional de Desenvolvimento Cient\'{\i}fico e Tecnol\'{o}gico (CNPq) -- [150223/2025-0]. This work by A.~F.~Santos is partially supported by National Council for Scientific and Technological Development - CNPq project No. 312406/2023-1. E. M. Siquieri thanks CAPES for financial support.

\section*{Data Availability Statement}

Data associated with this study consist of the analytical expressions and numerical figures presented in the manuscript. No additional dataset is required to reproduce the analytical results. The code used to generate the numerical results and figures is available from the corresponding author upon reasonable request.

\bibliography{main}

@article{Kostelecky2004,
  author = {Kosteleck{\'y}, V. Alan},
  title = {Gravity, {Lorentz} violation, and the {Standard Model}},
  journal = {Phys. Rev. D},
  volume = {69},
  pages = {105009},
  year = {2004},
  doi = {10.1103/PhysRevD.69.105009},
  eprint = {hep-th/0312310},
  archivePrefix = {arXiv}
}

@article{BluhmKostelecky2005,
  author = {Bluhm, Robert and Kosteleck{\'y}, V. Alan},
  title = {Spontaneous {Lorentz} violation, {Nambu--Goldstone} modes, and gravity},
  journal = {Phys. Rev. D},
  volume = {71},
  pages = {065008},
  year = {2005},
  doi = {10.1103/PhysRevD.71.065008},
  eprint = {hep-th/0412320},
  archivePrefix = {arXiv}
}

@article{BluhmFungKostelecky2008,
  author = {Bluhm, Robert and Fung, Shu-Hong and Kosteleck{\'y}, V. Alan},
  title = {Spontaneous {Lorentz} and diffeomorphism violation, massive modes, and gravity},
  journal = {Phys. Rev. D},
  volume = {77},
  pages = {065020},
  year = {2008},
  doi = {10.1103/PhysRevD.77.065020},
  eprint = {0712.4119},
  archivePrefix = {arXiv},
  primaryClass = {hep-th}
}

@article{BertolamiParamos2005,
  author = {Bertolami, Orfeu and P{\'a}ramos, Jorge},
  title = {The flight of the bumblebee: Vacuum solutions of a gravity model with vector-induced spontaneous {Lorentz} symmetry breaking},
  journal = {Phys. Rev. D},
  volume = {72},
  pages = {044001},
  year = {2005},
  doi = {10.1103/PhysRevD.72.044001},
  eprint = {hep-th/0504215},
  archivePrefix = {arXiv}
}

@article{Casana2018,
  author = {Casana, R. and Cavalcante, A. and Poulis, F. P. and Santos, E. B.},
  title = {Exact {Schwarzschild}-like solution in a bumblebee gravity model},
  journal = {Phys. Rev. D},
  volume = {97},
  pages = {104001},
  year = {2018},
  doi = {10.1103/PhysRevD.97.104001},
  eprint = {1711.02273},
  archivePrefix = {arXiv},
  primaryClass = {gr-qc}
}

@article{MalufNeves2021,
  author = {Maluf, R. V. and Neves, Juliano C. S.},
  title = {Black holes with a cosmological constant in bumblebee gravity},
  journal = {Phys. Rev. D},
  volume = {103},
  pages = {044002},
  year = {2021},
  doi = {10.1103/PhysRevD.103.044002},
  eprint = {2011.12841},
  archivePrefix = {arXiv},
  primaryClass = {gr-qc}
}

@article{XuLiangShao2023,
  author = {Xu, Rui and Liang, Dicong and Shao, Lijing},
  title = {Static spherical vacuum solutions in the bumblebee gravity model},
  journal = {Phys. Rev. D},
  volume = {107},
  pages = {024011},
  year = {2023},
  doi = {10.1103/PhysRevD.107.024011},
  eprint = {2209.02209},
  archivePrefix = {arXiv},
  primaryClass = {gr-qc}
}

@article{ZhuLi2026,
  author = {Zhu, Jie and Li, Hao},
  title = {Full classification of static spherical vacuum solutions to bumblebee gravity with general {VEVs}},
  journal = {Phys. Lett. B},
  volume = {876},
  pages = {140396},
  year = {2026},
  doi = {10.1016/j.physletb.2026.140396},
  eprint = {2511.03231},
  archivePrefix = {arXiv},
  primaryClass = {gr-qc}
}

@article{Nutma2014,
  author = {Nutma, Teake},
  title = {{xTras}: A field-theory inspired {xAct} package for {Mathematica}},
  journal = {Comput. Phys. Commun.},
  volume = {185},
  pages = {1719--1738},
  year = {2014},
  doi = {10.1016/j.cpc.2014.02.006},
  eprint = {1308.3493},
  archivePrefix = {arXiv},
  primaryClass = {cs.SC}
}

@article{Ovcharenko2026,
  author = {Ovcharenko, Hryhorii},
  title = {Exact {Kerr--Newman--(A)dS} and other spacetimes in bumblebee gravity: Employing a simple generating technique},
  journal = {Phys. Rev. D},
  volume = {114},
  pages = {064032},
  year = {2026},
  doi = {10.1103/4sp9-szhk},
  eprint = {2601.16037},
  archivePrefix = {arXiv},
  primaryClass = {gr-qc}
}

@article{AraujoFilho2023,
  author = {{Ara{\'u}jo Filho}, A. A. and Nascimento, J. R. and Petrov, A. Yu. and Porf{\'i}rio, P. J.},
  title = {Vacuum solution within a metric-affine bumblebee gravity},
  journal = {Phys. Rev. D},
  volume = {108},
  pages = {085010},
  year = {2023},
  doi = {10.1103/PhysRevD.108.085010},
  eprint = {2211.11821},
  archivePrefix = {arXiv},
  primaryClass = {gr-qc}
}

@misc{AraujoFilho2026,
  author = {{Ara{\'u}jo Filho}, A. A. and Heidari, N. and Bezerra, V. B. and Lobo, Francisco S. N.},
  title = {A general static and spherically symmetric black hole in traceless metric-affine bumblebee gravity},
  year = {2026},
  eprint = {2609.29675},
  archivePrefix = {arXiv},
  primaryClass = {gr-qc},
  url = {https://arxiv.org/abs/2609.29675},
  note = {Preprint, version 1}
}

@article{LiuWuWeiLiu2026,
  author = {Liu, Jia-Zhou and Wu, Shan-Ping and Wei, Shao-Wen and Liu, Yu-Xiao},
  title = {Exact black hole solutions in bumblebee gravity with lightlike or spacelike {VEVs}},
  journal = {Sci. China Phys. Mech. Astron.},
  volume = {69},
  pages = {270411},
  year = {2026},
  doi = {10.1007/s11433-026-2961-8},
  eprint = {2510.16731},
  archivePrefix = {arXiv},
  primaryClass = {gr-qc}
}

@article{KosteleckySamuel1989,
  author = {Kosteleck{\'y}, V. Alan and Samuel, Stuart},
  title = {Spontaneous breaking of {Lorentz} symmetry in string theory},
  journal = {Phys. Rev. D},
  volume = {39},
  pages = {683--685},
  year = {1989},
  doi = {10.1103/PhysRevD.39.683}
}

@article{BornInfeld1934,
  author = {Born, Max and Infeld, Leopold},
  title = {Foundations of the new field theory},
  journal = {Proc. Roy. Soc. Lond. A},
  volume = {144},
  number = {852},
  pages = {425--451},
  year = {1934},
  doi = {10.1098/rspa.1934.0059}
}

@article{CaiPangWang2004,
  author = {Cai, Rong-Gen and Pang, Da-Wei and Wang, Anzhong},
  title = {{Born--Infeld} black holes in {(A)dS} spaces},
  journal = {Phys. Rev. D},
  volume = {70},
  pages = {124034},
  year = {2004},
  doi = {10.1103/PhysRevD.70.124034},
  eprint = {hep-th/0410158},
  archivePrefix = {arXiv}
}

@article{LiLu2026,
  author = {Li, Zhi-Chao and L{\"u}, H.},
  title = {When bumblebee meets {NLED}: {Lorentz}-violating black holes and regular spacetimes},
  journal = {Eur. Phys. J. C},
  volume = {86},
  pages = {959},
  year = {2026},
  doi = {10.1140/epjc/s10052-026-16219-5},
  eprint = {2605.13963},
  archivePrefix = {arXiv},
  primaryClass = {gr-qc}
}

@article{ss:Will2014,
  author        = {Will, Clifford M.},
  title         = {The Confrontation between General Relativity and Experiment},
  journal       = {Living Reviews in Relativity},
  volume        = {17},
  pages         = {4},
  year          = {2014},
  doi           = {10.12942/lrr-2014-4},
  eprint        = {1403.7377},
  archivePrefix = {arXiv},
  primaryClass  = {gr-qc}
}

@article{ss:Casana2018,
  author        = {Casana, R. and Cavalcante, A. and Poulis, F. P. and Santos, E. B.},
  title         = {Exact {Schwarzschild}-like solution in a bumblebee gravity model},
  journal       = {Physical Review D},
  volume        = {97},
  pages         = {104001},
  year          = {2018},
  doi           = {10.1103/PhysRevD.97.104001},
  eprint        = {1711.02273},
  archivePrefix = {arXiv},
  primaryClass  = {gr-qc}
}

@article{ss:Iorio2012,
  author        = {Iorio, Lorenzo},
  title         = {Constraining the electric charges of some astronomical bodies in {Reissner--Nordstr\"om} spacetimes and generic {$r^{-2}$}-type power-law potentials from orbital motions},
  journal       = {General Relativity and Gravitation},
  volume        = {44},
  pages         = {1753--1767},
  year          = {2012},
  doi           = {10.1007/s10714-012-1365-0},
  eprint        = {1112.3520},
  archivePrefix = {arXiv},
  primaryClass  = {gr-qc}
}

@misc{ss:Fienga2016,
  author        = {Fienga, Agn{\`e}s and Laskar, Jacques and Manche, Herv{\'e} and Gastineau, Micka{\"e}l},
  title         = {Tests of {GR} with {INPOP15a} planetary ephemerides: estimations of possible supplementary advances of perihelia for {Mercury} and {Saturn}},
  year          = {2016},
  eprint        = {1601.00947},
  archivePrefix = {arXiv},
  primaryClass  = {gr-qc},
  url           = {https://arxiv.org/abs/1601.00947}
}

@article{ss:Pitjeva2013,
  author        = {Pitjeva, E. V. and Pitjev, N. P.},
  title         = {Relativistic effects and dark matter in the {Solar} system from observations of planets and spacecraft},
  journal       = {Monthly Notices of the Royal Astronomical Society},
  volume        = {432},
  pages         = {3431--3437},
  year          = {2013},
  doi           = {10.1093/mnras/stt695},
  eprint        = {1306.3043},
  archivePrefix = {arXiv},
  primaryClass  = {astro-ph.EP}
}

@article{ss:FiengaMinazzoli2024,
  author        = {Fienga, Agn{\`e}s and Minazzoli, Olivier},
  title         = {Testing theories of gravity with planetary ephemerides},
  journal       = {Living Reviews in Relativity},
  volume        = {27},
  pages         = {1},
  year          = {2024},
  doi           = {10.1007/s41114-023-00047-0},
  eprint        = {2303.01821},
  archivePrefix = {arXiv},
  primaryClass  = {gr-qc}
}

@article{ss:Iorio2011,
  author        = {Iorio, Lorenzo and Lichtenegger, Herbert I. M. and Ruggiero, Matteo Luca and Corda, Christian},
  title         = {Phenomenology of the {Lense--Thirring} effect in the {Solar System}},
  journal       = {Astrophysics and Space Science},
  volume        = {331},
  pages         = {351--395},
  year          = {2011},
  doi           = {10.1007/s10509-010-0489-5},
  eprint        = {1009.3225},
  archivePrefix = {arXiv},
  primaryClass  = {gr-qc}
}

@misc{ss:NASAPlanets,
  author       = {Williams, David R.},
  title        = {{Mercury Fact Sheet} and {Mars Fact Sheet}},
  howpublished = {NASA Space Science Data Coordinated Archive},
  url          = {https://nssdc.gsfc.nasa.gov/planetary/factsheet/mercuryfact.html},
  note         = {Mars data: \url{https://nssdc.gsfc.nasa.gov/planetary/factsheet/marsfact.html}. Accessed October 22, 2026}
}

@article{ss:Ishihara2016,
  author        = {Ishihara, Asahi and Suzuki, Yusuke and Ono, Toshiaki and Kitamura, Takao and Asada, Hideki},
  title         = {Gravitational bending angle of light for finite distance and the {Gauss--Bonnet} theorem},
  journal       = {Physical Review D},
  volume        = {94},
  pages         = {084015},
  year          = {2016},
  doi           = {10.1103/PhysRevD.94.084015},
  eprint        = {1604.08308},
  archivePrefix = {arXiv},
  primaryClass  = {gr-qc}
}

@article{ss:Bertotti2003,
  author  = {Bertotti, B. and Iess, L. and Tortora, P.},
  title   = {A test of general relativity using radio links with the {Cassini} spacecraft},
  journal = {Nature},
  volume  = {425},
  pages   = {374--376},
  year    = {2003},
  doi     = {10.1038/nature01997}
}

@article{sh:Perlick2022,
  author = {Perlick, Volker and Tsupko, Oleg Yu.},
  title = {Calculating black hole shadows: Review of analytical studies},
  journal = {Physics Reports},
  volume = {947},
  pages = {1--39},
  year = {2022},
  doi = {10.1016/j.physrep.2021.10.004},
  eprint = {2105.07101},
  archivePrefix = {arXiv},
  primaryClass = {gr-qc}
}

@article{sh:Novello2000,
  author = {Novello, M. and De Lorenci, V. A. and Salim, J. M. and Klippert, R.},
  title = {Geometrical aspects of light propagation in nonlinear electrodynamics},
  journal = {Physical Review D},
  volume = {61},
  pages = {045001},
  year = {2000},
  doi = {10.1103/PhysRevD.61.045001},
  eprint = {gr-qc/9911085},
  archivePrefix = {arXiv}
}

@article{sh:Gibbons2001,
  author = {Gibbons, G. W. and Herdeiro, C. A. R.},
  title = {{Born--Infeld} theory and stringy causality},
  journal = {Physical Review D},
  volume = {63},
  pages = {064006},
  year = {2001},
  doi = {10.1103/PhysRevD.63.064006},
  eprint = {hep-th/0008052},
  archivePrefix = {arXiv}
}

@article{sh:Abalos2015,
  author = {Abalos, Fernando and Carrasco, Federico and Goulart, {\'E}rico and Reula, Oscar},
  title = {Nonlinear electrodynamics as a symmetric hyperbolic system},
  journal = {Physical Review D},
  volume = {92},
  pages = {084024},
  year = {2015},
  doi = {10.1103/PhysRevD.92.084024},
  eprint = {1507.02262},
  archivePrefix = {arXiv},
  primaryClass = {gr-qc}
}

@article{sh:EHTSgr2022VI,
  author = {Akiyama, Kazunori and others},
  collaboration = {Event Horizon Telescope},
  title = {First {Sagittarius A* Event Horizon Telescope} Results. {VI}. Testing the Black Hole Metric},
  journal = {The Astrophysical Journal Letters},
  volume = {930},
  pages = {L17},
  year = {2022},
  doi = {10.3847/2041-8213/ac6756},
  eprint = {2311.09484},
  archivePrefix = {arXiv},
  primaryClass = {astro-ph.HE},
  note = {Published in 2022; arXiv deposit in 2023}
}

@article{sh:EHTM872019VI,
  author = {Akiyama, Kazunori and others},
  collaboration = {Event Horizon Telescope},
  title = {First {M87 Event Horizon Telescope} Results. {VI}. The Shadow and Mass of the Central Black Hole},
  journal = {The Astrophysical Journal Letters},
  volume = {875},
  pages = {L6},
  year = {2019},
  doi = {10.3847/2041-8213/ab1141},
  eprint = {1906.11243},
  archivePrefix = {arXiv},
  primaryClass = {astro-ph.GA}
}

@article{sh:Kocherlakota2021,
  author = {Kocherlakota, Prashant and Rezzolla, Luciano and Falcke, Heino and others},
  title = {Constraints on black-hole charges with the 2017 {EHT} observations of {M87*}},
  journal = {Physical Review D},
  volume = {103},
  pages = {104047},
  year = {2021},
  doi = {10.1103/PhysRevD.103.104047},
  eprint = {2105.09343},
  archivePrefix = {arXiv},
  primaryClass = {gr-qc}
}

@article{sh:EHTM872025,
  author = {Akiyama, Kazunori and others},
  collaboration = {Event Horizon Telescope},
  title = {Horizon-scale variability of {M87*} from 2017--2021 {EHT} observations},
  journal = {Astronomy and Astrophysics},
  volume = {704},
  pages = {A91},
  year = {2025},
  doi = {10.1051/0004-6361/202555855},
  eprint = {2509.24593},
  archivePrefix = {arXiv},
  primaryClass = {astro-ph.HE}
}

@article{sh:GRAVITY2020,
  author = {Abuter, R. and others},
  collaboration = {GRAVITY},
  title = {Detection of the {Schwarzschild} precession in the orbit of the star {S2} near the {Galactic} centre massive black hole},
  journal = {Astronomy and Astrophysics},
  volume = {636},
  pages = {L5},
  year = {2020},
  doi = {10.1051/0004-6361/202037813},
  eprint = {2004.07187},
  archivePrefix = {arXiv},
  primaryClass = {astro-ph.GA}
}

@article{sh:GRAVITY2024,
  author = {Abd El Dayem, K. and others},
  collaboration = {GRAVITY},
  title = {Improving constraints on the extended mass distribution in the {Galactic} center with stellar orbits},
  journal = {Astronomy and Astrophysics},
  volume = {692},
  pages = {A242},
  year = {2024},
  doi = {10.1051/0004-6361/202452274},
  eprint = {2409.12261},
  archivePrefix = {arXiv},
  primaryClass = {astro-ph.GA}
}

@article{GulluOvgun2022,
  author = {G{\"u}ll{\"u}, {\.I}. and {\"O}vg{\"u}n, A.},
  title = {{Schwarzschild-like black hole with a topological defect in bumblebee gravity}},
  journal = {Annals Phys.},
  volume = {436},
  pages = {168721},
  year = {2022},
  doi = {10.1016/j.aop.2021.168721},
  eprint = {2012.02611},
  url = {https://arxiv.org/abs/2012.02611},
  archivePrefix = {arXiv}
}

@article{DingHigher2023,
  author = {Ding, C. and Shi, Y. and Chen, J. and Zhou, Y. and Liu, C.},
  title = {{High dimensional AdS-like black hole and phase transition in Einstein-bumblebee gravity}},
  journal = {Chin. Phys. C},
  volume = {47},
  pages = {045102},
  year = {2023},
  doi = {10.1088/1674-1137/aca8f4},
  eprint = {2201.06683},
  url = {https://arxiv.org/abs/2201.06683},
  archivePrefix = {arXiv}
}

@article{DingBTZ2023,
  author = {Ding, C. and Shi, Y. and Chen, J. and Zhou, Y. and Liu, C. and Xiao, Y.},
  title = {{Rotating BTZ-like black hole and central charges in Einstein-bumblebee gravity}},
  journal = {Eur. Phys. J. C},
  volume = {83},
  pages = {573},
  year = {2023},
  doi = {10.1140/epjc/s10052-023-11761-y},
  eprint = {2302.01580},
  url = {https://arxiv.org/abs/2302.01580},
  archivePrefix = {arXiv}
}

@misc{Valtancoli2023,
  author = {Valtancoli, P.},
  title = {{Bumblebee gravity with cosmological constant}},
  year = {2023},
  eprint = {2308.05328},
  url = {https://arxiv.org/abs/2308.05328},
  archivePrefix = {arXiv},
  note = {Preprint}
}

@article{LiZhuTimelike2026,
  author = {Li, H. and Zhu, J.},
  title = {{Static spherical vacuum solution to bumblebee gravity with time-like VEVs}},
  journal = {Eur. Phys. J. C},
  volume = {86},
  pages = {2},
  year = {2026},
  doi = {10.1140/epjc/s10052-025-15229-z},
  eprint = {2506.17957},
  url = {https://arxiv.org/abs/2506.17957},
  archivePrefix = {arXiv}
}

@article{LiuWu2026,
  author = {Liu, J.-Z. and Wu, S.-P. and Wei, S.-W. and Liu, Y.-X.},
  title = {{Exact black hole solutions in bumblebee gravity with lightlike or spacelike VEVs}},
  journal = {Sci. China Phys. Mech. Astron.},
  volume = {69},
  pages = {270411},
  year = {2026},
  doi = {10.1007/s11433-026-2961-8},
  eprint = {2510.16731},
  url = {https://arxiv.org/abs/2510.16731},
  archivePrefix = {arXiv}
}

@article{BaileyMurrayWalter2025,
  author = {Bailey, Q. G. and Murray, H. S. and Walter-Cardona, D. T.},
  title = {{Bumblebee gravity: Spherically symmetric solutions away from the potential minimum}},
  journal = {Phys. Rev. D},
  volume = {112},
  pages = {024069},
  year = {2025},
  doi = {10.1103/nkbj-vbjf},
  eprint = {2503.10998},
  url = {https://arxiv.org/abs/2503.10998},
  archivePrefix = {arXiv}
}

@misc{ZhuLiExtended2026,
  author = {Zhu, J. and Li, H.},
  title = {{New exact vacuum solutions in extended bumblebee gravity}},
  year = {2026},
  eprint = {2604.09464},
  url = {https://arxiv.org/abs/2604.09464},
  archivePrefix = {arXiv},
  note = {Preprint}
}

@article{DingKerr2020,
  author = {Ding, C. and Liu, C. and Casana, R. and Cavalcante, A.},
  title = {{Exact Kerr-like solution and its shadow in a gravity model with spontaneous Lorentz symmetry breaking}},
  journal = {Eur. Phys. J. C},
  volume = {80},
  pages = {178},
  year = {2020},
  doi = {10.1140/epjc/s10052-020-7743-y},
  eprint = {1910.02674},
  url = {https://arxiv.org/abs/1910.02674},
  archivePrefix = {arXiv}
}

@article{JhaRahaman2021,
  author = {Jha, S. K. and Rahaman, A.},
  title = {{Bumblebee gravity with a Kerr-Sen-like solution and its Shadow}},
  journal = {Eur. Phys. J. C},
  volume = {81},
  pages = {345},
  year = {2021},
  doi = {10.1140/epjc/s10052-021-09132-6},
  eprint = {2011.14916},
  url = {https://arxiv.org/abs/2011.14916},
  archivePrefix = {arXiv}
}

@article{MalufMuniz2022,
  author = {Maluf, R. V. and Muniz, C. R.},
  title = {{Comment on {``Greybody radiation and quasinormal modes of Kerr-like black hole in Bumblebee gravity model''}}},
  journal = {Eur. Phys. J. C},
  volume = {82},
  pages = {94},
  year = {2022},
  doi = {10.1140/epjc/s10052-022-10039-z},
  eprint = {2202.01015},
  url = {https://arxiv.org/abs/2202.01015},
  archivePrefix = {arXiv}
}

@article{DingChen2021,
  author = {Ding, C. and Chen, X.},
  title = {{Slowly rotating Einstein-bumblebee black hole solution and its greybody factor in a Lorentz violation model}},
  journal = {Chin. Phys. C},
  volume = {45},
  pages = {025106},
  year = {2021},
  doi = {10.1088/1674-1137/abce51},
  eprint = {2008.10474},
  url = {https://arxiv.org/abs/2008.10474},
  archivePrefix = {arXiv}
}

@article{PoulisSoares2022,
  author = {Poulis, F. P. and Soares, M. A. C.},
  title = {{Exact modifications on a vacuum spacetime due to a gradient bumblebee field at its vacuum expectation value}},
  journal = {Eur. Phys. J. C},
  volume = {82},
  pages = {613},
  year = {2022},
  doi = {10.1140/epjc/s10052-022-10547-y},
  eprint = {2112.04040},
  url = {https://arxiv.org/abs/2112.04040},
  archivePrefix = {arXiv}
}

@article{ChenLiu2025,
  author = {Chen, Y.-Q. and Liu, H.-S.},
  title = {{Taub-NUT-like black holes in Einstein-bumblebee gravity}},
  journal = {Phys. Rev. D},
  volume = {112},
  pages = {084040},
  year = {2025},
  doi = {10.1103/wmhj-83x3},
  eprint = {2505.23104},
  url = {https://arxiv.org/abs/2505.23104},
  archivePrefix = {arXiv}
}

@article{AraujoMetricAffine2023,
  author = {Ara{\'u}jo Filho, A. A. and Nascimento, J. R. and Petrov, A. Yu. and Porf{\'i}rio, P. J.},
  title = {{Vacuum solution within a metric-affine bumblebee gravity}},
  journal = {Phys. Rev. D},
  volume = {108},
  pages = {085010},
  year = {2023},
  doi = {10.1103/PhysRevD.108.085010},
  eprint = {2211.11821},
  url = {https://arxiv.org/abs/2211.11821},
  archivePrefix = {arXiv}
}

@article{AraujoAxisymmetric2024,
  author = {Ara{\'u}jo Filho, A. A. and Nascimento, J. R. and Petrov, A. Yu. and Porf{\'i}rio, P. J.},
  title = {{An exact stationary axisymmetric vacuum solution within a metric-affine bumblebee gravity}},
  journal = {JCAP},
  volume = {07},
  pages = {004},
  year = {2024},
  doi = {10.1088/1475-7516/2024/07/004},
  eprint = {2402.13014},
  url = {https://arxiv.org/abs/2402.13014},
  archivePrefix = {arXiv}
}

@misc{AraujoGeneral2026,
  author = {Ara{\'u}jo Filho, A. A. and Heidari, N. and Bezerra, V. B. and Lobo, F. S. N.},
  title = {{A general static and spherically symmetric black hole in traceless metric-affine bumblebee gravity}},
  year = {2026},
  eprint = {2609.29675},
  url = {https://arxiv.org/abs/2609.29675},
  archivePrefix = {arXiv},
  note = {Preprint}
}

@article{DingGaussBonnet2022,
  author = {Ding, C. and Chen, X. and Fu, X.},
  title = {{Einstein-Gauss-Bonnet gravity coupled to bumblebee field in four dimensional spacetime}},
  journal = {Nucl. Phys. B},
  volume = {975},
  pages = {115688},
  year = {2022},
  doi = {10.1016/j.nuclphysb.2022.115688},
  eprint = {2102.13335},
  url = {https://arxiv.org/abs/2102.13335},
  archivePrefix = {arXiv}
}

@article{AraujoNoncommutative2026,
  author = {Ara{\'u}jo Filho, A. A. and Heidari, N. and Lobo, I. P. and Shi, Y. and Lobo, F. S. N.},
  title = {{The flight of the bumblebee in a non-commutative geometry: A new black hole solution}},
  journal = {Annals Phys.},
  volume = {490},
  pages = {170487},
  year = {2026},
  doi = {10.1016/j.aop.2026.170487},
  eprint = {2509.17867},
  url = {https://arxiv.org/abs/2509.17867},
  archivePrefix = {arXiv}
}

@article{LiuGuo2025,
  author = {Liu, J.-Z. and Guo, W.-D. and Wei, S.-W. and Liu, Y.-X.},
  title = {{Charged spherically symmetric and slowly rotating charged black hole solutions in bumblebee gravity}},
  journal = {Eur. Phys. J. C},
  volume = {85},
  pages = {145},
  year = {2025},
  doi = {10.1140/epjc/s10052-025-13859-x},
  eprint = {2407.08396},
  url = {https://arxiv.org/abs/2407.08396},
  archivePrefix = {arXiv}
}

@article{LessaSilva2023,
  author = {Lessa, L. A. and Silva, J. E. G.},
  title = {{Einstein-Bumblebee-dilaton black hole solution}},
  journal = {Eur. Phys. J. C},
  volume = {83},
  pages = {1035},
  year = {2023},
  doi = {10.1140/epjc/s10052-023-12201-7},
  eprint = {2308.14646},
  url = {https://arxiv.org/abs/2308.14646},
  archivePrefix = {arXiv}
}

@article{LessaSilva2024,
  author = {Lessa, L. A. and Silva, J. E. G.},
  title = {{Einstein-Bumblebee-dilaton black hole in Lifshitz spacetimes}},
  journal = {Eur. Phys. J. C},
  volume = {84},
  pages = {194},
  year = {2024},
  doi = {10.1140/epjc/s10052-024-12576-1},
  eprint = {2309.07269},
  url = {https://arxiv.org/abs/2309.07269},
  archivePrefix = {arXiv}
}

@article{AhmedKalaSilva2026,
  author = {Ahmed, F. and Kala, S. and Silva, E. O.},
  title = {{Charged black holes in bumblebee gravity with global monopole: Thermodynamics and shadow}},
  journal = {Annals Phys.},
  volume = {491},
  pages = {170532},
  year = {2026},
  doi = {10.1016/j.aop.2026.170532},
  eprint = {2604.00883},
  url = {https://arxiv.org/abs/2604.00883},
  archivePrefix = {arXiv}
}

@article{SekhmaniModMax2026,
  author = {Sekhmani, Y. and Liu, W. and Deng, W. and Boshkayev, K.},
  title = {{Quasinormal modes of massive scalar perturbations in slow-rotation bumblebee black holes with traceless conformal electrodynamics}},
  journal = {Phys. Dark Univ.},
  volume = {52},
  pages = {102358},
  year = {2026},
  doi = {10.1016/j.dark.2026.102358},
  eprint = {2510.16639},
  url = {https://arxiv.org/abs/2510.16639},
  archivePrefix = {arXiv}
}

@article{SekhmaniPhantom2025,
  author = {Sekhmani, Y. and Maurya, S. K. and Rayimbaev, J. and Altanji, M. and Ibragimov, I. and Muminov, S.},
  title = {{Exact static and slowly rotating phantom conformal nonlinear-electrodynamic black holes in Bumblebee gravity with Lorentz Violation and shadow phenomenology}},
  journal = {Phys. Dark Univ.},
  volume = {50},
  pages = {102116},
  year = {2025},
  doi = {10.1016/j.dark.2025.102116},
  url = {https://www.sciencedirect.com/science/article/abs/pii/S2212686425003097}
}

@article{BelchiorModMax2026,
  author = {Belchior, F. M. and Moreira, A. R. P. and Bouzenada, A. and Ahmed, F.},
  title = {{ModMax black hole surrounded by cloud of strings in bumblebee gravity}},
  journal = {Eur. Phys. J. Plus},
  volume = {141},
  pages = {959},
  year = {2026},
  doi = {10.1140/epjp/s13360-026-08166-7},
  url = {https://doi.org/10.1140/epjp/s13360-026-08166-7}
}

@misc{AraujoSimpsonVisser2026,
  author = {Ara{\'u}jo Filho, A. A.},
  title = {{An exact Simpson-Visser family in bumblebee gravity}},
  year = {2026},
  eprint = {2610.21701},
  url = {https://arxiv.org/abs/2610.21701},
  archivePrefix = {arXiv},
  note = {Preprint}
}

@article{FradkinTseytlin1985,
  author = {Fradkin, E. S. and Tseytlin, A. A.},
  title = {{Non-linear electrodynamics from quantized strings}},
  journal = {Phys. Lett. B},
  volume = {163},
  pages = {123--130},
  year = {1985},
  doi = {10.1016/0370-2693(85)90205-9},
  url = {https://doi.org/10.1016/0370-2693(85)90205-9}
}

@misc{Tseytlin1999,
  author = {Tseytlin, A. A.},
  title = {{Born-Infeld action, supersymmetry and string theory}},
  year = {1999},
  doi = {10.1142/9789812793850_0025},
  eprint = {hep-th/9908105},
  url = {https://arxiv.org/abs/hep-th/9908105},
  archivePrefix = {arXiv}
}

@article{Hoffmann1935,
  author = {Hoffmann, B.},
  title = {{Gravitational and electromagnetic mass in the Born-Infeld electrodynamics}},
  journal = {Phys. Rev.},
  volume = {47},
  pages = {877--880},
  year = {1935},
  doi = {10.1103/PhysRev.47.877},
  url = {https://doi.org/10.1103/PhysRev.47.877}
}

@article{GarciaSalazarPlebanski1984,
  author = {Garc{\'i}a D., A. and Salazar I., H. and Pleba{\'n}ski, J. F.},
  title = {{Type-D solutions of the Einstein and Born-Infeld nonlinear-electrodynamics equations}},
  journal = {Nuovo Cim. B},
  volume = {84},
  pages = {65--90},
  year = {1984},
  doi = {10.1007/BF02721649},
  url = {https://doi.org/10.1007/BF02721649}
}

@article{FernandoKrug2003,
  author = {Fernando, S. and Krug, D.},
  title = {{Charged black hole solutions in Einstein-Born-Infeld gravity with a cosmological constant}},
  journal = {Gen. Rel. Grav.},
  volume = {35},
  pages = {129--137},
  year = {2003},
  doi = {10.1023/A:1021315214180},
  eprint = {hep-th/0306120},
  url = {https://arxiv.org/abs/hep-th/0306120},
  archivePrefix = {arXiv}
}

@article{Dey2004,
  author = {Dey, T. K.},
  title = {{Born-Infeld black holes in the presence of a cosmological constant}},
  journal = {Phys. Lett. B},
  volume = {595},
  pages = {484--490},
  year = {2004},
  doi = {10.1016/j.physletb.2004.06.047},
  eprint = {hep-th/0406169},
  url = {https://arxiv.org/abs/hep-th/0406169},
  archivePrefix = {arXiv}
}

@article{Breton2002,
  author = {Bret{\'o}n, N.},
  title = {{Geodesic structure of the Born-Infeld black hole}},
  journal = {Class. Quant. Grav.},
  volume = {19},
  pages = {601--612},
  year = {2002},
  doi = {10.1088/0264-9381/19/4/301},
  url = {https://doi.org/10.1088/0264-9381/19/4/301}
}

@article{Novello2000,
  author = {Novello, M. and De Lorenci, V. A. and Salim, J. M. and Klippert, R.},
  title = {{Geometrical aspects of light propagation in nonlinear electrodynamics}},
  journal = {Phys. Rev. D},
  volume = {61},
  pages = {045001},
  year = {2000},
  doi = {10.1103/PhysRevD.61.045001},
  eprint = {gr-qc/9911085},
  url = {https://arxiv.org/abs/gr-qc/9911085},
  archivePrefix = {arXiv}
}

@article{Abalos2015,
  author = {Abalos, F. and Carrasco, F. and Goulart, {\'E}. and Reula, O.},
  title = {{Nonlinear electrodynamics as a symmetric hyperbolic system}},
  journal = {Phys. Rev. D},
  volume = {92},
  pages = {084024},
  year = {2015},
  doi = {10.1103/PhysRevD.92.084024},
  eprint = {1507.02262},
  url = {https://arxiv.org/abs/1507.02262},
  archivePrefix = {arXiv}
}

@article{GibbonsHerdeiro2001,
  author = {Gibbons, G. W. and Herdeiro, C. A. R.},
  title = {{Born-Infeld theory and stringy causality}},
  journal = {Phys. Rev. D},
  volume = {63},
  pages = {064006},
  year = {2001},
  doi = {10.1103/PhysRevD.63.064006},
  eprint = {hep-th/0008052},
  url = {https://arxiv.org/abs/hep-th/0008052},
  archivePrefix = {arXiv}
}

@article{EHTM872019,
  author = {Akiyama, K. and others},
  title = {{First M87 Event Horizon Telescope results. VI. The shadow and mass of the central black hole}},
  journal = {Astrophys. J. Lett.},
  volume = {875},
  pages = {L6},
  year = {2019},
  eprint = {1906.11243},
  url = {https://arxiv.org/abs/1906.11243},
  archivePrefix = {arXiv}
}

@article{EHTSgr2022,
  author = {Akiyama, K. and others},
  title = {{First Sagittarius A* Event Horizon Telescope results. VI. Testing the black hole metric}},
  journal = {Astrophys. J. Lett.},
  volume = {930},
  pages = {L17},
  year = {2022},
  eprint = {2311.09484},
  url = {https://arxiv.org/abs/2311.09484},
  archivePrefix = {arXiv}
}

@article{Kocherlakota2021,
  author = {Kocherlakota, P. and Rezzolla, L. and Falcke, H. and others},
  title = {{Constraints on black-hole charges with the 2017 EHT observations of M87*}},
  journal = {Phys. Rev. D},
  volume = {103},
  pages = {104047},
  year = {2021},
  doi = {10.1103/PhysRevD.103.104047},
  eprint = {2105.09343},
  url = {https://arxiv.org/abs/2105.09343},
  archivePrefix = {arXiv}
}

@article{XuEHT2023,
  author = {Xu, R. and Liang, D. and Shao, L.},
  title = {{Bumblebee black holes in light of Event Horizon Telescope observations}},
  journal = {Astrophys. J.},
  volume = {945},
  pages = {148},
  year = {2023},
  doi = {10.3847/1538-4357/acbdfb},
  eprint = {2302.05671},
  url = {https://arxiv.org/abs/2302.05671},
  archivePrefix = {arXiv}
}

@article{PerlickTsupko2022,
  author = {Perlick, V. and Tsupko, O. Yu.},
  title = {{Calculating black hole shadows: Review of analytical studies}},
  journal = {Phys. Rept.},
  volume = {947},
  pages = {1--39},
  year = {2022},
  doi = {10.1016/j.physrep.2021.10.004},
  eprint = {2105.07101},
  url = {https://arxiv.org/abs/2105.07101},
  archivePrefix = {arXiv}
}

@article{GRAVITY2024,
  author = {Abd El Dayem, K. and others},
  title = {{Improving constraints on the extended mass distribution in the Galactic center with stellar orbits}},
  journal = {Astron. Astrophys.},
  volume = {692},
  pages = {A242},
  year = {2024},
  eprint = {2409.12261},
  url = {https://arxiv.org/abs/2409.12261},
  archivePrefix = {arXiv}
}

@article{FiengaMinazzoli2024,
  author = {Fienga, A. and Minazzoli, O.},
  title = {{Testing theories of gravity with planetary ephemerides}},
  journal = {Living Rev. Rel.},
  volume = {27},
  pages = {1},
  year = {2024},
  doi = {10.1007/s41114-023-00047-0},
  eprint = {2303.01821},
  url = {https://arxiv.org/abs/2303.01821},
  archivePrefix = {arXiv}
}

@article{Filho:2023etf,
    author = "Filho, A. A. Ara{\'u}jo and Hassanabadi, H. and Heidari, N. and Kriz, J. and Zare, S.",
    title = "{Gravitational traces of bumblebee gravity in metric{\textendash}affine formalism}",
    eprint = "2305.18871",
    archivePrefix = "arXiv",
    primaryClass = "gr-qc",
    doi = "10.1088/1361-6382/ad1712",
    journal = "Class. Quant. Grav.",
    volume = "41",
    number = "5",
    pages = "055003",
    year = "2024"
}

@article{AraujoFilho:2025zaj,
    author = "Ara{\'u}jo Filho, A. A. and Heidari, N. and Lobo, Iarley P. and Bezerra, V. B.",
    title = "{Gravitational aspects of a new bumblebee black hole}",
    eprint = "2511.12839",
    archivePrefix = "arXiv",
    primaryClass = "gr-qc",
    doi = "10.1016/j.aop.2026.170469",
    journal = "Annals Phys.",
    volume = "489",
    pages = "170469",
    year = "2026"
}

@article{Filho:2024isd,
    author = "Filho, A. A. Ara{\'u}jo and Nascimento, J. R. and Petrov, A. Yu. and Porf{\'\i}rio, P. J.",
    title = "{Gravitational lensing by a Lorentz-violating black hole}",
    eprint = "2404.04176",
    archivePrefix = "arXiv",
    primaryClass = "gr-qc",
    doi = "10.1140/epjp/s13360-025-07040-2",
    journal = "Eur. Phys. J. Plus",
    volume = "140",
    number = "11",
    pages = "1117",
    year = "2025"
}

@article{AraujoFilho:2025hkm,
    author = "Ara{\'u}jo Filho, A. A.",
    title = "{How does non-metricity affect particle creation and evaporation in bumblebee gravity?}",
    eprint = "2501.00927",
    archivePrefix = "arXiv",
    primaryClass = "gr-qc",
    doi = "10.1088/1475-7516/2025/06/026",
    journal = "JCAP",
    volume = "06",
    pages = "026",
    year = "2025",
    note = "[Erratum: JCAP 02, E01 (2026)]"
}

@article{Lambiase:2023zeo,
    author = "Lambiase, Gaetano and Mastrototaro, Leonardo and Pantig, Reggie C. and Ovgun, Ali",
    title = "{Probing Schwarzschild-like black holes in metric-affine bumblebee gravity with accretion disk, deflection angle, greybody bounds, and neutrino propagation}",
    eprint = "2309.13594",
    archivePrefix = "arXiv",
    primaryClass = "gr-qc",
    doi = "10.1088/1475-7516/2023/12/026",
    journal = "JCAP",
    volume = "12",
    pages = "026",
    year = "2023"
}

@article{Heidari:2024bvd,
    author = "Heidari, N. and Macedo, Caio F. B. and Filho, A. A. Ara{\'u}jo and Hassanabadi, H.",
    title = "{Scattering effects of bumblebee gravity in metric-affine formalism}",
    eprint = "2407.05321",
    archivePrefix = "arXiv",
    primaryClass = "gr-qc",
    doi = "10.1140/epjc/s10052-024-13580-1",
    journal = "Eur. Phys. J. C",
    volume = "84",
    number = "11",
    pages = "1221",
    year = "2024"
}

@article{al2026shadow,
  title={Shadow-blind but lensing-sensitive: observational signatures of black holes in traceless metric-affine bumblebee gravity},
  author={Al-Badawi, Ahmad and Ahmed, Faizuddin and Sakall{\i}, Izzet},
  journal={Available at SSRN 7547938},
  year={2026}
}

@article{Breton2002Geodesics,
  author = {Bret{\'o}n, Nora},
  title = {Geodesic structure of the {Born--Infeld} black hole},
  journal = {Classical and Quantum Gravity},
  volume = {19},
  number = {4},
  pages = {601--612},
  year = {2002},
  doi = {10.1088/0264-9381/19/4/301},
  url = {https://doi.org/10.1088/0264-9381/19/4/301}
}

@article{Pugliese2011Circular,
  author = {Pugliese, Daniela and Quevedo, Hernando and Ruffini, Remo},
  title = {Circular motion of neutral test particles in {Reissner--Nordstr{\"o}m} spacetime},
  journal = {Physical Review D},
  volume = {83},
  pages = {024021},
  year = {2011},
  doi = {10.1103/PhysRevD.83.024021},
  eprint = {1012.5411},
  archivePrefix = {arXiv},
  primaryClass = {astro-ph.HE},
  url = {https://arxiv.org/abs/1012.5411}
}

@article{Cardoso2009Lyapunov,
  author = {Cardoso, Vitor and Miranda, Alex S. and Berti, Emanuele and Witek, Helvi and Zanchin, Vilson T.},
  title = {Geodesic stability, {Lyapunov} exponents, and quasinormal modes},
  journal = {Physical Review D},
  volume = {79},
  pages = {064016},
  year = {2009},
  doi = {10.1103/PhysRevD.79.064016},
  eprint = {0812.1806},
  archivePrefix = {arXiv},
  primaryClass = {hep-th},
  url = {https://arxiv.org/abs/0812.1806}
}

@article{Levin2008Periodic,
  author = {Levin, Janna and Perez-Giz, Gabe},
  title = {A periodic table for black hole orbits},
  journal = {Physical Review D},
  volume = {77},
  pages = {103005},
  year = {2008},
  doi = {10.1103/PhysRevD.77.103005},
  eprint = {0802.0459},
  archivePrefix = {arXiv},
  primaryClass = {gr-qc},
  url = {https://arxiv.org/abs/0802.0459}
}

@article{Schmidt2002Celestial,
  author = {Schmidt, Wolfram},
  title = {Celestial mechanics in {Kerr} spacetime},
  journal = {Classical and Quantum Gravity},
  volume = {19},
  number = {10},
  pages = {2743--2764},
  year = {2002},
  doi = {10.1088/0264-9381/19/10/314},
  eprint = {gr-qc/0202090},
  archivePrefix = {arXiv},
  url = {https://arxiv.org/abs/gr-qc/0202090}
}

@article{Fujita2009Bound,
  author = {Fujita, Ryuichi and Hikida, Wataru},
  title = {Analytical solutions of bound timelike geodesic orbits in {Kerr} spacetime},
  journal = {Classical and Quantum Gravity},
  volume = {26},
  pages = {135002},
  year = {2009},
  doi = {10.1088/0264-9381/26/13/135002},
  eprint = {0906.1420},
  archivePrefix = {arXiv},
  primaryClass = {gr-qc},
  url = {https://arxiv.org/abs/0906.1420}
}

@article{Claudel2001Photon,
  author = {Claudel, Clarissa-Marie and Virbhadra, K. S. and Ellis, G. F. R.},
  title = {The geometry of photon surfaces},
  journal = {Journal of Mathematical Physics},
  volume = {42},
  pages = {818--838},
  year = {2001},
  doi = {10.1063/1.1308507},
  eprint = {gr-qc/0005050},
  archivePrefix = {arXiv},
  url = {https://arxiv.org/abs/gr-qc/0005050}
}

@article{Bozza2002Strong,
  author = {Bozza, Valerio},
  title = {Gravitational lensing in the strong field limit},
  journal = {Physical Review D},
  volume = {66},
  pages = {103001},
  year = {2002},
  doi = {10.1103/PhysRevD.66.103001},
  eprint = {gr-qc/0208075},
  archivePrefix = {arXiv},
  url = {https://arxiv.org/abs/gr-qc/0208075}
}

@article{Gibbons2008GaussBonnet,
  author = {Gibbons, G. W. and Werner, M. C.},
  title = {Applications of the {Gauss--Bonnet} theorem to gravitational lensing},
  journal = {Classical and Quantum Gravity},
  volume = {25},
  pages = {235009},
  year = {2008},
  doi = {10.1088/0264-9381/25/23/235009},
  eprint = {0807.0854},
  archivePrefix = {arXiv},
  primaryClass = {gr-qc},
  url = {https://arxiv.org/abs/0807.0854}
}

@article{Casana2018Bumblebee,
  author = {Casana, R. and Cavalcante, A. and Poulis, F. P. and Santos, E. B.},
  title = {Exact {Schwarzschild}-like solution in a bumblebee gravity model},
  journal = {Physical Review D},
  volume = {97},
  pages = {104001},
  year = {2018},
  doi = {10.1103/PhysRevD.97.104001},
  eprint = {1711.02273},
  archivePrefix = {arXiv},
  primaryClass = {gr-qc},
  url = {https://arxiv.org/abs/1711.02273}
}

@article{Novello2000NLED,
  author = {Novello, M. and De Lorenci, V. A. and Salim, J. M. and Klippert, R.},
  title = {Geometrical aspects of light propagation in nonlinear electrodynamics},
  journal = {Physical Review D},
  volume = {61},
  pages = {045001},
  year = {2000},
  doi = {10.1103/PhysRevD.61.045001},
  eprint = {gr-qc/9911085},
  archivePrefix = {arXiv},
  url = {https://arxiv.org/abs/gr-qc/9911085}
}

\end{document}